\documentclass{ajour}
\def \D {\mbox{D}}
\def \d {\mbox{d}}
\def \tl{\tilde}
\def \div {\mbox{div}\,}
\def \rd {\displaystyle{\cdot}}
\def \c {\mbox{curl}\,}

\def \ts {\textstyle}  
\def\be {\begin{equation}}
\def\ee {\end{equation}}

\def\la {\langle}
\def\ra {\rangle}
\def\p  {\partial}
\def\bi {\bibitem}
\def\hs { \, -\, }
\def\case#1/#2{\textstyle\frac{#1}{#2} }
\def\ie{{\it { i.e. }}}

\def\boxit#1{\vbox{\hrule\hbox{\vrule\kern3pt{\vbox{\kern3pt#1\kern3pt}
\kern3pt\vrule}\hrule}}}
\begin{document}
\journame{Annals of Physics}
\articlenumber{}
\yearofpublication{1999}
\received{10 August 1999}

\authorrunninghead{Gebbie, Dunsby and Ellis}
\titlerunninghead{1+3 Covariant Cosmic Microwave Background anisotropies II}

\title{1+3 Covariant Cosmic Microwave Background anisotropies II: \\
The almost--Friedmann Lema\^{\i}tre model}
 
\author{Tim Gebbie$^1$, Peter K. S. Dunsby$^1$ and G. F. R. Ellis$^{1,2}$}

\affil{$1$ Department of Mathematics and Applied Mathematics\\
University of Cape Town, Cape Town 7701, South Africa.}

\affil{$2$ Astronomy Unit, Queen Mary \& Westfield College,\\
University of London, Mile End Road, London E1 4NS, United Kingdom}

\abstract{
This is the second of a series of papers extending the 1+3 Covariant 
and Gauge-Invariant treatment of kinetic theory to an
examination of Cosmic Microwave Background temperature anisotropies 
arising from inhomogeneities in the early universe. The first paper 
(Paper I) dealt with algebraic issues, representing anisotropies in 
a covariant and gauge invariant way by means of Projected Symmetric and 
Trace-Free tensors. 

Here we derive the mode form of the Integrated Boltzmann Equations, first, 
giving a covariant version of the standard derivation 
using the mode recursion relations, second, demonstrating the link to the 
the Multipole Divergence Equations and finally various analytic 
ways of solving the resulting equations are discussed.
  
A general integral form of solution is obtained for the equations 
with Thomson scattering. The covariant Friedmann-Lema\^{\i}tre multipole 
form of the transport equations are found near tight-coupling using 
the covariant and gauge-invariant generalization of the Peebles 
and Yu expansion in Thompson scattering
time. The dispersion relations and damping scale are then obtained from 
the covariant approach. The equations are integrated to give the 
covariant and gauge-invariant equivalent of the canonical scalar 
sourced anisotropies in the $K=0$ (flat background) case. 

We carry out a simple treatment of the matter dominated free-streaming 
projection, slow decoupling, and tight-coupling cases in covariant and 
gauge-invariant theory, with the aim both giving a unified transparent 
derivation of this range of results and clarifying the formal
connection between the usual approaches (for example Hu \& Sugiyama)
and the covariant and gauge-invariant like treatments for scalar 
perturbations (for example of Challinor \& Lasenby).
}

\section{Introduction} 
The present series of papers (Paper I \cite{GE}) establishes the 1+3 
covariant kinetic theory formalism of Ellis, Treciokas and Matravers 
\cite{ETMa,ETMb} in a form that makes possible an investigation
of the Cosmic Microwave Background anisotropies in the non-local context 
of emission of radiation at the surface of last scattering in the early 
universe, and its reception here and now (the `Sachs-Wolfe effect'
and its later extensions). These papers aims to provide the link 
between the exact (non-linear) theory \cite{MGE} and  the linearised 
threading formalism (this paper and to a lesser degree \cite{cl}), 
to the linearised foliation formalism, based on Bardeens' approach
\cite{Bardeen} to cosmological perturbations \cite{HS95a}.
 
The Ellis, Treciokas and Matravers papers introduced 
a covariant approach to kinetic theory based on a 1+3 covariant 
representation of Cosmic Microwave Background anisotropies in terms of Projected Symmetric 
Trace-Free (PSTF) tensors orthogonal to a preferred time-like vector 
field $u^a$ \cite{T1,P}. The benefits of the approach have been 
briefly summarized in Paper I \cite{GE} (see also Challinor \& Lasenby 
\cite{cl,cl2}) -- we seek clarity by providing a direct formal 
derivation of the standard results as well as providing the 
background necessary to include the non-perturbative
corrections discussed by Maartens, Gebbie and Ellis \cite{MGE}. 
In essence, we provide (a) clear definitions of the variables used, 
(b) 1+3 Covariant and Gauge-Invariant (CGI) variables and equations, 
(c) a sound basis from which to proceed to non-linear calculations 
(as introduced in \cite{MGE}), and (d) the possibility of using any 
desired coordinate and tetrad system for evaluating the variables and 
solving the equations in specific cases (because the general equations 
and variables used are covariant). 

This approach has been used in a previous series of papers 
\cite{MES,SME,MES1} to look at the local generation of anisotropies 
in freely-propagating radiation caused by anisotropies and 
inhomogeneities in any universe model, providing  a proof that 
near-spatial homogeneity in a region $U$ follows from radiation 
near-isotropy in that region (an `Almost-Ehlers, Geren and Sachs'
theorem, generalizing the important paper by Ehlers, Geren, 
and Sachs \cite{EGS}). By contrast, the present series looks at
non-local (integrated) anisotropy effects in the context of the 
standard model of cosmology -- given the observational justification
provided by the almost-Ehlers, Geren and Sachs theorem \cite{SAG}.
 
Paper I \cite{GE} of the present series of papers \cite{GE}) dealt with the
CGI irreducible representation of cosmic background radiation 
anisotropies by PSTF tensors, and their relation to observable 
quantities (specifically, the angular correlation functions), first 
within a general framework and then and specialized to almost-Friedmann-Lema\^{\i}tre
universe models \footnote{In order to be clear on the use of these 
names, Robertson-Walker refers to the Robertson-Walker {\it geometry} whatever 
the dynamics, while Friedmann-Lema\^{\i}tre refers to such a geometry which additionally 
obeys the Friedmann-Lema\^{\i}tre {\it dynamics} implied by imposing 
the Einstein Field Equations. An {\it almost-Friedmann-Lema\^{\i}tre universe} is a 
universe model governed by the Einstein field equations, whose
difference from a Friedmann-Lema\^{\i}tre\ universe is at most $O(\epsilon )\,$ in
terms of a small parameter $\epsilon $\thinspace \cite {BE}.} 
\cite{MES}, as well as dealing with multipole and mode expansions and 
the relation to the usual formalisms in the literature 
\cite{WS,GSS,HS95a,HS95b,EBW}. 

In this paper (Paper II), we use the CGI approach to study cosmic 
background radiation anisotropies in almost-Friedmann-Lema\^{\i}tre models in an analytic 
manner, by time-like integration of the almost-Friedmann-Lema\^{\i}tre differential 
relations. Our emphasis is on the canonical linearised model for 
cosmic background radiation anisotropies 
\cite{SW,W83,AS,S89,GSS,HS95a,HS95b,WHb}, systematically developing the CGI 
approach and providing a comprehensive and transparent
analytic link to the alternative analytic gauge-invariant treatments based on 
the Bardeen gauge-invariant  variables.  We develop these results both 
in mode and multipole form, emphasizing the different 
physical processes and assumptions and demonstrating how these are
dealt with in the CGI context. This requires the covariant mode form of 
the integrated Boltzmann equations, based on the recursion relations 
for almost-Friedmann-Lema\^{\i}tre mode functions \cite{GE}, enabling a direct mirroring 
of standard treatments based on Wilson's seminal paper 
\cite{W83,GSS,HS95a,HS95b,WHb}, except carried out in a CGI fashion, 
thus forming a sound basis for extension to non-linear effects.

Indeed one of the advantages of the CGI approach in the context of the 
generic multipole divergence equations is the ability to include
non-linear corrections to the almost-Friedmann-Lema\^{\i}tre treatment. Towards this end,
the relationship between the covariant mode formulations and the 
almost-Friedmann-Lema\^{\i}tre covariant multipole treatment are given with this in mind, 
based on the relations in Paper I \cite{GE}. A key point here is that 
there is nothing new about the linear formulation itself, however
recovering the standard analytic linear results from the 
the exact theory \cite{MGE} in a straightforward way, is new.

Moreover, the results presented here provide the foundation for 
a non-linear extension of this approach which is outlined in a 
paper by Maartens, Gebbie and Ellis \cite{MGE}. 

We emphasize that in our treatment, $\la \tau_{A_\ell}\tau^{A^\ell} \ra$ (the 
multipole form of the angular correlation function) is given for small
temperature anisotropies irrespective of the form of the 
geometry \cite{MGE}, making it the natural representation for the 
inclusion of non-linear dynamic effects, while the analysis for 
$|\tau_{\ell}|^2$ (the mode coefficient mean square) is specifically 
for almost-Friedmann-Lema\^{\i}tre models \cite{GE}. The non-linear extension of the
almost-Friedmann-Lema\^{\i}tre treatment given here will be based on the multipole-to-mode 
relations, leaving the use of mode expansion to the latest possible stage.

Our focus is on the era following spectral decoupling (near $500$ eV).      
A complex series of interactions take place at the various epochs of
the expansion of the universe. The kinetic equations developed in Paper
I and \cite{MGE} can represent almost any such interactions provided we use 
appropriate interaction (`collision') terms; the issue is how to obtain 
simplified models that are reasonably accurate in the various epochs. We
will consider only two kinds of representation here: namely 
\begin{enumerate} 
\item Thomson scattering, valid at late times when particle and photon 
numbers are conserved and the matter is non-relativistic (during
decoupling, an alternative approach is to use a visibility function).
 
\item An effective two-fluid approach, obtained by truncation of 
the hierarchy and valid at earlier times when strong interactions 
take place establishing equilibrium or close to equilibrium conditions 
between the components, i.e. when the interaction time-scale is much
less than the expansion time-scale; an alternative description is to use
a single imperfect fluid \cite{RT}. 

Both descriptions are valid when the matter is relativistic. 
The detailed form of interactions does not need to be represented 
in this case, because the state of the matter depends only on 
its equilibrium condition, characterized by its equation of state. 
\end{enumerate}
 
At some times either form is valid and they can then be
related to each other. We do not attempt here to give a description 
of earlier non-equilibrium eras when processes such as pair
production, nucleosynthesis, etc, occur, nor do we consider issues 
such as inflation and reheating after inflation, and the differences 
between the inflation sourced perturbations as opposed to
those based on other phase transitions. 
Thus our models will be valid after the end of any period of inflation 
that may have occurred and after strong non-equilibrium processes have 
ceased. During this era the processes which result from initial
fluctuations left over from earlier non-equilibrium epochs, determine the 
final cosmic background radiation anisotropy.

Specifically, we deal with four eras of interest. Going backwards in time 
from the present, they are, firstly, free streaming from last scattering to 
here and now, in the matter dominated almost-Friedmann-Lema\^{\i}tre context; secondly, slow 
decoupling during a matter dominated era, which is when the 
cosmic background radiation spectrum freezes out; thirdly, the late 
tight-coupling era after matter-radiation equality, during which 
structure formation begins; and fourthly, the
main tight coupling era after any inflationary epoch and before 
matter-radiation equality, during which acoustic modes occur 
in the tightly-coupled fluid, the initial matter perturbations having 
been seeded by earlier conditions (for example, inflation). We then
show how to put these CGI results together to determine the major 
features of the expected anisotropy spectrum. We develop sound models 
of the dominant effects in each of the eras we consider, but there
will always be a need for refinement of these models by taking into 
account further effects (in particular polarisation and the effect of 
neutrinos). 

Although the effect of the neutrinos is crucial, and can been subsumed into 
the gravitational variables. We do not provide nor discuss further the 
additional hierarchy of neutrino moment equations that would then need to be 
included. The exact massless neutrino evolution equations are given 
in a previous paper \cite{MGE} and it is easy to show that the resulting 
linearised equations are essentially the same as those for massless radiation 
without collisionals, hence in this paper we focus on the photon equations.

In more detail:
 
\begin{itemize}
\item {\it Free-Streaming} : We find the CGI integral solutions 
to the almost-Friedmann-Lema\^{\i}tre multipole divergence equations with no collision 
term, and use them to project the initial data from decoupling to 
the current sky. We explicitly do this for instant decoupling. 
Neither the Vishniac, Rees-Sciama, thermal Sunyaev-Zel'dovich nor 
lensing effects are considered here -- we will focus 
on the CGI model of the dominant processes, and these
further effects will introduce detailed modifications. However, a 
comprehensive understanding of such secondary higher order effects
relies on a derivation of the anisotropy effects given here.
 
\item {\it Slow-Decoupling} : Here we consider modification of the 
previous results when slow decoupling of the interactions due to 
Thomson scattering is taken into account. We consider the damped integral 
solution for slow recombination, and as an alternative description
modify the integral solutions appropriately with a visibility function 
carrying the functional dependence of the varying electron fraction, in 
a matter dominated context. Effectively, recombination is complete before 
the radiation decouples. This means that the surface of last scattering 
is found a little after the end of recombination \footnote{A note on 
nomenclature: By {\it decoupling} we have in mind the situation when the
interaction rate per particle becomes less that the expansion rate.
By {\it last scattering surface} we mean the surface upon which the diffusion 
scale is equal to the horizon scale, after which it is larger than that scale
and the free-streaming approximation is sufficient. The photons will decouple 
from the thermal plasma  near $0.2$ eV, and from the matter after 
recombination has effectively ended, near $0.3$ eV. Free-streaming is 
considered to be a good approximation from about $0.26$ eV.}. 
It is during this era that photon diffusion damping scale effects 
become important -- the damping scale is affected by the duration of 
this era. This will be investigated in the context of 
almost-Friedmann-Lema\^{\i}tre universes after matter radiation equality.
 
\item {\it Tight-coupling}: This is the key to the entire 
treatment. We give the CGI version of the tight-coupling 
approximation of Hu {\it et al} \cite{HS95a,HS95b,HWa2}. In the 
almost-Friedmann-Lema\^{\i}tre treatment, the slow decoupling and free-streaming era's will 
only `damp' and `project' the spectrum formed at the end of
tight-coupling onto the current sky \footnote{This useful feature of the 
almost-Friedmann-Lema\^{\i}tre models is due to the homogeneity and isotropy conditions in 
the background Friedmann-Lema\^{\i}tre universe and is not generic \cite{MGE}.}. 
 
We consider two different tight-coupling regimes.
\begin{itemize}
\item The {\it late tight coupling era} is separated, 
conceptually, from the early tight coupling era by matter-radiation 
equality, after which time the matter perturbations have
effectively decoupled and (CDM-based) structure formation begins. 
In this era, strong interactions have ceased and a Thomson scattering 
description can be used. We first carry out the near tight-coupling 
treatment of this era based on Peebles \& Yu \cite{PY,Kaiser}, 
and then reduce these equations to the covariant tight-coupling
equations equivalent to those of Hu {\it et al}
\cite{HS95a,HS95b,HWa2}. This provides the basis for understanding the
acoustic signatures in the temperature anisotropies within the CGI approach.
 
\item The {\it early tight coupling era} occurs between the matter 
and radiation eras (after strong interactions have ceased), when 
a Thomson scattering description will also be sufficient. This 
era is characterized by acoustic oscillations in a tightly coupled
fluid; for calculation convenience this can be represented as a single
dissipative fluid \cite{RT}, or for a slightly more sophisticated 
treatment by tightly coupled two-fluid models \cite{DBE92,dunsby}. 
We give a CGI derivation of the harmonic oscillator equation providing
the primary source terms in the standard model of Doppler peak
formation by acoustic oscillations.
\end{itemize}
\end{itemize}
 
We put these results together in sections 7 and 8, where the equations are 
integrated to give the CGI treatment in the $K=0$ (flat background)
case in terms of an integral solution. 

The primary sources of the temperature anisotropies (the acoustic and Doppler 
contributions near last scattering resulting in `Doppler Peaks' today) are 
demonstrated \footnote{following the analytic treatment of Hu \& Sugiyama 
\cite{HS95a,HS95b}}. This recovers the Sachs-Wolfe family of effects for 
flat background Robertson-Walker geometries, but derived from a CGI kinetic theory viewpoint 
as opposed to the photon-propagation description used in the original 
Sachs-Wolfe paper. 

The form of the angular correlation function is determined for the 
primary effects (although not given explicitly in terms of the 
matter power spectrum). The normalization to standard CDM (Adiabatic
CDM) is presented in terms of CGI variables. This demonstrates 
the basic effects in the CGI formalism, and links our approach to  
the standard literature, see for example \cite{WHb} and references 
therein, where further details of this era are given. 
 
It is important to note that the integrations considered here
are carried out along time-like curves, even though the cosmic
background radiation reaches us along null curves. These are 
alternative approaches that are equivalent in the context of 
linearisations about Friedmann-Lema\^{\i}tre models; differences will however occur 
if we include non-linear corrections. Briefly, the key point 
about cosmic background radiation integrations is that there are 
two ways in which to proceed:
Firstly a {\it null-cone integration}, following the radiation from last
scattering to the present day\footnote{which can be parametrized either
by a null cone parameter, a projected spatial coordinate, or a projected time
coordinate}, and secondly a {\it time-like integration}
along the matter flow lines (as here). In the latter case one is 
(at least implicitly) thinking of a small comoving box containing matter and 
radiation \cite{ERH} which is similar to all other small boxes at the 
same time, and where one has assumed that the radiation leaving is
exactly balanced by the radiation entering (from neighbouring boxes), 
whether in tight-coupling (when it is a local assumption) or in the 
free-streaming era (when it is a non-local assumption). In effect one
integrates behaviour in such a box in a small domain about our own
world-line from tight-coupling through decoupling to the present day;
to do this, one does not need to know about the behaviour of
null-geodesics (the integration is along time-like geodesics). Before 
decoupling the matter and radiation evolve as a unit while after they
need to be integrated separately (giving the corresponding transfer
functions in each case). Then this is related to observations by,
first, conceptually shifting copies of the small box at the time of
last scattering from our world line to all points where the past
null-cone intersects the surface of last scattering at that time; this
can be done because spatial homogeneity says that these boxes are
essentially the same (a Copernican assumption is used here, justified
by the almost-Ehlers-Geren-Sachs theorem \cite{MES,MES1}); second, by 
then relating distances on the last scattering surface to observed 
angles by using the area distance relation, relating physical 
distances at last scattering to angular size in the sky. 
 
The next paper in the series, Paper III \cite{DGE}, deals with the 
explicit relationship between null-cone and time-like integrations.
Further papers will look at non-linear extensions of the results given
here.  
\section{Linearised Covariant Mode Equations}
To study details of cosmic background radiation anisotropy generation
we need both a spatial Fourier decomposition, defining wavelengths 
of perturbations, together with the angular harmonic decomposition 
relating anisotropies to observed angles in the sky. The CGI versions 
of both decompositions were given in Paper I \cite{GE}, giving the 
relationship between the mode and multipole variables. 

The dynamic relations obeyed by these quantities, determining 
the cosmic background radiation spatial and angular structure, 
can be obtained from the Boltzmann equation in two ways: via 
multipole divergence equations or via the integrated Boltzmann 
equations. In each case the general equation
needs to be mode-analyzed to obtain the spatial structure.

In the first case, the almost-Friedmann-Lema\^{\i}tre multipole divergence equations are 
obtained by systematically linearising the non-linear multipole 
divergence equations for small temperature anisotropies given in 
Maartens, Gebbie and Ellis \cite{MGE}. The mode 
form of these equations (\ref{li_ibe_1}-\ref{li_ibe_l}) \cite{MGE}
can then be obtained by mode analysis, see
(\ref{PSTF-tau1}-\ref{PSTF-tau2}) below.

By contrast, the more common procedure is to directly construct the mode 
form of the integrated Boltzmann equations from the linearised
integrated Boltzmann equations by substituting (\ref{dEdv}) 
into (\ref{ibe}) and integrating over the energy shell with 
respect to $E^2 dE$ (For a more detailed relativistic kinetic 
theory description of these equations see \cite{MGE} and \cite{cl2}):
\begin{eqnarray}
\int_0^{\infty} E^2 d E \Big[ E (u^a+e^a) 
\nabla_a f - (\frac{1}{3}\Theta + A_a e^a + \sigma_{ab} e^a e^b )
E^2 \frac{\partial f}{\partial E} \Big]
\approx \int_0^{\infty} E^2 dE C[f]\;. 
\label{ibe-aflrw}
\end{eqnarray}
Upon using the CGI definition of directional
bolometric brightness \cite{MGE}:                                 
\begin{equation}
T(x)\left[1+\tau(x,e)\right]=\left[{4\pi\over r}\int E^3       
f(x^i,E,e^a)\d E \right]^{1/4}\,, \label{r28}
\end{equation}
the covariant equivalent of the standard formulae given in the 
Wilson-Silk approach \cite{W83} can be found, where as usual 
the partial energy derivative is removed by a integration by parts 
and application of the regularity conditions. This is the approach 
we develop now, showing later the relation to the approach based 
on the multipole divergence equations. 
\subsection{The Mode Equations} 
The optical depth $\kappa$ is given in terms
of the Thompson scattering cross-section $\sigma_T$, the 
number density of electrons $n_e$, and the fraction of electrons 
ionized $x_e$:
\begin{eqnarray}
 \kappa(t_0,t) = \int_{t}^{t_0} \sigma_T n_e(t') 
 x_e(t') d t' = \int \dot{\kappa} d t'\;,
\end{eqnarray}
where $\eta$ is the conformal time ($dt = a d\eta$)
in the Friedmann-Lema\^{\i}tre background.
Starting with the almost-Friedmann-Lema\^{\i}tre integrated Boltzmann equations
(\ref{ibe-aflrw}) for the temperature anisotropies (\ref{r28}) with 
the collisional term for isotropic scattering in terms of the optical 
depth, and the expansion replaced by substituting from 
the radiation energy conservation equation (\ref{EFE-mc},\ref{mono_temp}), 
we obtain the linearised integrated Boltzmann equations:
\begin{eqnarray}
 - {\dot \tau} \approx e^a \D_a \tau - \ts{1 \over 3} \D_a \tau^a 
 + ({\D_a \ln T} + A_a) e^a + \sigma_{ab} e^a e^b 
 - {\dot \kappa (e^a v_a^B - \tau)}\,. \label{li_ibe}
\end{eqnarray}
We take a mode expansion (see Paper I \cite{GE}) for $\tau$ (the temperature 
anisotropy), $A_a$ (the acceleration), 
$\D_a (\ln T)$ (the spatial-temperature perturbation), 
$\sigma_{ab}$ (the shear), and the gradient of the radiation 
dipole, $\D^a \tau_a$, based on solutions $Q(x)$ of the scalar
Helmholtz equation: 
\begin{equation}
\D^a \D_a Q=- {k^2 \over a^2} Q 
\end{equation}
in the (background) space sections, where the $Q$'s are
covariantly constant scalar functions (i.e. to linear order 
$\dot{Q} \approx 0$) corresponding to a wavenumber $k$. The 
physical wavenumber is defined by $k_{phys}(t)=k/a(t)$.   
These functions define tensors $Q_{A_{\ell}}(k^\nu ,x^i)$ that are 
PSTF (in the case of scalar perturbations they are given by 
the PSTF covariant derivatives of the eigenfunctions 
$Q$) and so allow us to define the {\it mode functions} \cite{GE}: 
\begin{eqnarray}
Q_{A_{\ell}}=\left({- {k \over a}}\right)^{-{\ell}}\D_{\la A_{\ell}
\ra}Q\,~~~\mbox{and}~~~  
G_{\ell}[Q]= O^{A_{\ell}} Q_{A_{\ell}}\;,  
\label{SQdef-modef}
\end{eqnarray}
where $O^{A_{\ell}}=e^{\la A_{\ell} \ra}$ is the trace-free part 
of $e^{A_{\ell}}$. We can expand any given function $f(x,e)$ in 
terms of these functions, thus for scalar perturbations (see
(\ref{scalar-potential1} -\ref{scalar-potential2}),
the mode coefficients of the temperature anisotropy are 
constructed as follows: 
\begin{eqnarray}
\tau (x,e)=\sum_{\ell=1}^\infty \,\sum_k\,\tau _{\ell}(t,k)\,G_{\ell}[Q]\;.
\label{expn} 
\end{eqnarray}
Note that $\tau_0$ is identically zero, because (\ref{r28}) 
defines the temperature $T$ gauge-invariantly as the all-sky average in the 
real (lumpy) universe (it is not defined in terms of a background model). 
We can then write:
\begin{eqnarray}
\D^a \tau_a = \sum_k {k \over a} \tau_1(k,t) Q\,,~~~~ 
\D_a (\ln T) + A_a= \sum_k [\ts{k \over a} \delta T(k,t)  + A(k,t) ] Q_a\,,
\end{eqnarray}
and
\begin{eqnarray}
\sigma_{ab} = \sum_k \sigma(k,t) Q_{ab}\,.
\end{eqnarray}
The mode coefficient of the velocity of the baryons $v_B$ relative 
to the reference frame $u^a$ is given by 
$$v_a^B = \sum_k v_B(k,t) Q_a\,.
$$
The equations are linearised at $O(v^2)$,
$O(\epsilon v)$ and $O(\epsilon^2)$ \cite{MGE}, so we can
use, for example, ${u}_B^a \approx u^a + v_B^a$ to give the baryon 
relative velocity.

The equations here are gauge-invariant (given relative to a 
unique physically-based choice of the 4-velocity vector $u^a$) and valid
for any choice of mode functions, but the detailed result of their 
translation back into the spacetime representation 
$\tau (x,e)$ via (\ref{expn}) will depend on the harmonic functions 
$Q(x^i)$ chosen\footnote{
In effect there are two major choices, namely plane wave solutions and
spherical solutions; the former occur naturally in galaxy formation studies
and the latter in observational analysis, so the relation between the two
(see Paper I \cite{GE}) is a central feature of analyzing 
null-cone observations.}.
 
We substitute these expansions into the linearised integrated 
Boltzmann equations and then use the recursion relation 
\cite{W83,GSS,HS95a,GE} for mode functions $G_{\ell}[Q](e^a,x^i)$ 
with wave-number $k$ : 
\begin{eqnarray}
e^a \D_a[G_{\ell}[Q]]=\frac{k}{a} \Big[ {{\frac{{\ell}^2}
{(2{\ell}+1)(2{\ell}-1)}}\left( {1-{\frac{K}{k^2}}(\ell^2-1)}\right) 
G_{\ell-1}[Q] - G_{\ell+1}[Q]}\Big]\;,
\label{mode_recursion}
\end{eqnarray}
where $K$ is the curvature constant of the background space sections. 

With the mode decompositions of each term in (\ref{li_ibe}) 
for each wave number\footnote{
There should be a summation over wave numbers in the following equations.
However we follow the established custom in omitting this summation and any
explicit reference to the assumed wave number $k$.}, on using the recursion 
relation (\ref{mode_recursion}) and separating out the different harmonic 
components,  the almost-Friedmann-Lema\^{\i}tre mode equations are found from (3)\footnote{
Note that these equations are valid for any choice of $Q$, including
both spherical and plane wave harmonic functions.}: 
\begin{eqnarray}
-\dot{\tau}_{\ell} &\approx & \frac{k}{a} \left[ {{\frac{(\ell+1)^2}{(2
\ell+3)(2\ell+1)}}\left( 
{1-{\frac K{k^2}}((\ell+1)^2-1)}\right) \tau _{\ell+1}-\tau _{\ell-1}}\right] 
+\dot{\kappa}\tau _{\ell}\;, \ell \geq 3, \label{mibec-1} \\
-\dot{\tau}_2 &\approx& \frac{k}{a} \left[ {\frac 9{35}\left( {1-{\frac{8K}{k^2}}}
\right) \tau _3-\tau _1}\right]+ [\sigma] +\dot{\kappa}\tau _2,
\label{mibec-2} \\
-\dot{\tau}_1 &\approx & \frac{k}{a} \left[ {\frac 4{15}\left( 
{1-{\frac{3K}{k^2}}}\right) \tau _2}\right] 
+ [\ts{k \over a} \delta T + A]  - \dot{\kappa}(v_B-\tau _1)\;.
\label{mibec-3} 
\end{eqnarray}
The above equations demonstrate the up and down cascading effect 
whereby lower order terms generate anisotropies in the higher order
terms, and vice versa, in a wavelength-dependent way; curvature 
affects the down-cascade but not the up one. These equations can be 
compared to the equations of Hu \& Sugiyama, in particular 
(eqn. 6, p. 2601) \cite{HS95b}. They are identical if we use the 
Newtonian frame (discussed in the following sections), and so have 
the same physical content, but are more general since they are 
valid with respect to a general frame $u^a$. 
\subsection{From Multipole Equations to Mode Equations}
The relationship between the angular harmonic and mode expansions 
are given in Paper I \cite{GE}. We start by writing the CGI harmonic 
coefficients $\tau_{A_\ell}$ in terms of the mode functions 
(\ref{SQdef-modef}):
\be
 \tau_{A_\ell} = \sum_k \tau_{\ell}(k,t) Q_{A_{\ell}} \approx \sum_k
\tau_{\ell}(t,k) (-{k \over a})^{-\ell} \D_{\la A_{\ell} \ra} Q\;.
\label{multi-mode}
\ee
Then the angular harmonic expansion for $f$ becomes the mode expansion 
(\ref{expn}). On taking the multipole integrals of $f$ as in 
Paper I \cite{GE}, they too are then mode-expanded by
(\ref{multi-mode}); so the linearised divergence relations 
for these multipoles given in
\cite{MGE}  become mode equations, equivalent to the almost-Friedmann-Lema\^{\i}tre mode 
equations (\ref{mibec-1}-\ref{mibec-3}) derived above.

In detail: The almost-Friedmann-Lema\^{\i}tre multipole divergence equations are \cite{MGE}: 
\begin{eqnarray}
&\;& -\left( { {\dot{T} \over T}+ \case1/3 \Theta }\right) \simeq +\case1/3
\D^c\tau _c, \label{mono_temp} \\
(-\dot{\tau}_a) &\simeq& \D_a \ln T +A_a+\case2/5\D^c\tau_{ac}
-\sigma _T n_e (v^B_a-\tau _a), \label{li_ibe_1} \\
(-\dot{\tau}_{ab}) &\simeq& \sigma _{ab}+ \D_{\la a} \tau _{b \ra}
+\case3/7 \D^c\tau_{abc}+\sigma _T n_e \tau _{ab}, \label{li_ibe_2} \\
(-\dot{\tau}_{A_{\ell}}) &\simeq& \D_{\la a_{\ell}} \tau _{A_{\ell-1} \ra}
+{\frac{(\ell+1)}{(2\ell+3)}} \D^c\tau _{A_{\ell} c}
+\sigma _Tn_e\tau _{A_{\ell}}\;. \label{li_ibe_l}
\end{eqnarray}
Now the following identities are used (dropping the k-summation):
\begin{eqnarray}
O^{A_{\ell}} \D^c \tau_{A_{\ell} c} &\approx& \tau_{\ell+1} {(\ell+1) 
\over (2 \ell+1)} 
\left({ + {k \over a} }\right) \left[{ 1 - {K \over k^2} \ell(\ell+2)} \right]
 O^{A_{\ell}} Q_{A_{\ell}}\,, \label{PSTF-tau1}\\
 O^{A_{\ell}} \D_{\la a_{\ell}} \tau_{ A_{\ell-1} \ra} &\approx& \tau_{\ell-1}
 \left({ - {k \over a}} \right) O^{A_{\ell}} Q_{A_{\ell}}\;, 
\label{PSTF-tau2}
\end{eqnarray}
where the first relation arises from the use of the identity           
\footnote{This has also been derived by Challinor and Lasenby \cite{cl2} 
and is found from the PSTF recursion relations and the generalized 
Helmholtz equation (which is in turn found from the constant
curvature Ricci identity) \cite{GE}.}:
\[
\D^c \D_{\la c A_{\ell} \ra} Q = {(\ell+1) \over (2 \ell +1)} \left( {
- {k^2 \over a^2}} \right) \left[ {1 - {K \over k^2} \ell(\ell+2)} 
\right] \D_{\la A_{\ell} \ra} Q\;.
\]                                                                    
These are substituted directly into the multipole equations after taking a
mode expansion of those equations and then dropping the k-space
summation, leading again to the equations (\ref{mibec-3}). The point to 
note is that while one does not explicitly need the multipole equations 
in order to find the almost-Friedmann-Lema\^{\i}tre mode equations (which can be derived
from the linearised integrated Boltzmann equations as shown above), 
in order to examine non-linear effects one can obtain the necessary 
equations by proceeding as here from the non-linear multipole
divergence equations, to obtain higher approximations of the mode 
equations and the mode-mode couplings.
\subsection{The Einstein Equations}
The key quantities which link the radiation evolution through to the
matter in the spacetime geometry are the shear $\sigma_{ab} =
u_{\la a;b \ra}$, the acceleration $A_a=u_{a;b} u^b$ and the expansion  
$\Theta$. These couple the multipole divergence equations to the 
Einstein field equations (which are given in Appendix G, 
see (\ref{h-constraint})-(\ref{Friedmann-Lemaitre_R})). 

The shear and acceleration, are related to the electric part of
the Weyl tensor $E_{\la ab\ra}$, the anisotropic pressure 
$\pi_{\la ab \ra}$ and matter spatial gradients (see
(\ref{h-constraint}) -(\ref{Friedmann-Lemaitre_R})) while the CGI spatial 
gradient of the expansion is linked to the divergence of the shear,
heat flux vector $q_a$ and the vorticity vector $\omega_a$:
\begin{eqnarray}
- \case1/2 (\rho+p) \sigma_{ab} &\approx& (\dot{E}_{ab} + \case1/2
\dot{\pi}_{ab} ) + 3 H ( E_{ab} + \case1/2 \pi_{ab} ) - H \pi_{ab} 
-\left\{ {\case1/2 \D_{\la a} q_{b \ra} } \right\}, ~~\label{s-1}\\
(\rho + p) A_a &\approx& - \D_a p - \D^b \pi_{ab} - \left\{ {
\dot{q}_a + 4 H q_a} \right\}, \label{a-1}\\
\case1/3 \D_a \Theta &\approx& \case1/3 (\D^b \sigma_{ab}) 
- \left\{ {\case1/3 q_a + \c \omega_a} \right\} \label{exp-1}\;. 
\end{eqnarray} 
These equations are valid for general almost-Friedmann-Lema\^{\i}tre perturbations.
In the restricted case of scalar perturbations, we set the vorticity to 
zero\footnote{We can obtain scalar equations even when the vorticity is 
not zero, by taking the total divergence of these equations; we will
not pursue that case here.} and non-zero quantities can be written in 
terms of potentials \cite{stewart}. In particular
\begin{eqnarray}
E_{ab} &\approx& \D_{\la a} \D_{b \ra} \Phi_E  
= \case1/2 \D_{\la a} \D_{b \ra}( \Phi_A - \Phi_H), \\
\pi_{ab} &\approx& \D_{\la a} \D_{b \ra} \Phi_{\pi} 
= - \D_{\la a} \D_{b \ra} (\Phi_H + \Phi_A)\;,
\end{eqnarray}
where the potentials $\Phi_A$ and $\Phi_H$ are analogous to  
the GI potentials used by Bardeen \cite{BDE92}. The following useful 
combinations can be found :
\begin{eqnarray}
E_{ab} - \case1/2 \pi_{ab} \approx \D_{\la a} \D_{b \ra} \Phi_A,~~\mbox{and}~~
E_{ab} + \case1/2 \pi_{ab} \approx - \D_{\la a} \D_{b \ra} \Phi_H.
\label{B_potentials}
\end{eqnarray}
Using the Einstein field equations, the total flux, $q_a$, can also 
be expressed covariantly in terms of these potentials:
\begin{eqnarray}
H q_a &\approx& \D^b \D_{\la a} \D_{b \ra} \Phi_H 
- \case1/3 \D_a \rho, \label{flux_phi} \\
H \D_{\la a} q_{b \ra} &\approx& \D_{\la a} \D_{b \ra} 
\left[ {\case2/3 ( \D^2 \Phi_H + (\rho - 3 H^2) \Phi_H)
 - \case1/3 \rho} \right]\;. \label{PSTF_flux_phi} 
\end{eqnarray}
This then allows us to write the shear and acceleration in terms of the 
scalar potentials and perturbation variables:
\begin{eqnarray}
\case1/2 (\rho + p) \sigma_{ab} &\approx& (\D_{\la a} \D_{b \ra}
\Phi_H)^{\dot{}} + 3 H \D_{\la a} \D_{b \ra} \Phi_H - H \D_{\la a} 
\D_{b \ra}(\Phi_H + \Phi_A) 
+ \left\{ {\case1/2 \D_{\la a} q_{b \ra}} \right\}\;,
\label{shear} \\
(\rho + p)  A_a &\approx& - \D_a p 
- \D^b \D_{\la a} \D_{b \ra} (\Phi_H + \Phi_A) 
-\left\{ {\dot{q}_a + 4 H q_a} \right\}\;. \label{acceleration}
\end{eqnarray}
\subsection{Frame Transformations and Gauge Fixing} \label{frame-gauge}
There is freedom associated with the choice of reference velocity
$u^a$, which we call a {\it frame choice}. This is to be distinguished 
from the choice of coordinates in the realistic universe model, which 
can be done independently of the choice of $u^a$. It is equivalent to
the choice of time-like world-lines mapped into each other by the 
perturbation gauge chosen (i.e. the mapping between the background 
model and the realistic lumpy universe model, see for example Bruni
and Ellis \cite{BE}), but is independent of the choice of time surfaces in 
that mapping. Given a particular covariantly defined choice for 
this velocity, the frame choice is physically specified and the 
equations are covariantly determined and gauge invariant under the 
remaining gauge freedom \footnote{Gauge fixing requires in addition a 
specification of the correspondence of time surfaces in the realistic 
and background models (effectively specified by determining the choice
of surfaces of constant time in the realistic universe model) and of 
points in initial space-like surfaces.}. In simple situations this choice
will be unique, however in more complex situations several choices of 
this velocity are possible, each leading to a somewhat different CGI 
description.

When we restrict ourselves to a particular frame in order to 
simplify calculations, we can straightforwardly make the appropriate
simplifications in the general equations to see what the implications are
(for example setting $q^a = 0$ for the {\it energy frame}, the quantities 
in the braces in equations (\ref{shear},\ref{acceleration}) above 
vanish). However it is also useful to explicitly transform from one frame to 
another and examine the resulting effect on dynamic and kinematic quantities.

Under a frame transformation $\tilde u^a \approx u^a + v^a$, $|v^a| \ll 1$ 
(restricting our analysis to non-relativistic relative velocities
\footnote{It is important to recall that gauge invariance is only guaranteed 
if the choice of velocity $u_a$ coincides {\it exactly} with its value
in the background spacetime. This is not difficult
in practice, as appropriate physically defined frames ${\tilde u}_a$ will
necessarily obey this condition because of the Robertson-Walker symmetries.}),
the following relations \cite{MGE} hold: 
\begin{eqnarray}
{\tilde\sigma}_{ab} &\approx& \sigma_{ab} + \D_{\la a} v_{b \ra}, 
\label{s-trans}\\  
{\tilde{A}}_a &\approx& A_a + \dot{v}_a + H v_a, \label{a-trans} \\
{\tilde{\Theta}} &\approx& \Theta + \div v, \label{exp-trans} \\
{\tilde{q}}_a &\approx& q_a - (\rho + p) v_a, \label{q-trans} \\
{\tilde{\omega}}_a &\approx& \omega_a - \case1/2 \c v_a\,. \label{vort-trans} 
\end{eqnarray}
The quantities $\rho$, $p$, $\pi_{ab}$, $E_{ab}$ and $H_{ab}$, 
remain unchanged to linear order in almost Friedmann-Lema\^{\i}tre universes 
(e.g. ${\tilde \pi}_{ab} \approx \pi_{ab}$ and 
${\tilde E}_{ab} \approx E_{ab}$), and the 
temperature anisotropies ($\tau_{A_\ell}$) for 
$\ell > 1$ are similarly invariant for the small 
velocity transformations. 
The baryon and radiation (dipole) relative velocities change 
according to:
\begin{eqnarray}
\tilde v_a^{B} \approx v_a^B - v_a\;, \\
\tilde \tau_a \approx \tau_a - v_a\;. \label{tau-trans}
\end{eqnarray}
Also, the projection tensor $h_{ab}$ changes if we boost to 
the frame $\tilde u_a$ giving ${\tilde h}_{ab}$, hence any spatial 
gradients need to be modified and ${\tl{\D}}_a$ will be the 
totally projected derivative in that frame. The consequence is that 
the perturbation variables change accordingly: Thus for any species 
I, \cite{MGE}
\begin{eqnarray}
{\tl{\D}}_a \ln \rho_I \approx \D_a \ln \rho_I 
- 3 H v_a (\rho^I + p^I)/ \rho^I\;. 
\end{eqnarray}
For example $I=R$ and $I=B$ give the equations for radiation and 
baryons respectively, implying:
\begin{eqnarray}
{\tl {\D}}_a \ln T \approx\D_a \ln T - v_a H\,, ~~~
{\tl {\D}}_a \ln \rho_M \approx \D_a \ln \rho_M - 3 H v_a\,. 
\label{pert-trans}
\end{eqnarray} 
These equations allow us to determine the required transformation to obtain 
desired properties of a particular choice $\tilde u^a$. 
The almost-Friedmann-Lema\^{\i}tre multipole divergence equations
(\ref{li_ibe_1}-\ref{mono_temp}) 
are valid in any frame; in particular, if a frame transformation 
is performed as above, they can be given in terms of the resulting 
variables in the new frame, ${\tilde u}_a$, with whatever restrictions result. 

While various choices of $\tilde u^a$ are available in a 
multi-fluid description of the early universe \cite{MGE}, 
there are three particularly useful choices. 
\begin{enumerate}
\item The {\it energy frame}: ${\tilde q}_a =0$ is preferred 
when dealing with two coupled particle species, as in the two 
fluid scenario \cite{dunsby}. This is useful as the Einstein field 
equations are simplified to a form which takes on a similar structure 
to the matter dominated equations, and even in the strong interaction 
case may be expected to be unaffected by collisions because of 
energy-momentum conservation (this choice is dealt with in more detail 
below in the context of scalar perturbations).

\item The {\it zero acceleration frame} (or CDM frame): $\tilde u_a
 = u^{C}_a\approx u_a + v_a^{C}$.
{}From the CDM velocity equation \cite{cl2,MGE} and (\ref{a-trans}) we then
find: $\dot{v}_a^{C} + H v_a^{C} + A_a \approx 0$~~$\Rightarrow$~~
${\tilde A}_a \approx 0$ \cite{MB,cl2,MGE}. This choice is 
particularly useful in multi-species situations, as this frame 
will be geodesic right through tight-coupling, slow decoupling 
and into the free-streaming era.

\item The {\it Newtonian frame}: ${\tilde u}_a = n_a$ in which 
the vorticity and shear of the reference frame vanishes: 
${\tilde \sigma}_{ab} \approx\D_{\la a} n_{b \ra} =0$, when 
such a frame can be found. This frame is only consistent 
in restricted cases \cite{HvEE}, but is particularly useful in 
making comparisons with much of the analytic literature 
\cite{HS95a,HS95b,GSS} and in making connections with the local
physics in terms of Newtonian analogues in Eulerian coordinates. 
For example, the matter shear can be then thought of in terms of distortion
due to the relative velocities (\ref{s-trans}): $\sigma_{ab} \approx 
-\D_{\la a} v^N_{b \ra}$. 

\item The {\it constant expansion frame}: $\tl{\D}_a \tilde \Theta =0$.
This choice is sometimes useful when discussing perturbations on small
scales.
\end{enumerate}

These various choices will simplify the equations in significant ways,
and enable us to recover many of the standard results. It should be
noted however, that the covariant equations we have given above are 
general and do not require either gauge or coordinate restrictions to
be meaningful, and the covariant quantities have in them a 
natural invariant geometric meaning.  We will therefore retain 
the covariant form and do not restrict ourselves to any particular 
frame nor gauge choice for most of this paper, however we retain the
freedom to make such a choice when useful. If and when we do pick 
a particular frame, this will be explicitly stated along with 
the reason for doing so.
\subsubsection{The Newtonian Frame Link  
to the Bardeen Variables} \label{newtg}
Here we demonstrate the direct link between our variables in the scalar case,
and those used in the Newtonian gauge, in terms of the Bardeen variables. 
{}From (\ref{s-trans}) and (\ref{a-trans}) we find easily that for 
$\tilde{u}^a = n^a$ where $\D_{\la a} n_{b \ra}=0$, $n_a = u_a + v_a^N$ 
(the consistency of this choice is discussed in \cite{HvEE}):
\begin{eqnarray}
0 &\approx& \sigma_{ab} - \D_{\la a} v_{b \ra}^N, \label{n-shear} \label{n-t1}\\
{\tilde A}_a &\approx& {A}_a + \dot{v}_a^N + H v_a^N, \label{n-t2}\\
{\tl {\D}}_a \ln T &\approx& \D_a \ln T - H v_a^N, \label{n-t3}\\ 
{\tilde \tau}_a &\approx& \tau_a - v_a^N, \label{n-t4}\\
{\tilde \Theta} &\approx& \Theta + \div v^N, \label{n-t5}\\
{\tilde q}_a &\approx& q_a + (\rho + p) v_a^N. \label{n-t6}
\end{eqnarray}
The effect of this frame transformation is to modify the $\ell=1$ and $2$
multipole divergence equations (\ref{mono_temp}-\ref{li_ibe_2}):
\begin{eqnarray}
- \dot{\tl{\tau}}_a &\approx& {\tl {\D}}_a \ln T + {\tilde A}_a 
+ \case2/5 \D^c \tau_{ab} - \sigma_T n_e(v_a^B -\tau_a), \\
- \dot{\tau}_{ab} &\approx& \D_{\la a} {\tilde \tau}_{b \ra} 
+ \case3/7 \D^c \tau_{abc} + \sigma_T n_e \tau_{ab}\;.
\end{eqnarray}
The $\ell >2$ equations (\ref{li_ibe_l}) remain unchanged, however the 
field equations as well the perturbation equations need to modified, if 
necessary using the transformation relations (\ref{a-trans} -\ref{tau-trans}).
For example, (\ref{mono_temp}) becomes
\begin{eqnarray}
({\tl {\D}}_a \ln T)^{\dot{}} + H ({\tl {\D}}_a \ln T + 
{\tl {A}}_a) \approx - \case1/3 {\tl {\D}}_a {\tilde \Theta} - 
\case1/3 \D_a (\D^c \tau_c), \label{tempc}
\end{eqnarray}
which can easily be checked to be invariant under the frame transformations
${\tilde u}_a \approx u_a + v_a$. 
{}From the shear evolution equation (\ref{EFE-dots}) :
\begin{eqnarray}
\D_{\la a} {\tilde A}_{b \ra} \approx E_{ab} - \case1/2 \pi_{ab}, 
~~\Rightarrow~~ {\tilde A}_a \approx \D_a \Phi_A\;,
\end{eqnarray}
the momentum constraint (\ref{mom-flux1}) and the propagation 
equation for the electric part of the Weyl tensor (\ref{E-dot}) one 
finds respectively that:
\begin{eqnarray}
\case1/3 \D_{\la a} {\tl {\D}}_{b \ra} {\tilde \Theta} 
\approx - \case1/2 \D_{\la a} {\tilde q}_{b \ra}, ~~~~
\case1/3 \D_{\la a} {\tl {\D}}_{b \ra} {\tilde \Theta} 
\approx \D_{\la a} \D_{b \ra} \dot{\Phi}_H 
- H \D_{\la a} \D_{b \ra} \Phi_A\;. \label{ng-expansion}
\end{eqnarray}
This gives the scalar monopole equation for the 
temperature perturbation:
\begin{eqnarray}
\D_{\la a} \D_{b \ra} (\ln T)^{\dot{}} \approx - \D_{\la a} \D_{b \ra} 
\dot{\Phi}_H - \case1/3 \D_{\la a} \D_{b \ra} (\D^c \tau_c)\;. 
\end{eqnarray}
It is important to note here is that in the shear-free frame we 
can interpret the acceleration directly in terms of the 
$\Phi_A$ potential, in other words, in terms of its Newtonian analog,
while $\Phi_H$ can be interpreted as a curvature perturbation. 
In terms of the potentials used by \cite{HS95a,D96} one can
identify $\Psi = \Phi_H$ and $\Phi=\Phi_A$.

The above formulation is useful in linking the covariant work to the usual 
GI treatments. So for example we take the mode expansion of 
the potentials, one finds on dropping the k-index on the right, 
\begin{eqnarray}
 A_{\alpha} &\approx& ( \Phi_{A | \alpha} + V_{S|\alpha}' + H V_{S|\alpha} ) =
 ({V'_S }^{(0)} + {a' \over a} V_S^{(0)} - k \Phi_A) Y_{\alpha}^{(0)}, \\
 \sigma_{\alpha \beta} &\approx& a (\D_{\la \alpha} \D_{\beta \ra} V_S) 
 = - a k V_S^{(0)} Y_{\alpha \beta}^{(0)},
\end{eqnarray}
where the prime ($'$) denotes the time-derivative with respect 
to the conformal time, $Y$ are the eigenfunctions 
of ${Y^{|\alpha}}_{|\alpha} =  - k^2 Y$ and following Kodama and Sasaki 
\cite{KS,BDE92}, the bar ($_{|\alpha}$) denotes spatial derivatives 
with respect to surfaces of constant curvature in the background. 
Furthermore, one can identify $V_S$ as a relative velocity.
\subsubsection{The Energy Frame}
In order to be clear on the consequences and subtleties involved in 
fixing the frame, here we give the source terms
in the {\it energy frame} ($\tilde{u}^a = u_a^E \Rightarrow 
\tilde{q}_a = 0$) for scalar perturbations. 
The important point is that this is a physical
frame, uniquely defined by the local physics.
The equations (\ref{shear}) and (\ref{acceleration}) then take on the form :
\begin{eqnarray}
(\rho + p) {\tilde \sigma}_{ab} &\approx& 
2 (\D_{\la a} \D_{b \ra} \Phi_H)^{\dot{}} + 4 H \D_{\la a} \D_{b \ra} \Phi_H 
- 2 H \D_{\la a} \D_{b \ra} \Phi_A + \D_{\la a} v_{b \ra}^E, \\
(\rho+p) {\tilde A_a} &\approx& 
- \D_a p - \D^b \D_{\la a} \D_{b \ra} (\Phi_H+ \Phi_A) + \dot{v}^E_a + H v^E_a.
\end{eqnarray}
Many CGI treatments use this frame \cite{dunsby}, and have the
advantage that the equations take on a form which is similar 
to those for the matter dominated case, but can still be used 
near to matter-radiation equality.
\subsubsection{Matter Domination}
During matter domination (we have in mind the CDM dominated case) there is 
a unique physically relevant frame defined by the matter 4-velocity, hence 
without restricting the almost-Friedmann-Lema\^{\i}tre universe further there is natural 
frame, $u^a$, in which the variables will be gauge invariant and the above 
relations hold.  

In this frame, equations (\ref{shear}) and (\ref{acceleration}) become 
\begin{eqnarray}
 a^3 \rho_{\!_M} \sigma_{ab} \approx - 2 (a^3 \D_{\la a } \D_{b \ra} 
 \Phi_H)^{\dot{}},~~
 \mbox{and}~~ A_a &\approx& 0\;.
\end{eqnarray} 
Here $\rho \approx \rho_{\!_M}$ is now the density of the matter content 
only. The key point is that to retain a consistent linearisation scheme 
as well as retaining gauge invariance, we now have three smallness 
parameters:
$v$ (non-relativistic relative velocities), $\eta$ (radiation-baryon 
ratio is at least $10^{-2}$), $\epsilon$ (the universe is almost Friedmann-Lema\^{\i}tre
when $\epsilon$ is at least $10^{-5}$) \cite{MES,MGE,SAG}. It
follows that $\rho_{\!_R}$ (the radiation energy density) is now $O(\eta)$  
and we can drop all terms at 
least $O(\eta \epsilon)$, $O(\epsilon^2)$ and $O(\eta^2)$ such as, 
for example, $p \sigma_{ab} \approx 0$ or $\case4/3 \rho_{\!_R} \tau_{ab} \simeq 
\pi_{ab} \approx 0$. This is how the anisotropic pressure is 
eliminated to the order of the calculation in this scheme. 

The link to the matter distribution in the spacetime comes through
the mode coefficients $\sigma(k,t)$ (of the shear), $A(k,t)$ (of the 
acceleration), and $\delta T(k,t)$ (of the temperature perturbation). 
A mode analysis leads to a particular solution of the linearised 
Einstein field equations 
due to scalar modes as in (\ref{scalar-potential1} - \ref{scalar-potential2}): 
\begin{eqnarray}
&& E_{ab} \approx \bar \Phi Q_{ab} =  {k^2 \over a^2} \Phi_H(t,k) Q_{ab},\\     
&& \sigma_{ab} \approx - \ts\frac23 (H_0^2 \Omega_0)^{-1} k^2 
(a \Phi_H(k,t))^{\cdot} Q_{ab}\,, \\
&&\D_a \ln \rho_{\!_M} \approx \ts\frac23 k 
(H_0^2 \Omega_0)^{-1} \Phi_H(k,t) \left[{k^2-3K}\right] Q_a\,.
\label{tf-Newtonian}
\end{eqnarray}
For a matter dominated open model ($K \neq 0$) where $a_0=+1$ we have 
$H_0^2 \approx K / (\Omega_0 -1)$. 

If we add the {\it adiabatic assumption} (see Appendix \ref{sec-adb}) we find
\begin{eqnarray}
\D_a \ln T \approx + \case1/3 \D_a \ln \rho_M\,,     
\end{eqnarray}
where we have used that $\rho_{\!_B} \approx 3 H_0^2 \Omega_0 a^{-3}$ in the 
background. One can then put the mode coefficients, in the matter 
dominated scalar adiabatic almost-Friedmann-Lema\^{\i}tre models, into the form :  
\begin{eqnarray} 
&&\delta T(k,t) \approx \frac13 (H^2_0 \Omega_0)^{-1} (a \Phi_H)   
\left[ {\ts\frac23(k^2 - 3 K)}\right]\,,\\
&& A(k,t) \approx 0 \approx v_{\!_B}(k,t),\\
&&\sigma(k,t) \approx - \frac23 (H_0^2 \Omega_0)^{-1} (a \Phi_H)^{\dot{}}
\left[ {k^2} \right]. 
\label{adb_sclr_Friedmann-Lemaitre}
\end{eqnarray}
The first expression gives the direct effect of the gravitational potential
on the cosmic background radiation anisotropies 
(Sachs-Wolfe effect), and the third the effect of 
the time variation  of the potential on these anisotropies.  
These are investigated in detail in later sections.
The matter dominated Einstein field equations are at 
least $O(\epsilon \eta)$ and fix the form
of the shear, the acceleration and the temperature 
perturbations $D_a \ln T$ as they enter the integrated 
Boltzmann equations (which is how the geometry enters into these
equations).  The hierarchy itself is $O(\epsilon)$ and although the   
radiation variables do not enter the almost-Friedmann-Lema\^{\i}tre (matter) 
Einstein field equations, they 
remain non-zero, and therefore there are still temperature anisotropies, 
$\tau_{A_\ell}$. This is an important but subtle point -- 
matter domination implies the radiation moves as a test field 
over the geometry. 
\subsection{Linearisations, Approximations and Scales} \label{sec-linear}
In this section we discuss the various linearisation 
and approximations (already mentioned in the last 
section), that will be used in this paper. 
\subsubsection{Almost-Friedmann-Lema\^{\i}tre Linearisation}
Here we drop all terms that are at least ${\cal O}(\epsilon^2)$. 
The implication of this is that one can only consider small velocity boosts,
large ones would break the linearity about the Friedmann-Lema\^{\i}tre background -- hence 
we include $v^2 = |v^a v_a| \ll 1$ as an almost-Friedmann-Lema\^{\i}tre limit dropping the
additional terms that are at least ${\cal O} (v \epsilon,v^2)$.

The important subtlety here is that $\epsilon_\ell$ is in fact the 
temperature anisotropy smallness parameter as related to the temperature
moment mean squares $|\tau_{A_\ell}| \propto \epsilon_{\ell}$. The almost-Friedmann-Lema\^{\i}tre
limits on the geometry, $\epsilon$, (which define $\sigma_{ab}$, $A_a$ and
$D_a \Theta$ (for example) as ${\cal O}(\epsilon)$ in appropriate dimensionless
units \cite{SME,SAG} are related to $\epsilon_{\ell}$ via the 
almost-Ehlers-Geren-Sachs theorem. In other words, limits on the 
temperature anisotropies, $\epsilon_\ell$, put bounds on the size of 
the smallness parameter $\epsilon$, given that a weak
Copernican principle holds. Furthermore, the limits on $\epsilon$ 
in turn place consistency limits on the size of the $v/c$ boosts 
that are applicable (here in units of $c=1$). Thus at least 
almost-Friedmann-Lema\^{\i}tre means keeping terms that are at most :
\begin{equation}
\mbox{Almost-Friedmann-Lemaitre}~~ \approx {\cal O}(\epsilon,v).
\label{almost-Friedmann-LemaitreRW}
\end{equation}
\subsubsection{Matter Dominated Linearisation}
This is based on the radiation-baryon ratio, 
$\eta \propto \ts{\rho_R \over \rho_M}$.
We keep every ${\cal O}(\eta)$ but in the almost-Friedmann-Lema\^{\i}tre case of matter 
domination we then drop everything that is at least 
${\cal O}(\eta v, \eta \epsilon, \epsilon v,\eta^2, \epsilon^2, v^2)$.
We then have that matter dominated almost-Friedmann-Lema\^{\i}tre means keeping 
terms that are at most :
\begin{equation}
\mbox{Matter dominated almost-Friedmann-Lemaitre} ~~\approx {\cal
  O}(\epsilon,v,\eta). \label{almost-Friedmann-Lemaitre}
\end{equation} 
\subsubsection{Expansion in Thompson Scattering Time}
We will introduce a perturbative scheme in the Thompson scattering
time, $t_c = (\sigma_T  n_e)^{-1}$, and will consider terms up to
${\cal O}(t_c^3)$ during the tight-coupling calculation -- such an 
expansion will be used to generate equations near to tight-coupling, 
the limiting case being when $t_c =0$. Additionally an equivalent 
scheme can be constructed in terms of the differential optical depth
$\kappa'$. This scheme is useful in the slow-decoupling era, 
\ie in  expansions where $\kappa'$ and $(\kappa')^2$ are 
sufficiently small to be ignored when compared to terms of order
$\kappa$. This approximation allows one to additionally 
consider the case when $\kappa' e^{-\kappa} \ll e^{-\kappa}$. 
\subsubsection{Small and Large Scales}
We will find it convenient to introduce the notion of small and
large scales. We will do this in two heuristically equivalent ways. The
first scheme is based on the parameter $\epsilon_H$, where the Hubble
expansion is of order $\epsilon_H$, and is used when considering
situations outside the Hubble flow; thus in the almost-Friedmann-Lema\^{\i}tre
small-scale case one would ignore all terms at least ${\cal O}
(\epsilon_H^2,\epsilon^2,v^2,v \epsilon_H, v \epsilon,\epsilon
\epsilon_H)$. This scheme is useful since it can be used 
without a mode expansion. It is ideal for making qualitative statements
without the details which arise when introducing mode functions; specifically
avoiding the complexity of mode-mode coupling in the small scale non-linear
situation. By small-scale almost-Friedmann-Lema\^{\i}tre we have in mind keeping only terms 
that are at most:
\begin{equation}
\mbox{Small-scale almost-Friedmann-Lema\^{\i}tre}~~\approx {\cal O}(\epsilon_H,\epsilon,v)\;.
\label{small_scale} 
\end{equation}
A second and more precise scheme is based on the Hubble scale 
$\lambda_H=a_E/k_H$ defined near the time of emission (E), allowing one 
to use $k/k_H > 1$ and $k/k_H <1$ as characterizing large and small
scales respectively. 
\section{Covariant Integral Solutions}
This section has three aims: (i) Reproducing the integral solution of the
free-streaming mode equations and modifying them in order to take into
account Thomson scattering, using the CGI variables \cite{HS95b,SeZalb}. 
We carry out a time-like integration, instead of a null-cone
integration corresponding to the original Sachs-Wolfe paper 
\cite{SW}, restricting  ourselves to scalar perturbations 
with adiabatic modes only and assuming for the most part a $K=0$ 
almost-Friedmann-Lema\^{\i}tre background universe. (ii) Showing in the CGI 
formulation how fluctuations at last scattering time result 
in measurable cosmic background radiation anisotropies. (iii) Demonstrating
how the solution can be related to standard formalisms by choosing
specific frames; in particular we consider the Newtonian frame 
based on a shear free congruence.  

The basic equation we are concerned with in this 
section is the integrated Boltzmann equation (\ref{li_ibe}). 
In covariant form it is given by
\begin{eqnarray}
\dot{\tau}(x,e) + e^a \D_a \tau(x,e) + {\cal B}(x,e) \simeq
C[x,e]\;, \label{IBE-aflrw}
\end{eqnarray}
where the {\it gravitational source term}, ${\cal B}$, and {\it Thompson 
source term}, $C$, for damping by Thompson scattering are respectively 
given by: 
\begin{eqnarray}
{\cal B} = -\ts{1 \over 3} \D^a \tau_a  + (\D_a \ln T 
+ A_a) e^a + \sigma_{ab} e^a e^b,~~~ C[x,e] \approx \dot{\kappa} 
(e^a v_a^B - \tau)\,,
\label{IBE-source}
\end{eqnarray}
and in the almost-Friedmann-Lema\^{\i}tre situation in mind\footnote{One can compare
this to the formulation of Durrer \cite{D96,DZ96} (eqns 3.5 and 18).
To see how this is linked to the
Bardeen potentials $\Psi$ and $\Phi$, we can use 
$E_{ab} = \case1/2 \D_{\la a} \D_{b \ra} (\Psi - \Phi)$ \cite{DZ96}. Notice 
that Durrer's integral solutions take on the Sachs-Wolfe form and can be
compared with the treatments in \cite{dunsby,Mag}, while ours follow the
form of \cite{W83,SeZalb}.}, the Einstein field equations give:
\begin{eqnarray}
3 \rho_{\!_M}^{-1} (\div E)_a \approx\D_a \ln \rho_{\!_M}, ~~  
 \sigma_{ab} \approx -2 [ (\rho_{\!_M}^{-1} E_{ab})^{{\dot{}}}  
- \c (\rho_{\!_M}^{-1} H_{ab})]\,.
\end{eqnarray}
The mode expanded form of this equation for the flat case 
($K=0$) can be written in the compact form as follows:
\begin{eqnarray}
\tau _\ell^{\prime }+k\left[ {{\frac{(\ell+1)^2}{(2\ell+1)(2\ell+3)}}
\tau _{\ell+1}-\tau_{\ell-1}}\right] +\kappa ^{\prime }\tau
_\ell\simeq {\cal S}_B\;,
\label{modeibe}
\end{eqnarray}
where
\begin{eqnarray}
{\cal S}_B= - \left[ {a{\cal B}_0\delta _{\ell0}+(a{\cal B}_1
+\kappa^{\prime }v_B)\delta _{\ell1}+a{\cal B}_2\delta _{\ell2}}\right]\;.
\label{source-calB}
\end{eqnarray}
This combines (\ref{mibec-1}-\ref{mibec-3}) in a single
equation for $\tau _\ell(\eta ,k)$ (see \cite{GE,MGE}), where 
$\tau = \sum_\ell \tau_\ell G_\ell[Q]$ is
written in terms of conformal time $\eta $ ($dt = a d \eta$), 
rather than proper time $t$. It is valid for all $\ell \ge 0$, with 
$\tau_0=0$ a solution as required, consistent with the definitions we 
introduced above (${\cal B}_0$ cancels the dipole term on the right in 
this case).

In what follows, we will deal with the integral solutions to (\ref{modeibe})
given the source terms (\ref{source-calB}). 
The paradigm is to match an (almost-Friedmann-Lema\^{\i}tre) era of free-streaming 
to one of tight-coupling. We will construct the homogeneous solution
(without gravity or scattering) first, then include the 
gravitational effects to construct the free-streaming solution 
(\ie  after decoupling to the present day) and finally include 
Thompson scattering to find the integral solution including
scattering (which can be used during slow decoupling or to 
include effects of reionization). 
Diffusion damping is included in this full solution with Thompson 
scattering, which in general has to be solved numerically, however 
it is helpful to introduce various analytical approximations 
for the different stages described by the solution;
this will be done later, where the visibility function 
approximation is used and the damping scale derived. Additional effects, 
such as the anisotropic correction and polarization correction, have to
be dealt with separately. 
\subsection{Integral Solutions (Flat Almost-Friedmann-Lema\^{\i}tre Case)}
Here we wish to find the general solution to (\ref{modeibe}) without 
collision terms, i.e. with $\kappa^{\prime} = 0$, integrating along time-like
curves using conformal variables. In order to do this we first find the 
solution to the homogeneous version of the above equation 
(i.e. for no gravity and no scattering), second, an integral solution 
of the inhomogeneous equation with gravity taken into account from 
the homogeneous solution, and third, the general solution for 
free-streaming. In the next sections we consider the effect 
of Thompson scattering ($\kappa \neq 0$), and the transition from 
tight coupling to free-streaming.
 
The approach here is similar to the Seljak-Zaldariagga treatment 
\cite{SeZalb,SZ97a}, however, they have taken the Sachs-Wolfe like 
formulation of the integrated Boltzmann equations, which is an
integration down the null cone, and integrated out the angular 
dependence over Legendre polynomials (angular averaging) in order 
to construct the mode coefficients; then the conformal radial 
distance, $\chi $, is written in terms of the conformal time $\eta$, 
leading to an integral solution dependent only on the conformal time. 
Thus, formally they have carried out a null-integration. By contrast, 
what is carried out here is in effect an integration of the 
integrated Boltzmann equations down the matter world-lines, thus this 
is a time-like integration, onto an initial surface (`last scattering').
The corresponding initial data near our past world line on that 
surface can then be related to data on the intersection of the past 
null-cone with the surface by means of a suitable homogeneity assumption. 
The time-like nature of the integration is often not made particularly 
clear in the literature, but solutions of the generic multipole divergence 
equations (from which the almost-Friedmann-Lema\^{\i}tre mode hierarchy of temperature 
anisotropies are derived) are usually based on time-like integrations 
in the relativistic kinetic theory \cite{MGE}.

There is no effective difference between the time-like and null-integrations
at linear order. This is because, at linear order, one needs to only integrate 
up one null geodesic, and then decompose the resulting temperature into
its various moments. This is equivalent to making the multipole decomposition
first and integrating these up one time-like world-line. 
This equivalence does not hold in the exact case, so when trying to
include the effects of non-linearity, assuming such an equivalence 
could lead to misleading results.
\subsubsection{Finding the Homogeneous Solution}
The $\ell =0,1$ and $2$ multipole divergence equations and hence 
mode form of the integrated Boltzmann equations are exceptional, 
given that $\tau_0=0$. The point of the integral solutions is to cast 
the exceptional equations $\ell=0,1,2$ into a form that allows 
analytic investigation. We now find the covariant homogeneous solutions. 
 
Consider the homogeneous equation (valid for $\ell \ge 1$) : 
\begin{equation}
{\tau _\ell^{(0)}}^{\prime }+k\left[ {{\frac{(\ell+1)^2}{(2\ell+1)
(2\ell+3)}}\tau_{\ell+1}^{(0)}-\tau _{\ell-1}^{(0)}}\right] =0\;,  
\label{homog1}
\end{equation}
for the background $K=0$ case\footnote{Or for the open or 
closed cases by following \cite{HS95a} \cite{GSS}.}
without damping. The functions 
\begin{equation}
\tau _\ell^{(0)}(k,\eta )=(2\ell+1)\beta _\ell^{-1}j_\ell(k\eta )  
\label{solution}
\end{equation}
are solutions of (\ref{homog}), if the coefficients $\beta _\ell$ obey the
recursion relations \cite{GE}: 
\begin{equation}
(\ell+1)\beta _\ell=(2\ell+1)\beta _{\ell+1},\;\beta _\ell(2\ell-1)
=\ell\beta _{\ell-1}\;. 
\label{recur}
\end{equation}
This can be shown by multiplying (\ref{homog1}) through by
$\alpha_\ell = \beta_\ell(2\ell+1)^{-1}$ and comparing the result with the 
recursion relation for spherical Bessel functions.
If the function $j_0(k\eta )$ satisfies the equation (\ref{homog1}) 
for $\ell=0$ (there are of course no terms with $\ell<1$) then the 
rest of the equations ($\ell\geq 1)$ will be satisfied because of the 
recursion relations :
\begin{eqnarray}
-(2\ell+1) (\alpha_\ell \tau_\ell)' \simeq k [ (\ell+1) (\alpha_{\ell+1} 
\tau_{\ell+1}) - \ell (\alpha_{\ell-1} \tau_{\ell-1})]\;.
\end{eqnarray}
The freedom in $\beta_\ell(k)$ occurs in $\beta_0(k)$ 
and $\beta_1(k)$. Given that the Bessel function is finite at the 
origin: $j_\ell(0) = \delta_{\ell0}$,
these can be chosen to satisfy $\beta_0 = \beta_1 = +1$, the rest are 
generated through the recursion relations on $\beta_\ell$
and then determine the solution $\tau_\ell^{(0)}(k,\eta )$.
The arbitrarily specifiable initial data is later fixed by introducing
an integral solution ((\ref{no2.1}) below) containing arbitrary functions
$C_A(\eta)$ (see (\ref{no2.2})) which are determined by the Einstein 
field equations through 
${\cal B}_I(\eta)$. 

The corresponding mode functions are 
\begin{eqnarray}
\tau ^{(0)}(x,e)=\sum_{\ell=1}^\infty \beta _\ell^{-1}(2\ell+1)
\,j_\ell(k\eta)O^{A_\ell}Q_{A_\ell}|_{\!_{\sf FLAT}}\;,
\label{modeff}
\end{eqnarray}
(cf. (\ref{expn})) 
and the corresponding multipole coefficients can then be found: 
\begin{eqnarray}
\tau _{A_\ell}|_{\!_{\sf FLAT}}\simeq \beta _\ell^{-1}(2\ell+1)\,j_\ell(k\eta )
Q_{A_\ell}|_{\!_{\sf FLAT}}\;.
\end{eqnarray}
Notice that this differs by a factor of $i^{-\ell}$ from Wilson
\cite{W83} since we are using plain mode functions instead of
plane waves, although these can be easily related. Note there is no 
explicit mode mixing in this approximation, but such mixing is implicitly 
determined by the recursion relations (\ref
{recur}). This shows that we should be careful with any truncation
procedure we propose (see Appendix E). This procedure can be 
easily extended to the open case using the recursion relations for the open 
mode functions (see Appendix K).
\subsubsection{Construction of the Integral Solution}
Given that we have the solution $\tau _\ell^{(0)}(k,\eta )$ to the homogeneous
equation of the form (\ref{homog}), now consider the equation with given
gravitational source terms, but still without damping: 
\begin{eqnarray}
~\tau _\ell^{\prime }+k\left[ {{\frac{(\ell+1)^2}{(2\ell+1)(2\ell+3)}}
\tau _{\ell+1}-\tau_{\ell-1}}\right] = - \left[ { 
a{\cal B}_1\delta _{\ell1}+a{\cal B}_2{\ \delta _{\ell2}}} \right].~~~  
\label{inhomog}
\end{eqnarray}
What is important to notice here, is that this equation is valid 
for $\ell \geq 1$, not $\ell \geq 0$ as in (\ref{modeibe}); 
indeed $\tau_0 = 0$. 
We need to find a particular solution to this equation.
 
We proceed as follows. Consider the ansatz in terms of $A_{\ell}$ using 
$\delta \eta= \eta- \eta'$, along with the Liebnitz rule for 
differentiation of integrals:
 \begin{eqnarray}
\tau_\ell^P(\eta) = \int_0^{\eta} d \eta' A_\ell(\eta,\eta')~~\Rightarrow
 ~~ {\p\tau_l^P \over \p\eta}(\eta) = \int_{0}^{\eta} d \eta' {\p \over \p
\eta} A_\ell(\eta,\eta') + A_\ell(\eta,\eta)\;. 
\label{no2.1}
\end{eqnarray}
Now we define the kernel, $A_{\ell}$, as in \cite{W83,SZ97a}:
\begin{equation}
{A_\ell(\eta ,\eta ^{\prime })
=C_0(\eta)\tau _\ell^{(0)}(\delta \eta)+C_1(\eta){\frac \partial
{\partial \eta ^{\prime }}}\tau _\ell^{(0)}(\delta \eta)
+C_2(\eta){\frac{\partial ^2}{%
\partial {\eta ^{\prime }}^2}}\tau _\ell^{(0)}(\delta \eta)}\;.  
\label{no2.2}
\end{equation}
It can then be shown from (\ref{no2.1}) and (\ref{no2.2}) that :
\begin{eqnarray}
{\tau _\ell^{\prime }}^P+k\left[
  {{\frac{(\ell+1)^2}{(2\ell+1)(2\ell+3)}}
\tau_{\ell+1}^P-\tau _{\ell-1}^P}\right]={\cal S}_C\;,
\end{eqnarray}
where
\begin{eqnarray}
{\cal S}_C=C_0(\eta )\tau _\ell^{(0)}(0)+C_1(\eta
){\tau_\ell^{(0)}}^{\prime }
(0)+C_2(\eta ){\tau _\ell^{(0)}}^{\prime \prime }(0)\;,
\label{int_soln_undamped}
\end{eqnarray}
given $\tau _\ell^P(\eta )$ as in (\ref{no2.1}-\ref{no2.2}). Thus 
(\ref{inhomog}) is satisfied by our ansatz provided the coefficients 
$C_0(\eta ),C_1(\eta ),$ and $C_2(\eta )$ in the integral solution are 
found in terms of the CGI variables 
${\cal B}_0(\eta)$, ${{\cal B}_1(\eta),}$ and ${\cal B}_2(\eta)$ 
determined by the Einstein field equations. 
This will be considered next, when we put the parts of the
solution together to obtain (\ref{C-B-coef}).
\subsubsection{Inclusion of Damping}
Here we extend the previous solution (\ref{inhomog}),
where the relationship between the coefficients in the integrated 
Boltzmann equations (\ref{LEIB001}) 
can be read off from (\ref{int_soln_undamped}),   
including damping through $\kappa ^{\prime }$.  
We notice that if $\tau _\ell^{(0)}$ is a solution to (\ref{homog}), then 
\begin{equation}
\tau _\ell^{*}(\eta )=e^{-\kappa (\eta )}\tau _\ell^{(0)}(\eta )
\end{equation}
will be a solution to 
\begin{equation}
{\tau _\ell^{*}}^{\prime }+k\left[ {{\frac{(\ell+1)^2}{(2\ell+1)(2\ell+3)}}
\tau_{\ell+1}^{*}-\tau _{\ell-1}^{*}}\right] +\kappa ^{\prime }
\tau _\ell^{*}\simeq 0\;.
\end{equation}
Similarly we find that for the integral solution (\ref{no2.1}),
of (\ref{inhomog}), the expressions
\begin{equation}
\tau_\ell^{*P}(\eta )=\int_0^\eta d\eta ^{\prime }e^{-\kappa (\eta ^{\prime
})}A_\ell(\eta ,\eta ^{\prime })~~\mbox{or}~~\;\tau _\ell^{*P}(\eta )
=e^{-\kappa (\eta)}\int_0^\eta d\eta ^{\prime } 
A_\ell(\eta ,\eta ^{\prime })\;,  
\label{no3}
\end{equation}
will be particular solutions to (\ref{modeibe}), given the correct
choice of $C_0$, $C_1$ and $C_2$. Hence we can modify the solutions
of the previous section to include Thompson scattering by simply including
the damping terms as in these equations. Hence the extended equations
include the special case of free streaming, when for some 
interval of time $\kappa = 0$; thus they can extend all the way 
from late tight coupling to the present day, if we include a 
suitably time-dependent scattering coefficient $\kappa$.

In more detail: we have that $\tau_\ell^{(0)}(\eta_*) =0$. This is simply
due to the fact that during tight-coupling there are no higher
moments, just the monopole. 
Here we assume that free-streaming begins after some 
$\eta_*$. The slow decoupling solution will modify this assumption. As
before we have (now using $\delta \eta^* = \eta- \eta_*$)
\begin{eqnarray}
\tau_\ell^{(0)}(\eta) = (2\ell+1) \beta_\ell^{-1} 
j_\ell(k \delta \eta^*)
\Rightarrow
\tau_\ell^{*P}(\eta) \simeq \int_0^{\delta \eta^*} 
d \eta' A_\ell(\eta, \eta')\;,
\label{homog}
\end{eqnarray}
where the initial data is now given by $C_I(\eta'- \eta_*)$; notice
that we do not introduce an additional $\eta_*$ as we will be using the
solution of $\tau^{(0)}$ already including the initial conditions
\footnote{We could have used $\tau_\ell^{(0)}(\delta \eta) \rightarrow
\tau_\ell^{(0)}(\delta \eta+ \eta_*)$ along with the original 
homogeneous solution unchanged.}. Once again we have the integral 
solution $\tau_\ell^P(\eta)$ integrated from $0$ to $\delta \eta^*$ 
such that the anisotropies are now determined by $\tau_\ell^P(\eta_0)$.
Evaluating  the integral from $0$ to $\Delta \eta_*$ 
(for $\Delta \eta = \eta_0 -\eta'$), we find 
\begin{eqnarray}
\tau_l(\eta_0) &=& \int_{\eta_*}^{\eta_0} d \eta' e^{-\kappa(\eta'-\eta_*)}
\left[ C_0(\eta'- \eta_*) \tau^{(0)}_\ell(\Delta \eta)\right.\nonumber\\
&+&\left. C_1(\eta'-\eta_*){\tau_\ell^{(0)}}'(\Delta \eta) +
  C_2(\eta'-\eta_*) {\tau_\ell^{(0)}}''(\Delta \eta)
\right]\;, 
\label{damp-soln-1}
\end{eqnarray}
where the damping term now enters explicitly. 
\subsubsection{The Complete Solution}
We can now construct the general solution to (\ref{modeibe}) 
with $\kappa^{\prime} \neq 0$ by putting
the previous results together. The homogeneous seed solution $\tau
_\ell^{(0)}(\eta )$ is given by (\ref{solution}). The particular integral
solution $\tau _\ell^{P}(\eta )$ is given, in terms of $\delta \eta =
\eta -\eta^{\prime }$ by (\ref{no3}). The general solution is then given by 
\begin{equation}
\tau _\ell(\eta )=e^{-\kappa (\eta )}\tau _\ell^{(0)}(\eta )
+\tau _\ell^{*P}(\eta )\;.
\label{no1}
\end{equation}
Substituting this into the general equation (\ref{modeibe}) and using
the radial eigenfunctions evaluated at zero (in particular $j_\ell(0)=\delta
_{\ell0}$ along with the recursion relation) gives 
\begin{eqnarray}
C_0 \delta _{\ell0}+C_1 {\ \beta _1^{-1}k\delta _{\ell1}} 
+C_2k^2 {5\beta _2^{-1}\left( {\frac 2{15}\delta _{\ell2}+\frac 13\delta
_{\ell0}}\right) } = - \left[ {
(a{\cal B}_1+\kappa ^{\prime}v_B)\delta _{\ell1}
+a{\cal B}_2\delta _{\ell2}} \right],  
\label{find_coef}
\end{eqnarray}
relating the functions determining the solution to the time-dependent
coefficients in the equation.
{}From (\ref{find_coef}) the functions in the integral
solution are found in terms of the dynamical CGI variables:
\begin{eqnarray}
-C_0(\eta )\simeq 
+\frac 52a{\cal B}_2,~~~-C_1(\eta )\simeq +\frac
1k\left( {a{\cal B}_1+\kappa ^{\prime } v_B}\right)\;,~~-C_2(\eta )\simeq 
+\frac 1{k^2}{a{\cal B}}_2\;.  
\label{C-B-coef}
\end{eqnarray}
For scalar perturbations, the term $C_2$ is effectively the coefficient of
the shear; the term $C_1$ is the coefficient of the gradient of the
temperature and the acceleration and the coefficient $C_0$ also
contributes to the shear (see (\ref{efe2.1b})). Note that these
are CGI with respect to $u^a$. 

These functions are all evaluated at a time $\eta$, 
which takes all values from $\eta_d$, the time of decoupling, to $\eta_0$, 
the time of observation. In the null-cone formulation of the 
problem this input of new information corresponds to the way the null 
geodesics from the point of emission to the observer keep crossing new 
matter and hence encounter new information. Because we are 
integrating on a time-like curve, this information is represented here as 
varying with time along that curve; and in some simple circumstances, the 
values at later times are determined fully by the values at earlier times 
(as happens, for example, in the original Sachs-Wolfe case: $K = 0$,
$p = 0$, and only growing scalar modes are considered). 
\subsection{Integration by Parts}
In order to deal easily with the initial data it is now useful to 
write the general solution for $K=0$ in terms of the present time, 
$\eta _0$, and the initial time, $\eta _*$, by integrating with
respect to conformal time and defining 
$\Delta \eta_* =\eta _0-\eta _*$\footnote{The relationship between the 
conformal time $\eta $ and the radial distance $\chi $ 
is $d\chi =-d\eta$ so $\chi = \eta _0 -\eta $ which follows for the 
homogeneity and isotropy in the background.}. We could fix the conformal
time by setting $\eta_0=0$ here. In order to recover the results
of \cite{cl,SZ97a} one would take $\eta_* \rightarrow 0$, however, we
would like to recover the results as close to \cite{HS95a} and so retain
their conventions where possible.

Notice that from $d\tau^{(0)} /dt=0$, we have $\tau ^{(0)}(x(\eta ),e(\eta ))
=\tau ^{(0)}(x(\eta_*),e(\eta _*))$, 
and $j_{\ell}(0)= \delta_{\ell0}$ (we have chosen the solution 
to be finite at origin). We choose the initial conditions 
$\tau_{\ell}(\eta_*)=0$ \cite{W83} and $\tau_0(\eta)=0$ and using the 
parameter freedom in the homogeneous solution, set $\beta_0(k)=+1$ 
and $\beta_1(k)=+1$.
The homogeneous solution is now fixed as in (\ref{homog}), 
for $\delta \eta^* = \eta -\eta_*$ and this in turn sets the integral 
solution to (\ref{no3}) \cite{W83}:
\begin{eqnarray}
\tau_\ell^{*P}(\eta) = \int_0^{\delta \eta^*} e^{- \kappa(\eta')}
 A_\ell(\eta, \eta') d \eta'\;,
\end{eqnarray}
where we still have the freedom of setting the initial data for the integral
solution from the $C_I(\eta)$'s which are fixed by the Einstein 
field equations. Putting this all together we find 
\begin{eqnarray}
\tau_\ell(\eta_0) = \tau^{(0)}_\ell(\eta_*) + \tau_\ell^{*P}(\eta_0) 
= \tau_\ell^{*P}(\eta_0) = \int_0^{\Delta \eta_*} e^{-\kappa} 
A_l (\eta_0, \eta') d \eta'\;.
\end{eqnarray}
 
On changing the integration to from $\eta_*$ to $\eta_0$ in 
(\ref{damp-soln-1}), integrating by parts, and using the initial conditions 
(once again $\tau^{(0)}_{\ell}(\eta_*) = 0$) we find  : 
\begin{eqnarray}
\tau_\ell(\eta_0) &\simeq& \left[ {C_2'(\eta_*) - C_1(\eta_*) } \right]
e^{-\kappa} \tau^{(0)}_\ell(\eta_0) - C_2(\eta_*) e^{-\kappa}
{\tau_\ell^{(0)}}'(\eta_0), \nonumber \\
&+& \int_{\eta_*}^{\eta_0} d \eta' e^{-\kappa} \left[ {
C_0(\eta') - C_1'(\eta') + C_2''(\eta')} \right] \tau^{(0)}_\ell(\eta_0
+ \eta_* - \eta'), \nonumber \\
&+& \int_{\eta_*}^{\eta_0} d \eta' (\kappa' e^{-\kappa}) \left[ {C_1(\eta') 
-2 C_2(\eta')} \right]
\tau^{(0)}_\ell(\eta_0 + \eta_* - \eta'), \ldots \nonumber \\
&+& \int_{\eta_*}^{\eta_0} d \eta' \left( {(\kappa')^2 - \kappa''} \right)
e^{-\kappa} C_2(\eta') \tau^{(0)}_\ell(\eta_0 + \eta_* - \eta')\;.
\label{int_soln_damping}
\end{eqnarray}
The initial data for the solution $\tau _\ell^{(0)}(k,\eta )$ is the set of
constants $C_I(\eta_*)$ which are determined by  ${\cal B}_A(k)
=\left\{ {\cal B}_0(k),{\cal B}_1(k),{\cal B}_2(k)\right\} $; these
must be matched to the initial distribution function on an appropriate
initial surface $\Sigma$ (for example, the `surface of last scattering'
which can be covariantly and gauge invariantly defined). This then determines 
the solution up to the present day (and after). We are free to chose 
any $Q$'s as long as they solve the Helmholtz equation in the 
background. The choice of Q then explicitly determines $G_\ell[Q]$,
for example we are free to choose Q to be the spherical or plane-wave
basis. In practice we naturally use two sets of mode functions $G_\ell[Q]$, 
matching those for the null-cone (given in a spherical basis) to those in 
some initial surface (given in terms of a plane-wave basis). The matching 
of these two sets of harmonics is then given by the relations usually 
written into the construction of the mode coefficients (see (\ref{modeff})). 
This matching is based on mode functions $G_l[Q]$ in the Robertson-Walker background,
which is acceptable because of the homogeneity assumption. By using 
$G_\ell[Q]$ we do not actually need the explicit form of the $Q$'s.
 
Equation (\ref{int_soln_damping}) shows (r.h.s. of the first line) 
how major parts of the cosmic background radiation anisotropy are 
determined directly from the set of initial conditions (at last 
scattering, for the freely propagating radiation). The integrated 
effect arises through the coefficients $C_I(\eta)$ as integrated down 
time-like geodesics in the remaining terms on the right hand side. In general 
there is a non-linear coupling through the field equations between 
the matter, the radiation and the acceleration and shear terms that 
arise in the integrated part. The situation is much simpler when this 
back-reaction can be neglected; for this reason it is convenient to 
consider the case of matter domination, during which the radiation 
can be considered as a test-field propagating on the background 
determined by the matter content.
However we can also consider the general set of linearised field 
equations (see Appendix \ref{almost-Friedmann-Lemaitre-efe}) and the coupling to the 
radiation via the source terms, first the gravitational source, 
${\cal B}$, and second, the scattering source, $C[\tau]$, 
(\ref{IBE-source}), in (\ref{IBE-aflrw}). In the following sections 
we look at the various approximations that can be applied at different
epochs.
\section{Free-streaming}
Using the integral solution (\ref{int_soln_damping}) we construct
the almost-Friedmann-Lema\^{\i}tre free-streaming projection of the initial conditions
near last scattering to here and now (the determination of these initial
conditions is demonstrated in latter sections) and the integrated
secondary contributions arising during the period after last-scattering 
until now (we have dropped the baryon relative velocity effect using the 
instantaneous decoupling assumption):
\begin{eqnarray}
&~&
{\frac{\tau _\ell(\eta _0)\beta _\ell}{(2\ell+1)}}\simeq \left[ {\frac
1k[a{\cal B}_1](\eta_*) 
- \frac 53\frac 1{k^2}[a{\cal B}_2]^{\prime}(\eta_*) 
-\frac 1k[a{\cal B}_2](\eta_*){\frac \partial {\partial \eta _0}}}\right] 
j_\ell(k\Delta \eta_*) 
\nonumber \\
&-&
\int_{\eta _*}^{\eta _0}d \eta \left\{ {\frac 56a{\cal B}_2
-\frac 1k(a{\cal B}_1)^{\prime }+\frac 53\frac 1{k^2}(a{\cal B}_2)^{\prime
\prime }}\right\} j_\ell(k\Delta \eta )\;,  
\label{freestreaming}
\end{eqnarray} 

Where we used as final conditions : 
\begin{equation}
[a{\cal B}_1]^{\prime}(\eta _0)=[a{\cal B}_2](\eta _0)=0\;.  
\label{init1}
\end{equation}
The first term on the right, the ${\cal B}_1$ term, will generate the
acoustic primary effect on the anisotropies, the second term is the Doppler
contribution due to the radiation dipole (the baryon velocity 
contribution which would arise through $C_1$ (\ref{C-B-coef})), the third and 
fourth terms give the effect of any shear, near last scattering 
(through the initial conditions of ${\cal B}_2$). 
The remaining terms represent the integrated Sachs-Wolfe effect.

The above equation will be modified in the following section 
to include slow decoupling, but first we demonstrate how to recover 
the basic Sachs-Wolfe effect. 
\subsection{The Almost-Friedmann-Lema\^{\i}tre Sachs-Wolfe Effect}
We now find the solutions corresponding to the matter dominated,
free-streaming era, with adiabatic modes only, using the Newtonian
frame treatment (see section \ref{newt-sachs-wolfe}). Using the field 
equations from \cite{MGE},
or from the Appendix \ref{almost-Friedmann-Lemaitre-efe} and \ref{tf-Newtonian},
we can find the source terms $a {\cal B}_I(k,t)$ for the free-streaming 
projection : eqns. \ref{efe2.1a} through \ref{efe2.3a} in Appendix 
\ref{app-source-terms}.  This applies to the case of instant decoupling.

During matter domination the dipole is negligible, so we ignore it. 
The shear contribution is small on large scales, hence we can also 
ignore it.

On substituting these equations into the flat almost-Friedmann-Lema\^{\i}tre integral solution
(\ref{int_soln_damping}) with $K=0$ in the source terms 
(eqns. \ref{efe2.1a} through \ref{efe2.3a}) we can find the 
free-streaming almost-Friedmann-Lema\^{\i}tre solutions for the temperature anisotropies
(\ref{freestreaming}).
Using $(\bar \Phi \rho_{\!_M}^{-1})^{\prime} \sim 0$ \footnote{ 
In fact one has that for an Eistein-de Sitter background : 
$[\bar \Phi \rho_{\!_M}^{-1}](\eta_*) = - (3 H_0^2 \Omega_0)^{-1}  
D_* [k^2 \Phi_A(k,0)]$ and $
(\Omega_0 D_*/a_*)^2 \approx \Omega_0^{1.54}$}
we find the CGI kinetic theory equivalent of the Sachs-Wolfe formula 
for cosmic background radiation anisotropies in terms of matter 
inhomogeneities at last scattering at various wavelengths, together 
with an integral term. In various cases (see \cite{HS-astro} 
for references to more general treatments), in particular the 
matter-dominated spatially flat solutions
with only growing scalar modes, the integral terms vanish and we
obtain:
\begin{eqnarray}
{\frac{\tau _\ell^{SW}(\eta _0,k)
\beta _\ell}{(2\ell+1)}}\approx 
\frac23 [\bar\Phi \rho_{\!_M}^{-1}](\eta_*) j_\ell(k\Delta \eta_* ).
\label{tau_l_SW}
\end{eqnarray}
This gives the (approximate)
projection of the large scale potential 
inhomogeneities at last scattering onto the sky today; 
the mean-squares $|\tau _\ell|^2$ can then constructed, using 
the results from Paper I \cite{GE}. 
The effect arise from the terms $\D_a \ln T \approx \frac13 \D_a \ln \rho 
\approx \rho_{\!_M}^{-1} \D^b E_{ab}$, 
having used the adiabatic assumption.

This recovers the standard Sachs-Wolfe result \cite{SW,dunsby} on
large scales, where the potential fluctuations are just due to
primordial initial inhomogeneities that are unchanged by intervening
physics.

We discuss the origin of these fluctuations in later sections -- 
they are given by solving these equations before decoupling, which for 
example implies the existence of acoustic oscillations. These potential 
fluctuations are what seed structure formation through the production of 
matter perturbations undergoing gravitational collapse beneath the Jeans 
scale. The matter perturbations effectively decouple near
matter-radiation equality, making the large scale temperature 
anisotropies the key link between the radiation anisotropies
now to the potential fluctuations then (near radiation decoupling), and so to
the matter power spectrum both on large and small scales today. 

Notice that these equations will hold for any choice of 4-velocity that is
close to the matter 4-velocity, i.e. there is still frame-freedom
associated with this freedom of choice. As has been remarked various times,
there are several possible physical choices for this 4-velocity (which will
all agree at late times); the interpretation of the physical meaning of the
cosmic background radiation anisotropy sources will change depending 
on this choice. The important difference about the derivation 
of the Sachs-Wolfe effect here as opposed to other treatments is that  
{\bf (i)} this result {\it is found in the total matter frame}, 
{\bf (ii)} the integration {\it is explicitly time-like}, 
so it is not treated mathematically as a projection along null rays 
but rather as the evolution of the anisotropies of radiation
in a small comoving box, as explained in the introduction. Thus 
the initial data here is not at the intersection of the past light 
cone and the last scattering surface, but rather at the intersection 
of the world line of the observer and the last scattering surface. 

This analysis can be compared to primary anisotropy source term of the 
gauge-invariant treatment used in \cite{HS95b}. The subtle difference between
the Bardeen variable gauge-invariant approach and the CGI approach used
here is that the Doppler source, which in their case arises through 
${\cal B}_0$, now enters through the integral terms 
only; there is no direct Doppler contribution at last scattering from 
the first term in the integrated solution.
\section{Slow Decoupling}
To deal with slow-decoupling, we return to the general damped solution, 
(\ref{int_soln_damping}), and introduce the slow decoupling approximation
\begin{equation}
\kappa^{\prime \prime} \ll1, ~~(\kappa^{\prime})^2 \ll 1\,,
\end{equation}
to find, on substituting in from the coefficient relations (\ref{C-B-coef}):
\begin{eqnarray}
{\tau_{\ell}(\eta_0) \beta_{\ell} \over (2 \ell +2)} &\simeq&
e^{-\kappa} \left[ {+ \frac{1}{k} [a {\cal B}_1](\eta_*) 
+ \frac{1}{k^2} \frac53 [a{\cal B}_2]^{\prime}(\eta_*)
- \frac{1}{k} [a {\cal B}_2](\eta_*) {\p \over \p \eta_0} } 
\right] j_{\ell}(k \Delta \eta_*) \nonumber \\
&\;& - \int_{\eta_*}^{\eta_0} d \eta' e^{-\kappa} \left\{ 
{ \frac56 a {\cal B}_2- \frac{1}{k} 
(a {\cal B}_1)^{\prime} + \frac53 \frac{1}{k^2} (a {\cal
B}_2)^{\prime \prime} }\right\} j_{\ell}(k \Delta \eta) \nonumber \\
&\;& + \frac{1}{k}(\kappa^{\prime} e^{-\kappa}) {v_B}(\eta_*)
j_{\ell}(k \Delta \eta_*) + \int_{\eta_*}^{\eta_0} 
d \eta' ( v_B^{\prime} \kappa^{\prime} + \kappa^{\prime \prime} v_B)
e^{-\kappa}) \frac{1}{k} j_{\ell}(k \Delta \eta) \nonumber \\
&\;& + \int_{\eta_*}^{\eta_0} d \eta' (\kappa' e^{-\kappa}) \left[ {
- \frac{1}{k} (a {\cal B}_1) + 2 \frac{1}{k^2} (a {\cal B}_2) 
} \right] j_{\ell}(k \Delta \eta)\;.
\label{slowdec_1}   
\end{eqnarray}  
We see that damping effects are controlled by $e^{-\kappa}$
and $\kappa^{\prime} e^{-\kappa}$. 
A further approximation would be to take 
$e^{-\kappa} \gg \kappa^{\prime} e^{-\kappa}$, 
so that we need only consider the free-streaming like
solutions, which we then convolve with the damping factor, 
defined as a combination
of the visibility function and the diffusion damping envelope -- this
is done later using the damping envelope as derived from the dispersion
relations in section (\ref{visible}). 
Later we will explicitly recover these equations in the Newtonian 
frame of section (\ref{newtg}).  
\subsection{Silk Damping}
Diffusion damping will occur and introduce a damping scale, the Silk 
scale, giving a cut-off in the matter perturbations, and there will be 
a corresponding diffusion damping effect in the photons. This is naturally 
included in our general damped solutions in terms of the exponential envelope 
implied by the equations, which can be demonstrated heuristically. 
 
The cut-off arising through photon diffusion occurs when the 
term involving $\kappa'$ in (\ref{modeibe}) dominates the other terms;
that is, when for any $\ell$, $k$ is large enough that 
\begin{eqnarray}
k \left[ { {\frac{1}{4}\tau _{\ell+1}-\tau_{\ell-1}}} \right] \approx
\frac34 k \tau_{\ell} \ll \kappa^{\prime}\tau _\ell\;,
\label{modeibe2}
\end{eqnarray}
the approximation assuming that the damped modes are roughly of the same 
magnitude (independent of $\ell$ when this condition is satisfied). 
This then implies an exponential decay in the relevant modes:
\begin{equation}
k \ll  \frac43 \kappa^{\prime} ~~\Rightarrow ~~
\tau _\ell^{\prime } \approx - \kappa ^{\prime }\tau _\ell~~
\Rightarrow ~~ \label{modeibe3}
\tau _\ell(\eta) \approx \exp(- \kappa ^{\prime}\eta)\tau _\ell(0)\;.
\end{equation}
Thus small scales will be heavily damped by this process and long wavelengths
unaffected, leading to a wavelength-dependent damping envelope.
The resulting cut-off in perturbation amplitude at a critical wavelength
at last scattering will result in a corresponding cut-off in cosmic
background radiation anisotropy amplitudes observed at a critical 
angular scale.

A more detailed examination undertaken later will show the explicit 
wavelength dependence of this cut-off effect. 
\subsection{The Visibility Function} \label{visible}
An alternative approach to the slow decoupling solution 
(\ref{slowdec_1}), is to argue that the dominant contribution during slow-decoupling arises from the visibility function defined by
${\cal V}(k,\eta) \approx \kappa' e^{-\kappa}$ as convolved with the 
free-streaming integral solution. The visibility function gives the 
probability of a photon last scattering during a small 
time interval $d \eta$. From Hu \& Sugiyama \cite{HS95a} it is useful to 
define the damping factor (now including diffusion damping, which will be
derived from the coupled baryon-photon equations 
in (\ref{dif-damp-correct}) ):
\begin{eqnarray}
{\cal D}(\eta_0,k) = \int_{\eta_*}^{\eta_0} d \eta {\cal V}(\eta,k) 
e^{-\left({k/k_D} \right)^2} \approx e^{-(k/k_D)^2}\;. \label{visibe}
\end{eqnarray}
The visibility function will model the changing ionization fraction,
this does not include the diffusion damping, which is added
in by hand above, through the damping scale. It should be realized here
that the Gaussian diffusion damping is {\it naturally included} in 
the original almost-Friedmann-Lema\^{\i}tre integral solution (\ref{int_soln_damping}). However,
given that we will use solutions that are first-order 
(in the scattering time) and then recover an explicit dispersion relation 
for the damping scale, at second-order in the scattering time, it is 
convenient to modify the damping factors such that they are re-written 
in terms of the visibility function \ref{visibe}. The second order damping
scale of the form used here is explicitly derived in section
\ref{sssec-dispersion}.  
 
Now, we can modify the free-streaming projection by including the damping
factor ${\cal D}$ and the baryon velocity effect (which must be 
put in from (\ref{int_soln_damping})). In this approximation, we 
can effectively drop the last two lines of (\ref{int_soln_damping}) 
except for the initial baryon velocity contribution, to obtain: 
\begin{eqnarray}
{\frac{\tau _\ell(\eta _0)\beta _\ell}{(2\ell+1)}} &\simeq&  
\left[ {C_2^{\prime}(\eta _*)-C_1(\eta _*)}\right] {\cal D}(\eta_0,k) 
j_\ell(k\Delta \eta_* ) 
\nonumber \\  &+& kC_2(\eta _*) {\cal D}(\eta_0,k) 
\left[ { {\ \frac \ell{(2\ell+1)}}j_{\ell-1}(k\Delta \eta_*)
- {\frac{(\ell+1)}{(2\ell+1)}}j_{\ell+1}(k\Delta \eta_*)}\right]
\nonumber \\
&+& \int_{\eta _*}^{\eta _0}d\eta {\cal V} e^{-(k/k_D)^2} \left\{ 
{C_0(\eta)-C_1^{\prime}(\eta)+
C_2^{\prime \prime }(\eta)}\right\} j_\ell(k \Delta \eta)\;,
\label{slowdecoupling}
\end{eqnarray}
where the coefficients $C_1$, $C_2$ and $C_3$ are given 
by (\ref{C-B-coef}). This has the effect of taking the previous 
more general solution (\ref{int_soln_damping}) and specializing it  
to the most important regime as far as decoupling is concerned, 
thus giving a major improvement on the sharp decoupling approximation,
while avoiding the complications of the complete integral solution 
given above. A similar correction is made using the visibility
function in the integrated part of the solution, in order to best 
deal with a changing ionization fraction, given that we will 
once again only be using almost-Friedmann-Lema\^{\i}tre solutions, that are either 
first order or zero-th order in the scattering times (discussed below).   
 \subsection{Slow Decoupling in the Conformal Newtonian Frame}
Here we cast the above derived solutions (\ref{freestreaming},\ref{slowdec_1},
\ref{slowdecoupling}) based on the integral solution (\ref{int_soln_damping}) 
into the CGI Newtonian frame (based on the shear-free frame described 
in section \ref{newtg}) for the case of scalar perturbations, in terms
of the Bardeen like scalar potentials $\Phi_H$ and $\Phi_A$.

The vanishing shear condition ${\tilde \sigma}_{ab} \approx 0$ implies
that  $\tilde C_0 (\eta)\approx 0$ and $ {\tilde C}_2(\eta)\approx 0$,
hence we can use these conditions directly to find the slow-decoupling
anisotropy solution for an almost-Friedmann-Lema\^{\i}tre model in the Newtonian frame:
\begin{eqnarray}
{\frac{\tau _\ell(\eta _0)\beta _\ell}{(2\ell+1)}} &\simeq&  
-{{\tilde C}_1(\eta _*)} {\cal D}(\eta_0,k) 
j_\ell(k\Delta \eta_* ) 
- \int_{\eta _*}^{\eta _0}d\eta {\cal V} e^{-(k/k_D)^2}  
{{\tilde C}_1^{\prime}(\eta)} j_\ell(k \Delta \eta)\;.
\label{slowdecoupling-ng-0}
\end{eqnarray} 
Using the results from section (\ref{newtg}) and the almost-Friedmann-Lema\^{\i}tre relations
(given in Appendix G), it can be shown that the key quantity of interest 
${\tilde B}_a \approx {\tilde {\D}}_a \ln T +\D_a \Phi_A$, in the case of
scalar perturbations, obeys the following relation in the Newtonian frame.
\begin{eqnarray}
D_{\la a} {\dot{\tilde {\cal B}}}_{b \ra} \approx\D_{\la a}\D_{b \ra} 
(\dot{\Phi}_A - \dot{\Phi}_H) - 2 H\D_{\la a}\D_{b \ra} \Phi_A 
- \frac13\D_{\la a}\D_{b \ra} (\D^c {\tilde \tau}_c) 
\label{ng-dot-B}
\end{eqnarray}

Now we need to find the mode coefficients for ${\cal B}$ in terms
of the quantities defined in the Newtonian frame. Writing
\begin{equation}
{\tilde {\D}}_a \ln T \approx \frac{k}{a} \delta \tl T Q_a,
~~\D_a \Phi_A \approx  \ts{k \over a} \Phi_A Q_a,  
~~ \mbox{and}~~ D_a \Phi_H \approx \ts{k \over a} \Phi_H Q_a\;, 
\end{equation}  
we find that 
\begin{eqnarray}
{\tilde {\cal B}}_a \approx \ts{k \over a}(\delta \tl T + \Phi_A ) Q_a,
~~\Rightarrow~~
{\tilde{\cal B}}_1 \approx \ts{k \over a}(\delta \tl T + \Phi_A)\;. 
\label{calb-ng}
\end{eqnarray} 
On mode expanding (\ref{ng-dot-B}) 
{\footnote{We will use $\D_{\la a} \D_{b \ra} \dot{\Phi}_A \approx
- \dot{\Phi}_A \D_{\la a} \D_{b \ra} Q$, $\D^a \tau_c \approx 
+ \ts{k \over a} \tau_1 Q$ and $\D_{\la a} {\dot{\cal B}}_{b \ra} \approx 
- \ts{a \over k}(\dot{\cal B}_1 + 2 H {\cal B}_1) \D_{\la a} \D_{b \ra} Q$.}}
and transforming to the conformal time derivative we obtain:
\begin{eqnarray}
(a {\tilde {\cal B}}_1)^{\prime} \approx - a^2 H {\tilde {\cal B}}_1 
+ k (\Phi_A^{\prime} - \Phi_H^{\prime}) - 2 H a k \Phi_A + \ts\frac13 k^2 
{\tilde \tau}_1.
\label{calb-dot-ng}
\end{eqnarray}
Now using (\ref{C-B-coef}) we find that:
\begin{eqnarray}
- {\tilde C}_1^{\prime}(\eta) - (\kappa' {\tilde v}_B)^{\prime} \approx +
\frac{1}{k} (a {\tilde{\cal B}}_1)^{\prime}. \label{calc-dot-ng-01}
\end{eqnarray}
We can now put this all together, first, from (\ref{C-B-coef}) 
and (\ref{calb-ng}) to find:
\begin{eqnarray}
- {\tilde C}_1(\eta) \approx \kappa' {\tilde v}_B + (\delta \tl T + \Phi_A).
\label{calc-ng}
\end{eqnarray}
and second from (\ref{calc-dot-ng-01}), (\ref{calc-ng}) and 
(\ref{calb-dot-ng}) to find:
\begin{eqnarray}
- {\tilde C}_1^{\prime}(\eta) \approx (\kappa^{\prime} 
{\tilde v}_B)^{\prime} + \left( {(\Phi_A^{\prime} - \Phi_H^{\prime})
- a H (\delta \tl T + \Phi_A) - 2 a H \Phi_A 
+ \ts\frac13 k {\tilde \tau}_1} \right)\;.
\end{eqnarray}
Substituting these results into the integral solution 
(\ref{slowdecoupling-ng-0}) we obtain:
\begin{eqnarray}
{\frac{ \tau _\ell(\eta _0)  \beta _\ell}{(2\ell+1)}} &\approx&  
\left[ {(\delta \tl T + \Phi_A) + \kappa^{\prime} {\tilde v}_B} \right](\eta_*,k) 
{\cal D}(\eta_0,k) j_\ell(k\Delta \eta_*) \nonumber \\
&+& \int_{\eta_*}^{\eta_0} d \eta {\cal V} e^{- (k/k_D)^2} 
\left\{ {(\kappa^{\prime} \tilde v_B^{\prime} + \kappa^{\prime \prime} \tilde
v_B) + \frac13 k {\tilde \tau}_1 } \right\} j_{\ell}(k \Delta \eta) \nonumber \\
&+& \int_{\eta _*}^{\eta _0} d \eta {\cal V} e^{-(k/k_D)^2} 
\left\{ {(\Phi_A^{\prime} - \Phi_H^{\prime}) - a H ( \delta \tl T + 3\Phi_A )
}\right\} j_\ell(k \Delta \eta)\;.
\label{slowdecoupling-ng}
\end{eqnarray} 
Here the second order terms (both in terms of the scattering time
and in the almost-Friedmann-Lema\^{\i}tre sense) have been dropped. We can then pull 
out the canonical solution when we ignore the Doppler contribution, 
the initial baryon relative velocity at last scattering 
(it is tightly coupled to the radiation velocity and is 
thus small already). We also ignore the intermediate scale 
integrated effect which contributes to the early-integrated Sachs
Wolfe effect. The result is:
\begin{eqnarray}
{\frac{\tau _\ell(\eta_0)\beta _\ell}{(2\ell+1)}}\approx  
\left[\delta \tl T + \Phi_A\right] {\cal D}(\eta_0,k)
j_\ell(k\Delta \eta_*) + \int_{\eta _*}^{\eta _0} d \eta {\cal V} 
e^{-(k/k_D)^2} \left[{\Phi_A^{\prime} - \Phi_H^{\prime}}\right] 
j_\ell(k \Delta \eta).~~~
\label{canon-temp}
\end{eqnarray}
This completes the recovery of the standard integral solution 
results using the 1+3 CGI approach at linear order -- we have
notationally suppressed the $k$-dependence of the temperature 
anisotropy ($\tau_{\ell}(\eta_0) \equiv \tau_{\ell}(\eta_0,k)$).
It corroborates the standard anisotropy derivations based on a 
3+1 hypersurface foliation, which uses the Bardeen formalism 
in the conformal Newtonian gauge.
\section{Late Tight-coupling}
Here we extend the Thomson scattering analysis of the previous
sections to include a simple model of late-tight coupling and hence
of fast decoupling. 

We aim to reproduce in covariant form the Peebles and Yu
{\it near-tight} coupling \cite{PY} and Hu and Sugiyama
{\it tight-coupling approximation} \cite{HS95b} treatments, valid for
the period of late tight-coupling, up to and including decoupling
\footnote{Here we are explicitly making a distinction between the
treatment \cite{HS95b} (what we call {\it tight-coupling approximation})
and that in \cite{PY} (what we call {\it near-tight} coupling. 
By {tight-coupling approximation} we mean that 
$\dot \kappa^{-1}$ is sufficiently small that it can be ignored 
(inducing a contribution of the order of magnitude (say) of at 
least $10^{-6}$) when multiplying quantities of linear
order such as the shear) The near-tight coupling includes the 
radiation quadrupole in the case of isotropic Thompson scattering 
(in the matter frame).}. Remember that we are ignoring the anisotropic 
and polarization effects as these can be corrected at the level of the 
damping scale. 
\subsection{Integrated Boltzmann Equation: Near-tight Coupling}
Here the almost-Friedmann-Lema\^{\i}tre integrated Boltzmann equations is used to construct a set
of multipole divergence equations that describe the radiation near
tight-coupling. These are the intermediate scale equations, valid in the
tight-coupling era. This is done by carrying out a CGI version of
perturbation theory in terms of the scattering time. 
\subsubsection{The Scattering-strength Expansion}
The solutions we have considered so far are linearised through a
small-parameter expansion in terms of the anisotropy parameter $\tau$. The
basic idea now, following the method of Peebles and Yu \cite{PY}, is that
additionally a second expansion is constructed in terms of the collision
parameter $t_c=({\sigma _Tn_e})^{-1}$, without truncating the Boltzmann 
hierarchy at the order of the calculation, and thus avoiding the problems 
inherent in exact truncation \cite{ETMa} (see Appendix E). We thus 
find the evolution equations for the energy density, momentum flux, and the 
anisotropic flux of the radiation close to tight coupling.
 
Consider the almost-Friedmann-Lema\^{\i}tre integrated Boltzmann equations (\ref{li_ibe}) and (\ref{IBE-aflrw}) for
isotropic (in the baryon frame) Thompson scattering (\ref{IBE-source}) 
\cite{MGE}; this is inverted to find :
\begin{eqnarray}
\tau (x^i,e^a)=v^a_B e_a- t_c \left[ {{\cal B}+\dot{\tau}
+e^a\D_a\tau }\right]\;.  
\label{pert_01}
\end{eqnarray}
We now systematically approximate (\ref{pert_01}) in terms of the
smallness parameter $t_c$. The right-hand side (the scattering term) is 
used to find the zero-th order collision-time correction to the total 
bolometric temperature, with corresponding temperature anisotropy
given by 
\begin{eqnarray}
\tau _{(0)}(x^i,e^a)\approx v^\alpha_B e_\alpha\;.
\end{eqnarray}
The equation is now perturbed about the zero order velocity
perturbation and one can then recover the first and second order 
corrections in $t_c$ to the zero-th order temperature anisotropies,
{\it {\ }}to find, $\tau _{(1)}$ and $\tau _{(2)}$ respectively, 
where the n-th. order correction is denoted by $\tau _{(n)}.$ 
We obtain an almost-Friedmann-Lema\^{\i}tre perturbative expansion in $t_c$ : 
\begin{eqnarray}
\tau _{(n)}\approx v^a _B e_a-t_c\left[ {{\cal B}+\dot{\tau}_
{(n-1)}+e^a\D_a\tau _{(n-1)}}\right]\;.  
\label{pert_02}
\end{eqnarray}
The tight-coupling limit is recovered when $t _c=0$. This 
treatment is then a consistent (in the sense of the truncation conditions
described in Appendix E) near-to-tight-coupling treatment 
in almost-Friedmann-Lema\^{\i}tre universes. (The first, in $n$, three temperature 
anisotropies, are given in Appendix C).
\subsubsection{Solid Angle Integration}
Now the temperature anisotropy is integrated over the solid sphere to ensure
the condition that there is no contribution to the bolometric average 
$T_{(b)}(x^i)$, 
\begin{eqnarray}
\int_{4\pi }\tau (x^i,e^a)d\Omega =0\;.  
\label{int_cond_I}
\end{eqnarray}
It should be clear why the second order correction to the temperature 
anisotropy is needed even though we intend to keep the expansion only to 
first order in $t_c$; the integrations over term the $e^av_a$ will vanish. 
Now by integrating $\tau _{(2)}$ (\ref{temp_2nd}) over the solid angle 
and using (\ref{int_cond_I}) and orthogonality of $O^{A_\ell}$ the gradient 
of the radiation flux is found: 
\begin{eqnarray}
\D_a\tau ^a\simeq \D_av^a_B-\alpha _c\left[ {(\D_av^a_B)^{\dot{}}
-(\D_a\tau ^a)^{\dot{}}+ \ts{1 \over 4}{ \D^2 (\ln \rho_R)}
+(\D_a\dot{v}^a_B)+(\D_aA^a)}\right]\;.
\label{monopole-gradient}
\end{eqnarray}
By taking spatial gradients of the radiation flux (\ref{radf_0013}) we
find on comparing with (\ref{monopole-gradient}), that in order for 
there to be no contributions to scalars, 
\begin{eqnarray}
(\div v)_B^{\dot{}}\approx (\div \tau)^{\dot{}}\;.
\end{eqnarray}
The above equation (\ref{monopole-gradient}) then becomes
\begin{eqnarray}
(\div \tau) \simeq (\div v)_B -t _c\left[ {\ts{1 \over 4}} {{\D^2 \ln \rho_R} +
(\D_a \dot{v}^a_B)+(\div A) }\right]\;.  
\label{mon-gradient}
\end{eqnarray}
\subsubsection{The Transport Equations}
Finally the individual PSTF multipoles are recovered at a given order 
\begin{eqnarray}
\tau _{A_\ell}=\Delta _\ell^{-1}\int_{4\pi }O_{A_\ell}\tau _{(n)}
(x^i,e^a)d\Omega. \label{int_cond_II}
\end{eqnarray}
Here $\Delta _\ell$ is defined as before in Paper I \cite{GE}.
The second order temperature multipoles are now found from (\ref
{int_cond_II}) and integrating $\tau_{(2)}(x,e)$ (\ref{temp_1st}) after 
the combination of direction vectors has been replaced by 
PSTF tensors (one can use \ref{pert_001}-\ref{pert_003}) : 
\begin{eqnarray}
\tau ^b &\approx &v^b_B-t _c\left[ { {\D^b {\ln T}}
+A^b+\dot{v}^b_B }\right] + t_c^2 \left[{ ( \D^b \ln T + A^b){}^{\dot{}} 
+\ddot{v}^b_B - \ts{1 \over 3} \D^b \D_c \tau^c} \right]\;, 
\label{radf_0013} \\
\tau ^{ab} &\approx &-t _c\left[ {\sigma ^{ab}+\D^{\la a}v^{b \ra}_B}\right] 
+ t_c^2 \left[{(\D^{\la a} v^{b \ra}_B)^{\dot{}} + \D^{\la a} 
( \D^{b \ra} \ln T + A^{b \ra}) + \D^{\la a} \dot{v}^{b \ra}_B 
+ \dot{\sigma}^{ab}}\right]\;,
\label{radf_0014} \\
\tau ^{abc} &\approx&  +t_c^2 \left[ { \D^{\la ab} v^{c \ra}_B } \right]\;, \label{radf_0014b}\\ 
\tau ^{A_\ell} &\approx &0~~~\forall ~\ell>3\;,  
\label{radf_0015}
\end{eqnarray}   
where we have dropped terms of ${\cal O}(t_c^3)$.
These are the key results of this section. They are the appropriate
transport equations for the late-tight coupling era, i.e. up to last
scattering, and are essentially equivalent to the
{\it causal transport equations} given by causal relativistic 
thermodynamics \cite{RT}. 

What we have shown here is that if we are interested in the behaviour
of the photon-baryon systems to first order in the scattering time, a
dissipative fluid approximation is sufficient to describe
the radiation (cf. the papers by Israel and Stewart \cite{Is}), and will
not result in an explicit truncation of the Boltzmann multipole hierarchy,
rather it gives a systematic approximation scheme where we can, to the
appropriate accuracy, ignore the third order and higher terms. This is
significant; one cannot merely drop the higher order moments and  
truncate to a fluid description, as the kinetic theory treatment fixes the
transport equations. Here we have consistently decoupled the $l<3$ multipole
equations from the rest of the hierarchy and the consistency of this 
decoupling is maintained through (\ref{monopole-gradient}) 
and (\ref{radf_0013} - \ref{radf_0015}). 

The solutions to these equations, which lead to acoustic oscillations 
during this period, will then affect the cosmic background radiation 
anisotropies by setting initial conditions for the free-streaming solution discussed in the previous section. We give a derivation of these results in the 
following section. 
\subsection{Late Tight-coupling and the Oscillator Equation}
Here we derive the CGI equivalent of the analytic tight-coupling 
approximation used by Hu \& Sugiyama \cite{HS95a,HS95b}. This approach 
uses the tight-coupling limit 
in order to get rid of the radiation quadrupole during late tight-coupling
\footnote{If there are any anisotropic contributions such as anisotropic 
scattering (in the matter frame) or large shear (from gravitational waves) 
at that time this sort of approximation should be considered with care - 
such phenomena would break tight-coupling.} and covariantly 
reproduces Hu and Sugiyama's conclusions about the cosmic background radiation anisotropy due to inhomogeneities, acoustic and Doppler sources (what
they call `primary sources'). This gives the `Sachs-Wolfe effect' due to
the Newtonian potential near last scattering, but not the 
`integrated Sachs-Wolfe effect' due to changing potentials after 
tight coupling (resulting from more complex matter models and/or spatial 
curvature). 
\subsubsection{Near Tight-coupling Equations}
We start with the near-tight coupling equations (\ref{radf_0013},
\ref{radf_0014}) and (\ref{radf_0015}) except rewritten to first order and 
in terms of the optical depth so as to be closer to the notation of the
better known treatments \cite{HS95b}:  
\begin{eqnarray}
\tau _a &\simeq &v_a^B-\dot{\kappa}^{-1}\left[ {\D_a(\ln T)+ A_a +\dot{v}_a^B}
\right]\;,  \label{L008} \\
\tau _{ab} &\simeq &-\dot{\kappa}^{-1}\left[ {\sigma _{ab}+\D_{\la a}
v_{b \ra}^B}\right]\;, \label{L0081} \\
\tau ^{A_\ell} &\simeq &0~~\forall \ell > 2\;.  \label{L0082}
\end{eqnarray}

Here we have assumed that the collisions are dominated by 
Thompson scattering and is therefore isotropic in the matter frame. 

The relative velocity of the matter with respect to the 
preferred reference frame
is $v_B^a \simeq u^a_B - u^a$. 
Rewriting (\ref{L0081}) in terms of the shear of the baryon frame, 
we have $\tau_{ab} \simeq -\dot \kappa^{-1} \D_{\la a} u^B_{b\ra }$, 
so the quadrupole is given entirely by the shear of the matter.

Notice that $\pi_{ab} = \rho_R \tau_{ab} \approx 0$, as the case of 
matter domination.  This condition is not sufficient to ensure 
that $\tau_{ab}$ can be ignored in equations when it appears 
on its own, even though the quadrupole is small. The key point, which
was discussed in section 2.5, is that there are four principal 
linearisations: the almost-Friedmann-Lema\^{\i}tre one at least $O(\epsilon^2)$, 
the almost-Friedmann-Lema\^{\i}tre radiation isotropy one $O(\epsilon \eta)$, $O(\eta^2)$,
(implying the previous by the almost-Ehlers-Geren-Sachs theorem), 
the non-relativistic assumption $O(v \eta)$, $O(v^2)$,
$O(v \epsilon)$, and the linearisation scheme based on the differential 
optical depth. Hence care must be taken when approximations are made
to the equations.
\subsubsection{Tight coupling: Momentum Equations}
The tight-coupling calculation is now carried out, assuming
(\ref{L008}-\ref{L0082}) hold. Consider once again the
radiation energy and momentum conservation equations 
($\ell=0$ and $\ell=1$ multipole divergence equations): 
\begin{eqnarray}
(\ln T)^{\dot{}}+\frac 13\Theta &\simeq & -\frac13 \div \tau\;,
\label{L001} \\
-\dot{\tau}_a &\simeq &A_a+\D_a(\ln T)-\dot{\kappa}(v_a^B -\tau _a) +
\frac25 \D^c \tau_{ac}\;,
\label{L002}
\end{eqnarray}
and the baryon energy and momentum conservation equations
\footnote{This is found to O[1] by substituting the matter energy 
conservation equation (dust part of \ref{EFE-ec}) into
matter momentum equation (dust part of \ref{EFE-mc}) all to O[1].}
\begin{eqnarray}
(\ln \rho )^{\dot{}}+\Theta &\simeq & - \D^a v_a^B,  \label{L003} \\
-{\dot{v}_a^B} &\simeq & + H v_a^B +A_a+R^{-1}\dot{\kappa}
(v_a^B-\tau _a)\;.  
\label{L004}
\end{eqnarray}
Here $\dot{\kappa}$ is the optical depth, and the radiation-baryon 
ratio in the real universe is given by (using the enthalpy 
$h= \rho+p$) by 
\begin{eqnarray}
R(x^i)= {h_B(x) \over h_R(x)} = {\rho_M + p_B \over \rho_R + p_R} 
\approx \frac 34{\frac{\rho (x^i)}{\rho_R (x^i)}}\Rightarrow 
\dot{R}\simeq H R\;.
\end{eqnarray}
This is related to the speed of sound
\footnote{By speed of sound we mean adiabatic sound speed :
$c_s^2 = {\dot{p}_0 \over \dot{\rho}_M + \dot{\rho}_R}$ 
$= \frac13 {\dot{\mu}_0 \over (\dot{\rho}_M + \dot{\rho}_R)}$ 
$= \frac13 {1 \over (\dot{\rho}_M/\dot{\rho}_R) +1}$
$= \frac13 {1 \over {(3 \rho_M / 4 \rho_R)} +1}$, for matter 
domination $c_s \approx 0$. } in the background via $c_s^2 = (1 /
3(R_0+1))$. The matter momentum equations, (\ref{L004}), give 
\begin{equation}
v_a^B\simeq \tau _a-\frac R{\dot{\kappa}}\left[ {\dot{v}_a^B + H v_a^B 
+A_a}\right]\;.  \label{L005}
\end{equation}
Substituting (\ref{L008}) into (\ref{L005}) and retaining  all
terms up to linear order (in the relaxation time) we obtain
\begin{equation}
v_a^B\simeq \tau_a - R \dot{\kappa}^{-1} [\dot{\tau_a} + A_a + H
\tau^a] + {\cal O}( \dot \kappa^{-2})\;.  \label{L005a}
\end{equation}
This is then substituted into (\ref{L002}) in order 
to remove the velocity terms, and with a little algebra, we find 
\begin{eqnarray}
- \dot \tau_a \simeq \dot u_a + {1 \over 1+R} \D_a \ln T + {\dot{R} \over
(1+R)} \tau_a  
- \frac25 {\dot \kappa^{-1} \over (1+R)} \D^b \D_{\la a} u^m_{b\ra }\,,
\end{eqnarray}
where the last term has been written in terms of the matter shear.
We can now consider the situation where $\dot \kappa^{-1}$ becomes
sufficiently small (but non-zero) so that the last term can be ignored.
This is possible as the matter shear is already at least first order.
We then find ({\it cf} \cite{HS95b} eqn. (B2 b)): 
\begin{eqnarray}
\dot{\tau}_a + {\frac{\dot{R}}{(1+R)}}\tau _a + {\frac 1{(1+R)}}\D_a(\ln
T) \simeq -A_a\;.  \label{mom-flux}
\end{eqnarray}
This is the momentum flux equation for the radiation and is a key result. 
It can be rewritten as 
\begin{equation}
\lbrack (1+R)\tau _a]^{\dot{}}+\D_a(\ln T)\simeq -(1+R)A_a\;,
\label{L009a}
\end{equation}
or on taking its divergence as
\begin{eqnarray}
[ a (1+R) (\D^a \tau_a) ]^{\cdot} + a (\D^2 \ln T) \simeq - (1+R) a 
(\D^a A_a)\;. \label{L009ab} 
\end{eqnarray}
\subsubsection{Spatial Gradients and the Oscillator 
Equation on Small Scales}
The basis of this derivation is the `small-scale' assumption which 
effectively means that on small enough scales we can ignore
the expansion (see section 2.5). This is just the statement 
that the scale of inhomogeneity is less than the Hubble scale
(\ref{small_scale}), so we drop all terms of 
$O(\epsilon \epsilon_H)$ (see section 2.5.4).
 
Our aim is to recover the standard oscillator equation
(the source equation for the acoustic oscillations)
using the 1+3 CGI formalism. Note however that we still have 
the freedom to set the relative velocity in the small boost 
equations (which we will do in the next section).
 
Taking the spatial gradient\footnote{Using the identity 
$(\D_a f)^{\cdot} \simeq \D_a \dot{f} - H \D_a f + A_a \dot{f}$ 
we find that $(\D_a \ln T)^{\cdot} 
\simeq \D_a \dot{\ln T} - H (\D_a \ln T + A_a)$ from the
almost-Friedmann-Lema\^{\i}tre $\ell=0$ multipole divergence equations and $H = {\dot{a} 
\over a}$ \cite{MGE}.} of the radiation energy
conservation equation, (\ref{L001}), we find 
\begin{equation}
- \frac 13\D_a(\D_c\tau ^c)\simeq (\D_a\ln T)^{\dot{}}+\frac 13\D_a 
\Theta + H (\D_a \ln T + A_a)\;,  \label{L009}
\end{equation}
and taking the divergence of the resulting equation above gives
\begin{eqnarray}
- \frac13 (\D^2 (\D_c \tau^c) \simeq (\D^2 \ln T)^{\cdot} + 2 H (\D^2 \ln T)
+\frac13 \D^2 \Theta + H (\div A)\;,
\end{eqnarray}
and this can be written as 
\begin{equation}
- \frac 13 ( a^2 \D^2 (\D_c \tau^c)) \simeq (a^2 \D^2 \ln T)^{\cdot}
+\frac 13(a^2 \D^2 \Theta ) + H (a^2 \div A)\;. 
\label{L010}
\end{equation}
This is analogous to equation (B3) in \cite{HS95b}. Then using 
(\ref{L009ab}) we find 
\begin{equation}
\lbrack (1+R) a \D_c(\D_a\tau ^a)]^{\dot{}}+ H (1+R) a \D_c(\D^a\tau
_a)+a\D_c(\D^2 \ln T)\simeq - (1+R)a \D_c (\div A).  \label{L011}
\end{equation}
Substituting (\ref{L009}) into (\ref{L011}) and using the fact that  
$H \D^a A_a \approx {\cal O}(\epsilon_H \epsilon)$, we obtain
\begin{eqnarray}
&- 3[a(1+R)(a\D_c\ln T)^{\dot{}}~]^{\dot{}}+a^2\D_c(\D^2 \ln T) \nonumber
\\ &~~~~~~~~~~~~~~~~\simeq 
 [a(1+R)(a\D_c\Theta )]^{\dot{}} - a(1+R)\D_c(\div A)\;. 
\end{eqnarray}
Finally, transforming to conformal time, $dt=ad\eta$ gives 
\begin{eqnarray}
&3[(1+R)(a\D_c\ln T)^{\prime }]^{\prime} - a^2\D_c(\D^a(a\D_a\ln T)) \nonumber
\\ &~~~~~~~~~~~~~~~~\simeq - [a(1+R)(a\D_c\Theta )]^{\prime} 
+ a^2(1+R)\D_c(\div A)\;.
\end{eqnarray}
On using the small-scale linearisation scheme described in section 
(\ref{small_scale}) we find, on dividing through by $3(1+R)$, the
oscillator equation:
\begin{equation}
(a\D_c\ln T)^{\prime \prime } \approx
 + {\frac{a^3}{3(1+R)}}\D_c(\D^2 \ln T) 
 - a^2 (\D_c \Theta)^{\prime} 
 + {\frac{a^2}3}\D_c(\D^a A_a)\;. 
\label{acoustic-source}
\end{equation}
This is the covariant harmonic-oscillator equation  which 
describes the acoustic modes \footnote{This can also be 
obtained from a two-fluid CGI description \cite{dunsby}, 
as well as from the imperfect-fluid description \cite{RT} 
-- the point here is that we have derived it from a self 
consistent kinetic theory approach, listing along the way the 
necessary physical approximation required to reduce it to the
standard acoustic oscillator form.}. We can compare it to the usual
gauge invariant result found in the Newtonian gauge by transforming 
to the shear-free frame $\tilde u_a=n^a$ where $\D_{\la a} n_{b \ra} =0$. 

We will investigate this equation in more detail 
in the next section and relate our results to those in the  
standard literature (which are expressed in the Newtonian gauge).

\subsubsection{The Newtonian Frame Oscillator Equation}

In this section we recover the harmonic oscillator
equation of Hu and Sugiyama in the Newtonian gauge \cite{HS95a} 
from the CGI formalism.
The difference between this and the previous section is that
here we apply the small scale approximation at the very end of
the calculation.

Using the Newtonian frame choice $n^a \approx u^a + v^a_N$, 
$\D_{\la a} n_{b \ra} =0$, equations (\ref{mom-flux}) and
(\ref{mono_temp}) become 
\begin{eqnarray}
\dot{\tilde \tau}_a + {\dot{R} \over 1+R} {\tilde \tau}_a + {1 \over 1+R}
{\tilde D}_a (\ln T) \approx - {\tilde A}_a \approx -\D_a \Phi_A\;, 
\label{ng-dot-tau}\\
( {\tilde D}_a \ln T )^{\dot{}} + H ({\tilde D}_a \ln T + {\tilde A}_a)
+ \frac13 {\tilde D}_a {\tilde \Theta} \approx - \frac13\D_a (D_c {\tilde
\tau}^c)\;. \label{ng-pertT}
\end{eqnarray}
These can be re-written and put into the following form by using the
almost-Friedmann-Lema\^{\i}tre Einstein field equations, together with the transformation 
relations given in Appendix G:
\begin{eqnarray}
(D_a {\tilde \tau}^a)^{\dot{}} + H (\D_a {\tilde \tau}^a) 
&\approx& - {\dot{R} \over 1+R} (\D_a {\tilde \tau}^a)  - {1 \over 1+R}
({\tilde D}^2 \ln T) - (D^2 \Phi_A)\;, \label{D-ng-dot-tau}\\
D_{\la a} {\tilde D}_{ b \ra} (\ln T)^{\dot{}} &\approx& 
-\D_{\la a}\D_{b \ra} \dot{\Phi}_H 
- \frac13\D_{\la a}\D_{b \ra} (D_c {\tilde \tau}^c)\;. 
\label{pstf-ng-pertT} 
\end{eqnarray}
Taking the time derivative of (\ref{pstf-ng-pertT}) and substituting
into equation (\ref{D-ng-dot-tau}) after first taking PSTF derivatives
we obtain the full equation for $\D_{\la a} \D_{b \ra} \ln T$:
\begin{eqnarray}
( \D_{\la a} \D_{b \ra} \ln T)^{\ddot{}} &+& \dot{H} ( \D_{\la a} \D_{b \ra} 
\Phi_A + 2 \D_{\la a} \D_{b \ra} \ln T) + H \left[{ ( \D_{\la a} \D_{b \ra}
\Phi_A)^{\dot{}} + 2 (\D_{\la a} \D_{b \ra} \ln T)^{\dot{}} } \right] 
\nonumber \\
&\approx& - (D_{\la a}\D_{b \ra} \Phi_H)^{\ddot{}} - 2 \dot{H} \D_{\la a}
\D_{b \ra} \Phi_H - 2 H ( \D_{\la a} \D_{b \ra} \Phi)^{\dot{}}
+ {1 \over 1+R}  \D_{\la a} \D_{b \ra} \D^2 \ln T\nonumber\\ 
&-& \D_{\la a} \D_{b \ra} D^2 \Phi_A 
+ \left[ {{\dot{R} \over (1+R)} - 3 H} \right]
\left( {\D_{\la a}\D_{b \ra} (\ln T)^{\dot{}} 
+ \D_{\la a} \D_{b \ra} \dot{\Phi}_H } \right)\;.
\end{eqnarray}
On dropping all terms $O(\epsilon \epsilon_H)$, we once again obtain the 1+3
covariant form of the small scale Newtonian frame oscillator equation
(without using a mode expansion):
\begin{eqnarray}
( \D_{\la a} \D_{b \ra} \ln T)^{\ddot{}}&+&{\dot{R} \over (1+R)}
\D_{\la a}\D_{b \ra} (\ln T)^{\dot{}}   
 + {1 \over 1+R}  \D_{\la a} \D_{b \ra} \D^2 \ln T \nonumber \\
&\approx& - (D_{\la a}\D_{b \ra} \Phi_H)^{\ddot{}} + 
{{\dot{R} \over (1+R)}}{\D_{\la a} \D_{b \ra} \dot{\Phi}_H }
 - \D_{\la a} \D_{b \ra} \D^2 \Phi_A\;. 
\end{eqnarray}
The techniques used to derive the above equation become useful latter
when dealing with the non-linear terms as they avoid the complication
of mode-mode couplings when understanding the qualitative features
of various effects \cite{MGE}. 

Upon using the mode decomposition definitions for the temperature 
perturbations, radiation dipole and
the scalar Newtonian and curvature perturbations respectively
{\footnote{We use, as before, $
 {\tilde D}_a \ln T \approx \ts\frac{k}{a} \delta {\tilde T} Q_a$ , 
 ${\tilde \tau}_a \approx {\tilde \tau}_1 Q_a$ ,
 $D_a \Phi_A =\ts\frac{k}{a} \Phi_A Q_a$ and  
 $\D_a\Phi_H \approx \ts{k \over a} \Phi_H Q_a$. }},
we can write the mode decomposition of (\ref{D-ng-dot-tau}) and 
(\ref{pstf-ng-pertT}) as:
\begin{eqnarray}
\dot{\tilde \tau}_1 &\approx& - {\dot{R} \over 1+R} {\tilde \tau}_1 
- {1 \over 1+R}  \frac{k}{a} \delta {\tilde T} - {\frac{k}{a}} \Phi_A, 
\label{mode-dipole} \\
\delta \dot{\tilde T} &\approx&  - H \Phi_A  - \dot{\Phi}_H 
+ \frac{k}{a} \frac13 {\tilde \tau}_1\;. 
\label{mode-mono}
\end{eqnarray}
Upon ignoring the expansion coupled term (i.e the small scale 
approximation) and including the curvature fluctuations, since this 
makes the resulting equations applicable up to the 
Jeans length (above which the matter would not be gravitationally
bound), and substituting the
second equation (\ref{mode-mono}) into (\ref{mode-dipole}) we
obtain the well know equation describing the acoustic oscillations in 
the radiation \cite{HS95a}:
\begin{eqnarray}
\delta {\tl T}'' + {R' \over 1+R} \delta {\tl T}' + k^2 c_S^2 \delta {\tl T}
 \approx - \Phi_H'' -
{R' \over 1+R} \Phi'_H - {k^2 \over 3} \Phi_A\;. \label{temp_osc_ng}
\end{eqnarray}
Here we have used conformal time (since we are now working in the
conformal Newtonian frame).
\subsubsection{The Dispersion Relations and Photon Damping
Scale}\label{sssec-dispersion}
In this section we derive the dispersion relations for small scale
anisotropies, where the focus is once again on developing generic
covariant equations in parallel to the usual gauge invariant  
treatments \cite{Kaiser,kosowsky,HS-astro}. 
To this end, we begin by iterating the baryon velocity 
equation in much the same manner as we iterated the integrated 
Boltzmann equations for the radiation.

We begin with the the baryon relative velocity equation (\ref{L004})
which is once again inverted in order for it to take the form 
in (\ref{L005}). This is then turned into the basis of an 
iteration scheme in terms of the scattering time:
\begin{equation}
{v_{(n)}}^a_B \approx \tau_a - {R \over \dot\kappa} \left[ 
{ \dot{v}^a_{(n-1)B} + H {v^a_{(n-1)}}_B + A^a } \right]\;. \label{v-iter}
\end{equation}
Using this equation and the zero-th order tight-coupling approximation, 
$v^a_{(0)B} \approx \tau_a$, the covariant second order baryon 
velocity equation is found: 
\begin{eqnarray}
v_a^B \approx \tau_a - { R \over \dot\kappa} \left[ {\dot{v}_a^B + H v_a^B 
+ A_a} \right] + {R^2 \over \dot{\kappa}^{2}} \left[{\ddot{v}_a^B + (\dot{H} 
+ H^2) v_a^B + \dot{A}_a + H A_a + 2 H \dot{v}_a^B}\right]\;, 
\label{v-2nd}
\end{eqnarray}
where we have dropped terms of ${\cal O}(\dot\kappa^{-3})$.
We now consider the small scale version of this equation, by ignoring
terms scaled by the Hubble parameter $H$ and  
effects due to the gravitational potentials (see (\ref{small_scale})):
\begin{eqnarray}
v_a^B \approx \tau_a - {R \over \dot\kappa} \left[ {\dot{\tau}_a + A_a} 
\right] + {R^2 \over \dot\kappa^2} \ddot{\tau}_a\;. 
\label{s_b_vel}  
\end{eqnarray} 
Expanding the transport equation for the second order radiation quadrupole
(\ref{radf_0014}) to first order and ignoring the shear contribution 
which is negligible in the almost-Friedmann-Lema\^{\i}tre small scale limit, we obtain
\begin{equation}
\tau^{ab} \approx - t_c \D^{\la a} v^{b \ra}_B \approx 
- \dot\kappa^{-1} \D^{\la a} \tau^{b \ra}\;, 
\label{tab-s}
\end{equation}  
where again we retain only first order terms. We now
substitute (\ref{s_b_vel}) and (\ref{tab-s}) into the radiation dipole
evolution equation (\ref{L002}) to find: 
\begin{eqnarray}
- \dot{\tau}_a = ( A_a + \D_a \ln T) - R (\dot{\tau}_a + A_a) +
R^2 \dot{\kappa}^{-1} \ddot{\tau}_a - \ts{2 \over 5} \dot{\kappa}^{-1} \D^c
{\D_{\la a} \tau_{c \ra}}\;, \label{rad-dipole-2nd}
\end{eqnarray}
which can be compared to the mode 
equation A-11 in Hu and Sugiyama \cite{HS-astro}). This is the 
key equation from which we will now proceed to recover the dispersion 
relations and hence standard damping scale results. The key-point here
is that the diffusion damping is second order in the scattering time, 
while the acoustic oscillations are first order.

We now take a covariant mode expansion of the necessary quantities: 
$\tau_a = \tau_1 Q_a$, $A_a = A_k Q_a$ and 
$\D_a \ln T = (\ts\frac{k}{a}\delta T) Q_a$, together with the
well known result (see Paper I \cite{GE} for the general 
case): $\D^b Q_{ab} = - \frac23 (a k)^{-1} (-k^2 + 3K) Q_a$.   
We consider only the flat case here, so we set $K=0$. Putting these
all into the second order radiation dipole equation (\ref{rad-dipole-2nd})
we obtain:
\begin{eqnarray}
- \dot{\tau}_1 \approx (A + (\ts\frac{k}{a}\delta T)) + R ( \dot{\tau}_1 + A) + R^2
\dot{\kappa}^{-1} \ddot{\tau}_1 + \ts{4 \over 15} \dot{\kappa}^{-1} 
{k^2 \over a^2} \tau_1\;. \label{rad_dip_mode_2nd}
\end{eqnarray}
Now we use the WKB approximation: 
\begin{equation}
 \tau_1 \propto \exp i \int (\omega / a) dt\;,
\end{equation}
and drop terms scaled by $H$ and $a \approx R$. This gives 
\begin{eqnarray}
- i (1+R)\ts{\omega \over a} \tau_1 \approx [(\ts\frac{k}{a}\delta T) + (1+R) A] 
- R^2 \dot\kappa^{-1} \ts{\omega^2 \over a^2} \tau_1 + \ts{4 \over 15} 
\dot{\kappa} ^{-1} \ts{k^2 \over a^2} \tau_1\;. \label{rad-WKB}  
\end{eqnarray}

Now, in order to deal with the terms arising from $A_a$ and $\D_a \ln T$ we
consider the covariant radiation monopole perturbation equation 
(\ref{mono_temp}):
\begin{equation}
(\D_a \ln T)^{\dot{}} + \frac13 \D_a \Theta + H ( \D_a \ln T + A_a) 
\simeq - \frac13 \D_a (\D_c \tau^c)\;. \label{monopole} 
\end{equation}
We find, on taking a mode expansion and applying the WKB approximation
again, dropping terms scaled by $H$ and the expansion gradient (the 
small scale approximation described in section 2.5), that
\begin{equation}
(\ts\frac{k}{a}\delta T) \approx  {\ts \frac13 {k^2 \over a}} 
(-i \omega) \tau_1\;. 
\label{lnT-WKB}
\end{equation}
Upon substituting (\ref{lnT-WKB}) into (\ref{rad-WKB}), factoring out
the dipole coefficient $\tau_1$, writing the differential optical in
terms of conformal time:  
$\dot{\kappa}^{-1} = (a \kappa^{\prime})^{-1}$ and multiplying through by 
$i \omega a$ we get:
\begin{equation}
 \omega^2 (1+R) + R^2 {\kappa^{\prime}}^{-1} \omega^2 (i \omega)
 - {\ts{4 \over 15}} {\kappa^{\prime}}^{-1} k^2 (i \omega) \approx  
 {\ts{1 \over 3}} k^2\;. \nonumber 
\end{equation} 
On re-arranging terms we finally obtain
\begin{equation}
\omega^2 \approx \ts{k^2 \over 3(1+R)} + k^2 ({i \omega 
{\kappa^{\prime}}^{-1}}) \left( {\ts{R^2 \over 3(1+R)} 
+ \ts{4 \over 15}} \right)\;. 
\end{equation}
Splitting $\omega$ up into its natural frequency $\omega_0$ and 
the diffusion damping term $\gamma$ and then solving the 
quadratic for the frequency $w$ we obtain
\begin{equation}
\omega \approx \pm \omega_0 + i \gamma\;,
\end{equation}
where
\begin{eqnarray}
\omega_0 \approx \ts{k \over \sqrt{3 (1+R)}} \approx c_s k, 
~~\mbox{and}~~
\gamma \approx  k^2 \left( \ts{{\kappa^{-1}}^{\prime} \over 6} \right) 
\left[ \ts{R^2 + \frac45 (1+R) \over (1+R)^2} \right] \approx k^2/k_D^2\;.
\label{dif-damp}
\end{eqnarray}
Here $k_D$ is the diffusion damping scale and $c_s$ the baryotropic speed 
of sound in the matter. We get oscillations when $k<k_D$ (see the next 
section), and diffusion damping when $k> k_D$, 
leading to a damping envelope.
 
The above covariant result is equivalent to the 
analytic small scale approximation scheme 
of \cite{HS95a,HS95b,HS-astro}, who have 
demonstrated its robustness and closeness to more precise numerical 
studies\footnote{For numerical integrations of the covariant scalar 
temperature anisotropy equations we would refer the reader to \cite{cl2}.}. 

This derivation can be easily corrected to include the effect 
of anisotropic scattering (which breaks the tight-coupling approximation) 
and polarization. This is done 
by correcting the scattering terms in the calculation of the damping
scale. We follow the approach of Kaiser  
\cite{Kaiser,HS95a,HS95b,HS-astro} and include the 
correction $f_2$ to find the modified damping factor:
\begin{equation}
\gamma^* \approx  k^2 \left( \ts{{\kappa^{-1}}^{\prime} \over 6} \right) 
\left[ \ts{R^2 + \frac45 f^{-1}_2 (1+R) \over (1+R)^2} \right] 
\approx k^2/{k^*}_D^2\;,
\label{dif-damp-correct}
\end{equation} 
where, first, for the anisotropic effect, $f_2 = \frac{9}{10}$, and 
second, to compensate for the polarization, $f_2 =
\frac{3}{4}$. Hence we get the diffusion damping envelope 
for $\tau_1$ when $k > k_D$.                    
\subsubsection{The Temperature Oscillations}
To examine the solution when $k < k_D$, we take a mode expansion, 
using the $Q_{A_\ell}$'s defined in Paper I \cite{GE}:
\begin{eqnarray}
 a(\D_c \ln T)^k = k (\delta T) Q_c \Rightarrow ~a\D_c(\D^a\D_a\ln T)
=(-{k^3 \over a})(\delta T) Q_c\;,
\end{eqnarray}
where the driving term is written in terms of a generic 
potential $\Phi_F$ defined by
\begin{eqnarray}
- \frac{a^3}{3} \D_c (\D^2 \Phi_F) \approx - a^2 (\D_c \Theta)^{\prime} 
 + {\frac{a^2}3}\D_c(\D^a A_a) ~~\mbox{and}~~ 
 a^3 \D_c (\D^2 \Phi_F) \approx + \frac{k^3}{3} \Phi_F Q_c\;.
\end{eqnarray}
{}From (\ref{acoustic-source}) we find the oscillator equation
(once again working in the Newtonian frame {\it cf.} \cite{HS95b} 
eqn. (B3) and equation (\ref{temp_osc_ng})). 

The solution at first order must be convolved with the damping 
envelope, found from the dispersion relations, in order to include
the damping cut-off. We have waited until the form of this damping
is known, so that it can be easily included by simply replacing the
natural oscillator frequency, $\omega_0$, with $\omega=\omega_0
+ i \gamma$, which now includes the damping. We use equation 
(\ref{acoustic-source}) in a manifestly gauge invariant form, 
to find upon mode expanding 
\begin{equation}
(\delta T)^{\prime \prime }+k^2c_s^2(\delta T) \approx 
- {\frac{k^2}3}\Phi_F\;,
\end{equation}
where the {\it sound speed} $c_s$ is given by $[3(1+R)]^{-\frac 12}$. 
This gives the well known solution (see for example \cite{HS95a,HWa2}
\begin{equation}
(\delta T)(\eta )\approx [(\delta T)(0)+(1+R)\Phi _F]\cos (kr_s)-(1+R)\Phi _F,
\label{acoustic}
\end{equation}
where the {\it sound horizon scale} is given by 
$r_s=\int c_sd\eta \approx c_s\eta $. 
This describes the source term for the acoustic oscillations with the
isocurvature term dropped
{\footnote{
The other part of this solution comes from $\sin (kr_s)$; this is
the isocurvature part ( $\delta T^{\prime }(0)\neq 0$), giving 
the full solution:
$$
(\delta T)(\eta )=\left[ {(kc_s)^{-1}(\delta T)^{\prime}(0)} \right] 
\sin (kr_s)+\left[ { (\delta T)(0)+(1+R)\Phi_F} \right]\cos (kr_s)
-(1+R)\Phi_F.
$$
In this paper we consider adiabatic perturbations only so 
$(\delta T)^{\prime}(0)=0$.}}. 
The last term on the right gives the
Sachs-Wolfe effect, due to the potential; the other terms give the adiabatic
acoustic oscillations.
 
By putting this back into the momentum flux equation (\ref{mom-flux})
in mode form we obtain
\begin{equation}
\tau_1'(\eta,k) \simeq - {1 \over 1+R} (k \delta T) - a k \Phi_F\;,
\end{equation}
and using the temperature anisotropy mode expansion that 
results when (\ref{L0081}) holds, the Doppler contribution 
to the temperature anisotropies at decoupling can be found. 
These are given by
\begin{equation}
\tau _1(\eta,k)\approx (- k^{-1}) 3[(a\ln T)(0)+(1+R)\Phi _u]c_s\sin (kr_s).
\label{Doppler}
\end{equation}
\section{Temperature Anisotropies from Integral Solutions}
In this section we derive the explicit form of the temperature
anisotropies, using the general $u^a$-frame integral solutions
and the tight-coupling approximation solutions. 
 
Before we do this, there are two important points 
that need to be discussed.  
Firstly, where does the high-$\ell$ cut-off come from?
The natural frequency of the system is $\omega_0$ and is set by
the sound speed in the tight-coupled system to first order in 
Thompson scattering time -- the oscillator equation. The diffusion 
damping is a second order effect whose correction is found by 
deriving the dispersion relations at second order in the Thompson
expansion for the baryon and photon momentum equations and using 
the WKB approximation to find the new oscillator frequency to 
 be : $\omega \approx \omega_0 + i \gamma$. 
The assumption that the anisotropies are sourced by these oscillations
in the radiation is the key to the physics; the initial conditions
before free-streaming are $\tau_{A_\ell} \approx 0$ for $\ell>2$
(where the anisotropic correction at $\ell=2$ is included as a 
correction to the damping scale). The free-streaming radiation 
transfer function (the spherical Bessel function in k-space) is then 
convolved with the initial power sourced by the oscillations in the 
average temperature and the dipole (essentially a cosine and sine 
function in k-space respectively). As free-streaming continues 
power is shifted up into the higher $\ell$'s, as the radiation 
transfer functions' maximum is near $\ell \propto k \Delta \eta$ 
(where $\Delta \eta$ is the elapsed time since last scattering).
This maximum moves into higher
$\ell$ for longer times to give one a sense for how the power is 
distributed through the $\ell \pm 1$ moment couplings in the 
almost-Friedmann-Lema\^{\i}tre integrated Boltzmann equations. Obviously at high-$\ell$ (which corresponds to 
high-$k$) the amount of power surviving the $(k/k_D)^2$ damping will 
be very small, and hence as time progresses the peak in the transfer
functions drops off for higher-k, thus giving the high-$\ell$ cut-off.

Given that in free-streaming there is no diffusion damping 
to cut-off the high-$k$ power, the truncation of the 
hierarchy is only consistent and meaningful if the 
significant anisotropy signal is sourced from the initial conditions
at low-$\ell$ near to tight-coupling.

Secondly, why do we find angular variations in the temperature 
anisotropies that are entirely due to oscillations in the dipole 
and monopole of the radiation -- particularly given that 
in relativistic kinetic theory one has time-like integrations 
\footnote{Only in the almost-Friedmann-Lema\^{\i}tre sub-case
can one think of the averaged relationship between the 
null-projection}?
 
This is important for non-linear extensions of the almost-Friedmann-Lema\^{\i}tre 
model \cite{MGE}, since care should
be taken when treating ensemble averaged quantities integrated down
the null-cone as being generally equivalent to those integrated 
down a single time-like world-line. 
The issue will become even more complex once the Gaussian
averaging necessary for the construction of the angular correlation
functions is relaxed. It is this, along with the weak Copernican
principle that makes possible the extension of the data here and now,
as integrated in a little box down a time-like world-line to 
last scattering, to global statements. 

Now we return to the issue of sourcing the temperature anisotropies
from oscillations in the dipole and monopole.
The integral solutions give the full almost-Friedmann-Lema\^{\i}tre solution to the radiation 
anisotropies given the appropriate initial data. 
Generically it is the transfer of power from low-$\ell$ initial data up 
into high-$\ell$, with the accompanying reverse transfer of 
power; the $\ell \pm 2$, $\ell \pm 1$ transfer of power where there 
is a wall at $\ell=0$ but none at high-$\ell$. 
There is strictly no $\ell_{max}$ unless the geometry is exactly
Robertson-Walker \cite{ETMb}. The tight-coupling approximation gives
the monopole and dipole at the $\ell=0$ initial wall. 
Diffusion damping in the initial power near last scattering as well as 
additional damping through slow decoupling cuts off the transfer of 
power. This power is sourced by those initial conditions
and therefore give the temperature anisotropies in terms of the dipole
and monopole alone. Additional integrated effects will
only modify the primary projection, effectively leaving
the cut-off unchanged.

\subsection{Temperature Anisotropies} 

The integral solution gives the projection of these conditions at 
last scattering onto the current sky. These results are covariant. 

The diffusion damping can be found from the dispersion relations arising 
from the coupled baryon-photon equations using the WKB method in 
tight-coupling which was described in section 6.2. We can now 
construct the primary source temperature anisotropies, 
by first considering the acoustic and Doppler 
contributions in the free-streaming projection (\ref{freestreaming}). 
Our solutions for ${\cal B}_0$ and ${\cal B}_1$ are first order in  
scattering time expansion hence to include the second order diffusion 
cut off, we multiply the solution through by $\exp{\gamma}$ 
(or $\exp{\gamma^*}$ if we want to include the corrections for the 
anisotropic and polarization effects).

We can re-write $\gamma$ or $\gamma^*$ in terms of the damping scale, 
$k _D$ ( {\it cf} \cite{HS95b} eqn (A7)-(A8)). 

\subsubsection{Sources of Temperature Anisotropies}

In the manifestly gauge invariant integral solution of the temperature
anisotropies in almost-Friedmann-Lema\^{\i}tre spacetime, using the slow decoupling
approximation with scalar perturbations (\ref{slowdec_1},\ref{slowdecoupling}),
 there is additional complexity of having to deal with shear terms. 
We therefore choose the Newtonian frame in which  
the shear contribution does not appear\footnote{Incidently this would also 
remove any problems we may have with the introduction of high-$\ell$ 
truncation as discussed in Appendix E}. 
We are then able to recover the canonical scalar treatment 
\cite{W83,S89,GSS,HS95a,HS95b}. This can be written out in terms 
of the scalar perturbation potentials $\Phi_A$ and $\Phi_H$, the temperature
perturbation, $\delta T$, the radiation dipole $\tau_1$ (through slow
decoupling) and the baryon relative velocity, $v_B$, all within the 
covariant approach (including Doppler contributions 
(\ref{slowdecoupling-ng})):
\begin{eqnarray}
{\beta_{\ell} \tau_{\ell}(\eta_0) \over (2 \ell+1)} 
\approx \underbrace{{\bf S}_{P}(k,\eta_*) j_{\ell} (k \Delta \eta_*)}_{\sf
 primary~sources} + \underbrace{\int_{\eta_*}^{\eta_0} d \eta 
\left( {{\bf S}_{DISW} + {\bf S}_{ISW}} \right) j_{\ell}(k\Delta \eta)}_{\sf
secondary~sources}\;.
\label{temp-anisotropy}
\end{eqnarray}
Here the source terms are given by:
\begin{eqnarray}
{\bf S}_{P}(\eta,k) &=& {\cal D}(\eta_0,k) \left[{ {(\tilde \delta T} + \Phi_A) 
+ \kappa^{\prime} v_B } \right],  \label{primary}\\
{\bf S}_{DISW}(\eta,k) &=& {\cal V} e^{-(k/k_D)^2} \left[ {\frac13 k 
{\tilde \tau}_1 + (\kappa^{\prime} \tilde v_B^{\prime} +
 \kappa^{\prime \prime} \tilde v_B)} \right] \label{doppler} \\
{\bf S}_{ISW}(\eta,k) &=& {\cal V}e^{-(k/k_D)^2} 
\left[{ [\Phi_A^{\prime} - \Phi_H^{\prime}](k,\eta) - a H 
[ {\tilde \delta T} + 3 \Phi_A](k,\eta)} \right]\;.
\label{secondary} 
\end{eqnarray}
In the first line, on the RHS, we have the Sachs-Wolfe and acoustic sourced
projection effects which are the primary sources. The next term describe
secondary Doppler effects during slow decoupling. The last term models
the integrated Sachs-Wolfe (ISW), the late-ISW and early-ISW effects 
respectively. We have used $\Delta \eta_* = \eta_0 - \eta_*$ 
(the conformal time difference between
the surface of last scattering and the time of reception, here and
now) and $\Delta \eta = \eta_0 - \eta$. 

In this way we have recovered the solution from the
exact equations in a systematic manner using the 1+3 CGI formalism
rather than recovering the solution from a perturbation
theory about a foliation of Robertson-Walker surfaces of homogeneity. Note that 
there is no monopole temperature anisotropy in the CGI approach while 
there is one in the canonical treatment. 

The angular correlation functions, $C_{\ell}$, for the general small
temperature anisotropy case have been derived in terms of a superposition of
homogeneous and isotropic Gaussian random fields with respect to the
temperature anisotropy multipoles to find the multipole mean-squares, 
${\la \tau_{A_{\ell}}\tau^{A_{\ell}} \ra}$, in terms of the an 
ensemble average (see Paper I \cite{GE}). The relationship
between the multipole mean-squares (in the 
almost-Friedmann-Lema\^{\i}tre case) and the mode coefficient 
mean-squares $|\tau_{\ell}|^2$ in terms of the Robertson-Walker mode 
functions $G_{\ell}[Q]$ are also given in Paper I \cite{GE}. We do 
not discuss these further here. It must be emphasized that
the multipole mean-squares are given for general geometries, 
while the mode means squares are only for almost-Robertson-Walker geometries. This
is why the application of the non-linear extension is possible; 
where the corrections to the almost-Friedmann-Lema\^{\i}tre standard model are calculated
using the multipole formalism of \cite{MGE}.

\subsubsection{Sachs-Wolfe Effect and the Acoustic Source}
\label{newt-sachs-wolfe}
In the Newtonian frame (\ref{primary}) we find that
the Sachs-Wolfe effect arises from a combination of the
$\tilde {\D}_a \ln T + \tilde A_a \approx \ts \frac{k}{a} (\delta \tl T 
+ \Phi_A) Q_a$. From the solution to the oscillator equation in the 
Newtonian frame (eqn. (\ref{temp_osc_ng})) along with 
$r_s \sim c_s \eta$ and ignoring the time evolution of the 
potential (for the equivalent canonical version see \cite{HS95a}) we find
\begin{equation}
\delta \tl T(\eta,k) + \Phi_A(\eta_*,k) \approx [\delta \tl T(0,k) + (1+R)
\Phi_A(0,k)] \cos (k r_s) - R \Phi_A(\eta_*,k).
\end{equation}
Upon taking the matter dominated limit ($R \approx {\cal O}(\epsilon^2)$) and 
then using the adiabatic assumption in tight-coupling  $\delta T(0,k) 
\sim \frac13 \Delta(0,k) \sim - \frac23 \Phi_{A}(0,k)$ 
(adiabatic flat CDM model) and finally taking  $r^*_s \sim 0$
(by using that $k \eta_* \ll 0$) we recover the usual results as in 
\cite{HS95a,HWa}
\begin{equation}
[\delta \tl T + \Phi_A](\eta_*,k) \approx + \ts\frac13 \Phi_A(0,k) 
(1+ 3 R_*) \cos(k r^*_s) + R_* \Phi_A(0,k) \sim \ts\frac13 \Phi_A(0,k).
\label{sachs-wolfe-newtonian}
\end{equation}
Here we have used that $\Phi_A(\eta_*,k) \approx \Phi_A(0,k)$ from
the Einstein de-Sitter result that the potential is constant if we drop the 
decaying mode. 
The physics of the Sachs-Wolfe effect as opposed to the acoustic 
oscillations is then quite clear. There are no oscillations just
an imprint due to an acceleration potential. Although the adiabatic
assumption is invariant to order ${\cal O}(\epsilon)$, the
relationship between the Electric part of the Weyl tensor
and the perturbations are not -- so we will always use the matter 
perturbation in the total frame where it can be easily related to 
the Newtonian potential and the temperature perturbation in the
Newtonian frame where the oscillator equation takes on a useful
form. In a similar manner we find the explicit form of the 
radiation dipole and acoustic oscillations in the slow decoupling 
era:
\begin{eqnarray}
&&\delta \tl T(k,\eta) \approx \ts \frac13 
\Phi_A(0,k) (1 +3 R) \cos (k c_s \eta) + (1+R) \Phi_A(0,k), ~\nonumber \\
&&\tl \tau_1 (k,\eta) \approx \Phi_A(0,k) 
(1 + 3 R) c_s \sin (k c_s \eta).
\end{eqnarray}

\subsubsection{Weak-coupling for the Small Scale Solution}

Here we briefly consider the integrated Sachs-Wolfe effect in connection with the
weak-coupling approximation. The idea is that the anisotropies 
fall with $\ell$ more rapidly than a simple projection would imply. 
What one has in mind is the situation where the anisotropy 
contributes across many wavelengths of the fluctuation allowing 
cancellations on small scales; the secondary sources, in particular 
the term $({{\Phi}^{\prime}}_A - {{\Phi}^{\prime}}_H)$,   
varies slowly on small scales \cite{HWa}. 
Specifically we consider the situation where
\begin{eqnarray}
\int_{\eta _*}^{\eta _0}d\eta {\cal V} e^{-(k/k_D)^2} 
\left\{{(\Phi_A^{\prime} - \Phi_H^{\prime})
}\right\} j_\ell(k \Delta \eta) \approx \sqrt{{\pi \over 2 \ell}}
{1 \over k} (\Phi^{\prime}_A - \Phi^{\prime}_H)(\Delta \eta_*) 
{\cal D}(k,\eta_0)\;,   
\end{eqnarray}
where we used $\int_0^{\infty} j_{\ell} (k \Delta \eta) d \eta = 
[\sqrt{\pi}/2 k] [\Gamma((\ell+1)/2)/\Gamma((\ell+2)/2)]$. The 
weak-coupling solution is implied by the assumption that:
\begin{equation}
(\ddot{\Phi}_H - \ddot{\Phi}_A ) \ll k (\dot{\Phi}_H - \dot{\Phi}_A)\;.
\end{equation}
This is nothing more than a useful approximation allowing the direct 
construction of analytic solutions. Some care should be taken when
using the weak-coupling approximation when trying to estimate the
early-ISW (the free-streaming analogue of the acoustic driving effect) 
and late-ISW (due to non-vanishing curvature or cosmological constant, 
which will dominate the expansion rate at latter time) effects. Ideally 
one should evaluate the slowly-varying function, which has been taken out
of the integral, at the $\ell$-th peak; $\eta_{\ell} 
= \eta_0 - ({\ell + \frac12}) / k$ rather than at 
$\Delta \eta_* = \eta_0 - \eta_*$. 

In the case of the early-ISW effect, since it satisfies neither the
tight-coupling nor weak-coupling criteria, our approximations schemes
here are not entirely appropriate; its decay time and wavelength 
are comparable \cite{HS95a,HS95b,HWa}. However we can use the 
weak-coupling in the case of the late-ISW effect because cancellation effects
lead to damping on small scales. 
The temperature perturbation decays, and hence 
the potential decays on the order of the expansion time near the end of 
the matter dominated era. The photons will free-stream across many 
wavelengths of the perturbations below the Hubble scale, leading to 
cancellation and damping effects.    

The important point about the late-ISW effect is that there will
be an imprint due to the exit from the matter dominated era into 
a $\Lambda$ or curvature dominated one.
In the context of the equations 
here, we can consider $\Lambda$-dominated effects by investigating
the evolution equations for the potentials in a $\Lambda$-dominated case.

In this case the effective expansion changes, given here for late-times 
(well after matter-radiation equality):
\begin{equation}
H^2 = a^{-3} \Omega_0 H_0^2 + \frac13 \Lambda
\end{equation}
for the $\Omega_0 + \Omega _\Lambda = +1$ model. 
This has however been dealt with in depth in the literature 
(see for example \cite{HWa,HS95b}). 

\subsubsection{The Mode Coefficients}  
  
The angular correlation functions, measured here and now ($x_0^i$) can then 
be computed using the the mode coefficients from Paper I \cite{GE}:
\begin{eqnarray}
C_{\ell} = \frac{2}{\pi} {\beta_{\ell}^2 \over (2 \ell+1)^2} \int_0^{\infty} 
{dk \over k} k^3 | \tau_{\ell}(k,\eta_0)|^2\;.
\label{angular-corr-fn}
\end{eqnarray}
The final step is to construct the temperature
anisotropy solutions in terms of the matter power spectra. We discuss
this in the next section. 
At linear order, there is no important difference
in our solutions from those found in the canonical 
treatment \cite{HS95a,HS95b}, however the advantage is that 
we can easily relate our formulation of the temperature anisotropies 
to the mean-squares of the multipoles ($\tau_{A_\ell}$) and 
hence to the mean-square of generalized 
temperature anisotropies, ($\Pi_{A_\ell}$) \cite{MGE}. This was not 
attainable in the canonical treatment. 

\subsubsection{The Multipole Coefficients}

The mean-square of the multipole moments are related to the almost-Friedmann-Lema\^{\i}tre
mode coefficients by (see Paper I \cite{GE})
\begin{equation}
\left\langle {\tau _{A_\ell}\tau ^{A_\ell}}\right\rangle \approx
{1 \over 2 \pi^2 } \beta _\ell\int k^2 dk|\tau _\ell(k,\eta_0 )|^2\;,
\label{t0_ang_powers}
\end{equation}
giving the {\it angular correlation function} (see Paper I \cite{GE}): 
\begin{equation}
C_\ell= \Delta_\ell (2\ell+1)^{-1} \left\langle {\tau _{A_\ell}\tau
^{A_\ell}}\right\rangle\;. \label{t1_ang_powers}
\end{equation}
This relates the matter fluctuation amplitude at last scattering to the
present day via the Newtonian like potential, which in turn may
be related to the matter power spectrum directly. 
A plethora of numerical studies of Doppler peak features exist in the 
gauge-invariant literature (see for example \cite{hswz} and
\cite{SEf}).

We will now summarize the standard picture of acoustic peak
characteristics before discussing the matter power spectrum.

\subsection{Some `Acoustic peak' Characteristics}

The key features of the standard model of cosmic background 
radiation primary sourced Doppler peaks are listed below 
\cite{HS95a,HS95b,HWa,HWa2,WHb,Jungmann96}. 
The standard model of acoustic peak formation has been given 
in its analytic form above.

\begin{enumerate}

\item The $j-th$ {\it peak positions} (as given in the flat
adiabatic case) is given by 
\begin{equation}
\ell_j \approx k_j \left| {r_{\theta}(\eta) } \right| 
= \left| {r_{\theta} \over r_s}\right| j \pi\;,
\end{equation}
where $r_{\theta}$ is  the comoving angular diameter and $r_{s}$ 
is the sound-horizon near decoupling.
Notice that $j_\ell(k \Delta \eta)$ peaks near $\ell \sim k \Delta \eta$,
where free-streaming projects this physical scale onto the angular scale 
$\theta \Delta \eta$ on the current sky. The first peak is dependent on 
the driving force which is model independent. It provides a way of 
fixing the angular diameter distance -- when using the almost-Friedmann-Lema\^{\i}tre 
assumptions.

\item The acoustic {\it relative peak spacings} are given by
\begin{equation}
k_{j+1} - k_j = k_A \approx {\pi \over r_s} \iff \Delta \ell \sim 
\ell_A = k_A r_{\theta}\;.
\end{equation}
The peak spacing is fixed by the natural frequency of the
oscillator: $\omega = k c_s$, which is independent of the driving force.
The factor $c_s$ is the photon-baryon sound speed.

\item The peak ratios arise from the angular power spectrum
ratio $C_{jk} = C_j /C_k$ found from $C_\ell$ in terms of $|\tau_\ell|^2$ or
$\la {\tau_{A_\ell} \tau^{A_\ell}} \ra$.

\item The {\it damping tail} provides yet another angular diameter 
distance test of curvature \cite{HWa} via the damping 
scale $k_D$. The peak spacings 
$\Delta \ell$ and the damping tail location $\ell_D$ depend only on the 
background quantities, they are robust to model changes, assuming that 
secondary effects do not overwhelm the signal. The diffusion scale near 
decoupling is the angular scale of the wavenumber 
$k_D \sim \sqrt{\dot{\kappa} / \Delta \eta} 
\sim \sqrt{a / \Delta \eta \dot{a}} \sim \Delta \eta ^{-1}$. 
Given that the damping tail of the acoustic oscillations takes on the 
form $e^{-(k /k_D)^2}$,  $k_D$ can be
found from (as derived previously):
\begin{equation}
k_D^{-2} \approx \frac16 \int_{\eta_*}^{\eta_d} 
d \eta {1 \over {\kappa^{\prime}}} 
{R^2 + \ts{4 \over 5} f^{-1}_2 (1+R) \over (1+R)^2}\;,
\end{equation}
for no anisotropic stress \cite{HWa,Kaiser}. This can be used to find
that the damping tail location is $\ell_D= k_D r_{\theta}$. 

\item The damping tail position to peak scale, $\ell_D /
\ell_A$, is a good measure of the number of peaks and gives an 
indication of the delay in recombination independently 
of the area distance.

\end{enumerate}

\section{The Matter Power Spectrum}

In this section we related the temperature anisotropies in the
case of the Sachs-Wolfe effect to the matter 
power spectrum using the 1+3 CGI formalism. We also deal with basics 
of the normalization of the cosmic background radiation angular power spectrum 
to the matter power spectrum. 

Up to this point everything we have derived has been 
found covariantly from general relativity and 
from relativistic kinetic theory, however on small scales 
we do not have a covariant analytic derivation of the 
transfer function, although it is well known for large scales, so
this is the only gap in our treatment \footnote{There is 
still some hope that the almost Friedmann-Lema\^{\i}tre Quasi-Newtonian models, 
will be useful in this regard \cite{HvEE} providing us 
a covariant derivation of the Harrison-Zel'dovich 
spectrum.}. 

The aim here is to try to predict the matter distribution
today from the cosmic background radiation spectrum. From 
the cosmic background radiation spectrum one finds the matter 
power spectrum (as fixed by the cosmic background radiation 
data today) from which the current 
distribution of matter is found (in terms of $\sigma_8$). 
 
Briefly, in the literature there have been two methods 
for normalizing the cosmic background radiation data in relation to COBE:
{\it firstly} the $\sigma(10^0)$ normalization where the
r.m.s temperature fluctuation was averaged over a $10^0$
FWHM beam, and {\it secondly} the $Q_{rms-PS}$\footnote{$Q_{rms-PS} \approx 
\ts{5 \over 4 \pi}C_2$ and $ C_{\ell}^{\!_{SW}} \sim \ts{6 \over \ell(\ell+1)}
C_2$.} normalization which
uses the best fitting amplitude for a $n=1$ Harrison-Zel'dovich (HZ)
spectrum quoted for the quadrupole, $\la{Q}\ra$. The first method 
had the advantage of being a model independent way of fitting the data 
-- it is observationally determined, however, because of this it 
does not provide the most accurate normalization for a specific model 
and care must be taken in order to properly deal with 
cosmic variance. The second approach is model dependent and 
works well only for the HZ spectrum \cite{EBW,BSW}.
 
Improved, more general normalization schemes have now been
developed and there currently seem to be two favoured ones, both
having variations of the HZ spectrum in mind. It is these two
schemes we now briefly consider here. We emphasis the
so called {\it CDM-models} with $\Omega_0=1$. These models
have a baryon fraction $\Omega_B$ with the rest of the matter
being made up of massive Cold Dark Matter (CDM). Of these the 
most easily dealt with is the so called {\it standard-CDM}
model, where initial fluctuations are assumed to have a Gaussian 
distribution and have adiabatic, scalar density fluctuations with a HZ 
spectrum on large scales: 
\begin{eqnarray}
 {P}(k) \propto k^{n-1}\;,~~~ n=1\;,~~~ H_0=50\mbox{km}^{-1}\mbox{Mpc}^{-1}\;,
 ~~~ \Omega_B=0.05\;. 
\end{eqnarray}
There are three other popular models: the {\it Adiabatic CDM model} of Peebles,
which is a version of {\it standard-CDM}; the {\it ICDM} model (which
is an isocurvature version of the {\it Adiabatic CDM model}; 
and lastly the $\Lambda${\it CMD model} which is the
large cosmological constant version of {\it standard-CDM}. More
recently Hu has introduce a Generalized Dark Matter model (GDM) 
\cite{HuGDM}.
 
\subsection{The Power Spectrum}

The Power Spectrum ${\cal P}(k)$ is fully specified by a {\it shape}
and {\it normalization} (see Appendix J for additional details 
and references). 
The normalization is fixed by the amplitude of the temperature fluctuations 
in the cosmic background radiation\footnote{It seems that models normalized to COBE which are 
fixed in both the amplitude at small scales and by the shape predicted by 
{\it standard-CMD}, are inconsistent with observations \cite{KSper,BSW}.}. 

There are two normalization scales, large and 
small (see Appendix J).  
On small scales the normalization is expressed in terms of $\sigma_8$, the
ratio of the r.m.s mass fluctuations to the galaxy number fluctuations
- both averaged over randomly located spheres of radius $8 \mbox{h}^{-1}
\mbox{Mpc}$; this is the variance of the density field in these spheres.
 
On large scales the power spectrum is 
found by fixing the shape function \cite{E-scot36} which is a
measure of the horizon scale at matter radiation equality. We look
at these now for the case of a flat matter dominated almost-Friedmann-Lema\^{\i}tre model
(the open case is considered in Appendix K).

In the 1+3 CGI approach the emphasis is on the
quantity $a \D_a \ln \rho_{\!_M} \approx a \delta_{\!_M} Q_a$. Here
$\delta_{\!_M}= a \delta_{\!_M}(0,k)$ and $\la \delta_{\!_M}(k,0) 
\delta_{\!_M}(k',0) \ra = (2 \pi)^3 {\cal P}(k) \delta(k-k') /k^2$
for $|\delta_{\!_M}(k,0)|^2 = {\cal P}(k)$.
It becomes preferable in the context here to use the variable
$\Delta$: $\ts\frac{k}{a}\Delta(k,t) = \delta_{\!_M}(k,t)$\footnote{
$\Delta(k)^2 \approx \ts\frac{a^2}{k^2} | \delta_{\!_M}(k)|^2 
= \ts\frac{a^2}{k^2} A k^{n+1}$} . We will,
however, be using the power spectrum for the gauge invariant 
acceleration potential because we consider the situation of
matter dominated adiabatic perturbations. In the general situation it
is best to use $\bar \Phi$. Note, in the matter dominated situation 
we can use that $\Phi_A = - \Phi_H$ and write everything in terms 
of either $\Phi_H$ or $\Phi_A$.

The relationship between the power spectrum of the acceleration
potential and the power spectrum of the matter perturbation is
\begin{eqnarray}
| \Phi_A(k,0) |^2 = P_{\Phi_A}(k),~~\mbox{and}~~
| \Delta(k,0) |^2 = P(k), \label{flatpower}
\end{eqnarray}
where these are then related to each other in the total frame via 
the constraints for the electric part of the Weyl tensor for the 
flat case:
\begin{eqnarray}
P_{\Phi_A}(k,0) = \left( {\ts\frac{3}{2} H_0^2 \Omega_0 \frac{D}{a}} 
\right)^2 k^{-4} P(k).
\label{phi-power-D}
\end{eqnarray}
Here  $\bar \Phi = - \ts{k^2 \over a^2} \Phi_A$ and as noted before:
\begin{eqnarray}
( \bar \Phi \rho_{\!_M}^{-1})(k,t) \approx -(3 H_0^2 \Omega_0)^{-1} D(t)
[k^2 \Phi_A(k,0)].
\label{phi-power-rel}
\end{eqnarray}
and $D \approx a \approx \eta^2$ for a flat almost-Friedmann-Lema\^{\i}tre dust model. We
will also use $a_0 =+1$.

\subsection{Relating the Power Spectrum to Temperature Anisotropies}

We now combine the CGI 
Sachs-Wolfe effect (due to potential fluctuations) and the matter
power spectrum to express the anisotropies in terms of the
matter power spectrum near decoupling. 

\subsubsection{Flat Matter Dominated Almost-Friedmann-Lema\^{\i}tre}
We first consider the flat almost-Friedmann-Lema\^{\i}tre model. Using (\ref{flatpower}) 
and (\ref{sachs-wolfe-newtonian}, \ref{primary} and \ref{temp-anisotropy}) 
(that is with $K=0$) we obtain  
\begin{eqnarray}
{\tau^{\!_{SW}}_\ell(\eta_0) \beta_\ell \over (2\ell+1)} \sim  + \ts\frac13
\Phi_A(k,\eta_*) j_\ell(k \Delta \eta)\;, \label{sw-again}
\end{eqnarray}
Here we find the angular power spectrum 
from (\ref{angular-corr-fn}), (\ref{sw-again}) and (\ref{phi-power-D}):
\begin{eqnarray}
C^{\!_{SW}}_{\ell} =  \frac{2}{\pi} \left({ H_0^2 \Omega_0 \ts\frac{D_*}{a_*} 
}\right)^2 \frac{1}{4} \int \frac{dk}{k^2} P(k) j_{\ell}^2( k \Delta \eta_*)\;. 
\end{eqnarray}
where $P(k)$ is given by either (\ref{power_sA}) or
(\ref{power_sB}). Now if use that $(\Omega_0 D_*/a_*)^2 \approx \Omega_0^{1.54}$
we find the standard result \cite{EBW}:
\begin{eqnarray}
C^{\!_{SW}}_\ell =  \frac{1}{2 \pi} H_0^4 \Omega_0^{1.54}
\int \frac{dk}{k^2} P(k) j_\ell^2( k \Delta \eta_*)\;. 
\label{flat-power}
\end{eqnarray}
The next step is to normalize the matter power spectrum to any given 
structure formation theory on both small and large scales, this is done 
in the next section  in the case of the {\it standard-CDM model}.
\subsection{Setting the CDM Normalization}
{}From the previous section we can now relate the angular
power spectrum for the matter dominated flat ($K=0$) almost-Friedmann-Lema\^{\i}tre model to
the current matter power spectrum on small and large scales
using the two different normalizations to the {standard-CDM model}. 
In what follows we will use the useful result:
\begin{eqnarray}
\int_0^{\infty} {dz \over z^m} j_{\ell}^2 (z) = {\pi \over 2^{m+2}}
{m! \over (m/2)!}{ (\ell - \frac{m}{2} - \frac{1}{2} )! \over 
(\ell + \frac{m}{2} + \frac{1}{2})!}\;. 
\label{soln-kj}
\end{eqnarray}
\subsubsection{Large Scales}
{}From (\ref{flat-power}) and $P(k) = A k^{n-1}$ (\ref{power_sA}) we obtain
\begin{eqnarray}
C_{\ell} =  {\frac{1}{2 \pi}} A H_0^4 \Omega_0^{1.54}
\int \frac{dk}{k^2} k^{n-1} j_l^2( k \chi)\;,
\end{eqnarray}
and on using (\ref{soln-kj}) for $m=2$ ($n=1$) we find that
\begin{eqnarray}
C_{\ell} =  {\frac{A}{2}} H_0^4 \Omega_0^{1.54}
{1 \over (2\ell + 3)(2 \ell +1) (2 \ell -1)}\;.
\label{Cl-CDM-lrgs}
\end{eqnarray}
Of more immediate interest is the angular correlation function
for matter below the horizon scale near decoupling, as this
is what is used to normalize the angular correlation function. 
\subsubsection{Small Scales}
For small scales one can use the $\sigma_8$ normalization 
via (\ref{flat-power}) and $P(k)=B T^2(k) k^n$ (\ref{power_sB}) and the 
parametrized transfer function (\ref{transfer-fn-efs}). This gives
\begin{eqnarray}
C_{\ell} \simeq  {\frac{1}{2 \pi}} H_0^4 \Omega_0^{1.54}
\int \frac{dk}{k^2} B k^n T^2(k) j_{\ell}^2( k \Delta \eta_*)\;.
\end{eqnarray}
For example using the {\it standard-CDM model} with $n=1$, $h=0.5$,
$\Omega_B=0.05$ and $\Gamma=0.48$ we can relate $B$ and
$A$ to the scale of fluctuations near horizon crossing and the
quadrupole: 
\begin{eqnarray}
B = 2 \pi^2 A = (6 \pi^{2/5}) 
\la {Q} \ra / T_0^2\;. 
\end{eqnarray}
To normalize to $\sigma_8$ we can choose $\sigma_8 \simeq 1.3$. 
This then allows us, in principle, to invert
the angular correlation function to find the matter 
power spectrum once the initial spectrum is known.

We are however more concerned with normalizing the radiation angular 
correlation function to CDM on horizon scales near decoupling. In 
this regard we once again begin with the matter power 
spectrum (we set $T(k)=1$) and on using (\ref{soln-kj}) 
for $m=1$ we find that:
\begin{eqnarray}
C_{\ell}^{(CDM)} =  \left[{ {\frac{1}{2 \pi}} H_0^4 \Omega_0^{1.54}
B {{\pi^{\frac12}} \over 8} }\right] {1 \over{\ell(\ell+1)}}\;. 
\label{Cl-CDM-smls}
\end{eqnarray}
\subsubsection{$C_\ell$ Normalized to Adiabatic CDM}
Finally we normalize the angular correlation function 
$C_{\ell}$ (found from \ref{slowdecoupling}) to the 
potential fluctuations normalized to Adiabatic CDM (\ref{Cl-CDM-smls}) above: 
\begin{equation}
D_{\ell} =[{C_{\ell} / C_{\ell}^{(CDM)}} ]
= \left[{ {\frac{1}{2 \pi}} H_0^4 \Omega_0^{1.54}
B {{\pi^{\frac12}} \over 8}}\right]^{-1} \ell (\ell+1)C_{\ell}\;.  
\end{equation}
The angular correlation function $C_{\ell}$ can be found
from the primary (\ref{primary}) and secondary (the integrated part of
\ref{slowdecoupling}) source that make up the total: 
$C_{\ell}=C_{\ell}^{(P)}+C_{\ell}^{(S)}$. This then allows 
one to remove that part of the angular correlation 
function arising from the standard Sachs-Wolfe potential fluctuations 
leaving a signal which is dominated by the photon primary and
secondary sourced physics. The convention is to 
use $D_{\ell}$ rather than $C_{\ell}$ \cite{Jungmann96}, so we  
have from (\ref{flat-power} and \ref{Cl-CDM-smls}) for the flat case:
\begin{eqnarray}
D_{\ell} = \ell(\ell+1) C_{\ell} = \int {dk \over k^2} {\cal P}^*(k) 
| {j_{\ell}(k \chi)} |^2 ~~~\mbox{using}~~{\cal P}^*(k)= \left( {k \over k_s} 
\right)^{n_s + \alpha \ln (k/k_s)}\;, 
\label{pstar}   
\end{eqnarray}
where $k_s$ is the normalization scale, $\alpha$ gives the deviation 
from the power law, and $n_s$ give the scalar power law index. The
angular power per $\ln \ell$ is $(\ell(\ell+1)/4 \pi) C_{\ell}$.

\section{Conclusions}
In this paper we have carried out a covariant analytic  
time-like integration reproducing the well known primary
effects generating the `acoustic' peaks measured here and now 
for an almost-Friedmann-Lema\^{\i}tre universe with adiabatic scalar perturbations.
We have also demonstrated how, in the CGI formalism, 
the angular correlation functions are constructed in terms of 
the matter power spectrum and normalized on large 
and small scales for {\it standard-CDM}, given appropriate 
approximations for the transfer functions. 
 
As pointed out initially, the aim of this paper was to 
clarify the link between the standard gauge-invariant 
and CGI treatments of cosmic background radiation anisotropies 
and provide a strong 
basis from which to tackle non-linear and gravitational 
wave effects using CGI methods. 

Some of the key outstanding issues are:
\begin{enumerate}
\item How does one deal with anisotropic scattering and anisotropic 
stresses before and during decoupling within the covariant approach, 
specifically in such a manner so that consistency is maintained 
when using general relativity, its covariant linearisations 
and relativistic kinetic theory.

\item The small anisotropy equations developed in 
\cite{MGE} with the application to spacetimes with 
arbitrary anisotropy and inhomogeneity have yet to be properly 
investigated; these become applicable when one ignores the Copernican 
principle that underlies the almost-Ehlers-Geren-Sachs theorem, 
which in turn provides 
the theoretical basis for using almost-Friedmann-Lema\^{\i}tre spacetime dynamics. An 
investigation of their consequences on the cosmic background 
radiation may provide an 
alternative method of testing the Copernican principle other than 
the Sunyeav-Zel'dovich effect or via the use of polarization maps. 
 
\item There is a need to find a working  non-Gaussian treatment 
from which one can  construct a generic characterization 
of observables here and now (other than the angular power spectrum 
alone (see Paper I \cite{GE}) and, second, finding an 
{\it ab initio} covariant analysis of transfer functions extending  
the post-Newtonian treatments which use periodic boundary conditions. 

\end{enumerate}

In the next paper in the series, Paper III \cite{DGE}, we 
hope to establish, in a complete fashion, the relationship  
between the null-cone integrations
(favoured in the literature) and the time-like integrations
(found in the exact relativistic kinetic theory).

\section{Acknowledgements}
We acknowledges the partial support of this work 
by the NRF and the University of Cape Town. We
would also like to thank the referee for taking the
time to provide helpful and interesting comments. TG
would also like to thank the Dept of Mathematics at
the University of the Witwatersrand for assistance.
 

\appendix{A}
\appendixtitle{Some Useful Almost-Friedmann-Lema\^{\i}tre relations}
Some useful covariant linearised differential 
identities are given below \cite{RT}: 
\begin{eqnarray}
\c\D_a\psi &=& -2\dot{\psi}\omega_a\;,
\label{a17} \\
\D^2\left(\D_a\psi\right) &\approx&\D_a\left(\D^2\psi\right) 
+{\ts{2\over3}}\left(\rho-3H^2\right)\D_a \psi+2\dot{\psi}\c\omega_a\;, 
\label{a18}\\
\left(\D_a\psi\right)^{\rd} &\approx& \D_a\dot{\psi}
-H\D_a\psi+\dot{\psi}A_a\;,
\label{a19}\\
\left(a\D_aJ^{B_{m}}{}_{A_\ell}\right)^{\rd} &\approx& 
a\D_a\dot{J}^{B_{m}}{}_{A_\ell}\;,
\label{a20}\\
\left(\D^2 \psi\right)^{\rd} &\approx& \D^2\dot{\psi}-2H\D^2 \psi
+\dot{\psi}\D^aA_a \,,\label{a21}\\
\D_{[a}\D_{b]}V_c &\approx&{\ts{1\over3}}\left(3H^2-\rho\right)
V_{[a}h_{b]c}\;, 
\label{a22}\\
\D_{[a}\D_{b]}S^{cd} &\approx& {\ts{2\over3}}\left(3H^2-\rho\right)
S_{[a}{}^{(c}h_{b]}{}^{d)} \,, \label{a23}\\
\D^a\c V_a &\approx& 0\;, 
\label{a24}\\
\D^b\c S_{ab} &\approx& {\ts{1\over2}}\c\left(\D^bS_{ab}\right)\;,
\label{a25}\\
\c\c V_a &\approx& \D_a\left(\D^bV_b\right)
-\D^2V_a+{\ts{2\over3}}\left(\rho-3H^2\right)V_a\;,
\label{a26}\\
\c\c S_{ab} &\approx& {\ts{3\over2}}\D_{\la a}\D^cS_{b\ra c}-\D^2S_{ab}
+\left(\rho-3H^2\right)S_{ab}\;,
\label{a27}\\
D^a \D^b\D_{\la a} V_{b \ra} &\approx& \frac23 \D^2 \div V + (\rho - 3 H ^2) 
\div V\;, 
\label{a28}
\end{eqnarray}
where the vectors and tensors are $O(\epsilon)$ and $S_{ab}=S_{\la ab\ra}$.

\appendix{B} 
\appendixtitle{Integrated Boltzmann Equation Relations}

Here we repeat some useful results from 
\cite{MGE}. The integrated Boltzmann equation is :
\begin{eqnarray}
\int_0^{\infty} L(f) E^2 dE = 
\int_0^{\infty} E^2 dE \left[ {p^a \p_a f - {\Gamma^a}_{bc} p^b p^c 
{\p f \over \p p^a} }\right]= \int_0^{\infty} E^2 dE\;. 
\label{ibe}
\end{eqnarray}
Here f is the single particle distribution function, C[f] the scattering 
correction. The Liouvile operator for the photons, $L(f) = {d f \over d v}$
and $d /dv = p^a \nabla_a$ is the null derivative. Hence we find 
\begin{eqnarray}
L(f(E,x^i,e^a))={d f \over dv} = p^a \nabla_a f + 
{d E \over dv} {\p f \over \p E}\;,
\end{eqnarray} 
as in \cite{MGE}, so that given the identity (see \cite{MGE})
\be
{\d E\over\d v}=-E^2\left[
{\ts{1\over3}}\Theta+A_ae^a+\sigma_{ab}
e^ae^b\right]\,,~ \mbox{and}~~ p^a = E(u^a+e^a)\,, \label{dEdv}
\ee
we have covariant derivation of the almost-Friedmann-Lema\^{\i}tre integrated Boltzmann
equations. 
\appendix{C}
\appendixtitle{Scattering Strength Expansion}

The almost-Friedmann-Lema\^{\i}tre integrated Boltzmann equation for Thompson scattering is
\begin{eqnarray}
 {\cal B}+\dot{\tau}+e^a\D_a\tau \simeq t_c^{-1}
\left[ {v^ae_a-\tau }\right]\;.
\end{eqnarray}
This enables us to find 
\begin{eqnarray}
\tau (x^i,e^a)=v^ae_a- t_c \left[ {{\cal B}+\dot{\tau}+e^a\D_a\tau }%
\right]\;.  
\label{pert_01a}
\end{eqnarray}
We now systematically approximate (\ref{pert_01a}) in terms of the
smallness parameter $t_c$:  
\begin{eqnarray}
\tau _{(n)}\approx v^ae_a-t_c\left[ {{\cal B}+\dot{\tau}%
{(n-1)}+e^a\D_a\tau _{(n-1)}}\right]\;.  
\label{pert_02a}
\end{eqnarray}
Up to second order the following anisotropies are recovered: 
\begin{eqnarray}
\tau _{(0)}(x^i,e^a) &\approx &v^ae_a\;, \\
\tau _{(1)}(x^i,e^a) &\approx &v^ae_a-t _c\left[ {{\cal B}+e^a\dot{v}
_a+e^ae^b\D_av_b}\right]\;,  
\label{temp_1st} \\
\tau _{(2)}(x^i,e^a) &\approx &v^ae_a-t _c\left[ {{\cal B}+e^a\dot{v}
_a+e^ae^b\D_av_b}\right] +t _c^2\left[ {\dot{{\cal B}}+e^a\ddot{v}_a}
\right.  \nonumber \\
&+&\left. {e^ae^b(\D_av_b)^{\dot{}}+e^c\D_c{\cal B}+e^ce^d\D_c{\dot{v}}
_b+e^ce^ae^b\D_c\D_av_b}\right]\;.  
\label{temp_2nd}
\end{eqnarray} 

In order to carry out the solid angle integration over the sky 
(\ref{int_cond_I}),
the following results will be useful. From the
normalization of the direction vectors, 
$e^a$, along with the recursive definition 
of $O^{A_l}$, it can be demonstrated that 
\begin{eqnarray}
e^ae^b\D_av_b &=&O^{ab}\D_av_b+\frac 13h^{ab}\D_av_b=O^{ab}\D_av_b+\frac
13\D_a v^a\;,  
\label{pert_001} \\
e^ae^b\D_a\dot{v}_b &=&O^{ab}\D_a\dot{v}_b+\frac 13\D_a\dot{v}^a\;,
\label{pert_002} \\
e^ae^be^c\D_a\D_bv_c &=&O^{abc}\D_a\D_bv_c-\frac 15\left( {\
O^a\D_{(a}\D_{b)}v^b+O^b\D_a\D^av_b}\right)\;.  
\label{pert_003}
\end{eqnarray}
Using the orthogonality conditions we find, first, from 
(\ref{IBE-source}): 
\begin{eqnarray}
\int_{4\pi }d\Omega O^{A_\ell}{\cal B}\simeq \delta _0^\ell\left[ 
{-\frac{4\pi} 3\D_a\tau ^a}\right] 
+\delta _1^\ell\left[ {\frac{4\pi }3\left( {\frac 14}
{\D^{a_1} \ln \rho_R}+A^{a_1}\right) }\right] +\delta _2^\ell\left[ {\frac{8\pi
}{15}\sigma ^{a_1a_2}}\right]\;,
\end{eqnarray}
and second,
\begin{equation}
\int_{4\pi }e^a\D_a{\cal B}d\Omega =+\frac{4\pi }3\left( {\frac 14{
\D^a\D_a \ln \rho_R}+\D_aA^a}\right)\;.
\end{equation}
The integration over the solid angle can now carried out resulting in
an equation for the gradient of the radiation flux.

\appendix{D}
\appendixtitle{Integral Solutions}

We use the following notation: 
$\Delta \eta_* = \eta_0 - \eta_*$, $\Delta \eta = \eta_0 - \eta$
and $\delta \eta = \eta - \eta'$ such that $\delta \eta_*= \eta -\eta_*$
and $\delta \eta_0 = \eta - \eta_0$.

Beginning with an integral ansatz of the form: 
\begin{equation}
\tau _\ell^P(\eta )=\int_0^\eta d\eta ^{\prime }\left[ {\ C_0(\eta ^{\prime
})\tau _\ell^{(0)}(\delta \eta)+C_1(\eta ^{\prime }){\frac \partial
{\partial \eta ^{\prime}}}\tau _\ell^{(0)}(\delta \eta)+C_2(\eta
^{\prime }){\frac{\partial ^2}{\partial {\eta ^{\prime }}^2}}\tau
_\ell^{(0)}(\delta \eta)}\right]\;,
\label{LEIB001}
\end{equation} 
Using the Leibnitz rule for differentiation of integrals we obtain:
\begin{eqnarray}
{\frac{\partial \tau _\ell^P}{\partial \eta }}(\eta ) &=&\int_0^\eta d\eta
^{\prime } {\p \over \p \eta} \left[ {\ C_0(\eta ^{\prime }) 
\tau _\ell^{(0)}(\delta \eta)+C_1(\eta ^{\prime })
{\tau _\ell^{(0)}}^{\prime }(\delta \eta)+C_2(\eta ^{\prime })
{\tau _\ell^{(0)}}^{\prime \prime }(\delta \eta)}\right]
\nonumber \\
&+&\left[ {C_0(\eta )\tau _\ell^{(0)}(0)+C_1(\eta ){\tau _\ell^{(0)}}^{\prime
}(0)+C_2(\eta ){\tau _\ell^{(0)}}^{\prime \prime }(0)}\right]\;.
\label{LEIB002}
\end{eqnarray}
If we hold $\eta ^{\prime }$ constant in
the partial derivatives, it follows from 
(\ref{homog}) and (\ref{LEIB001}) that 
\begin{eqnarray}
k\left[ {{\frac{(\ell+1)^2}{(2\ell+1)(2\ell+3)}}\tau _{\ell+1}^P(\delta \eta)-\tau
 _{\ell-1}^P(\delta \eta)}\right] =-{\frac \partial {\partial \eta }}\tau
 _l^P(\delta \eta)\;.
 \end{eqnarray}
This gives us
\begin{eqnarray}
&~&k\left[ {\ {\frac{(\ell+1)^2}{(2\ell+1)(2\ell+3)}}\tau _{\ell+1}^P(\delta \eta)-\tau
_{\ell-1}^P(\delta \eta)}\right]  \nonumber \\
&=&-\int_0^\eta d\eta ^{\prime }\left[ { 
C_0(\eta ^{\prime }){\p \over \p \eta}{\tau _\ell^{(0)}}(\delta \eta)
+C_1(\eta ^{\prime }){\p \over \p \eta}{\tau _\ell^{(0)}}^{\prime }(\delta
 \eta) +C_2(\eta ^{\prime }){\p \over \p \eta}{\tau _\ell^{(0)}}
^{\prime \prime }(\delta \eta)}\right]\;.  
\label{LEIB003}
\end{eqnarray}
Putting (\ref{LEIB001}), (\ref{LEIB002}) and (\ref{LEIB003})
together, we find :
\begin{eqnarray}
{\tau _\ell^{\prime }}^P+k\left[ {{\frac{(\ell+1)^2}{(2\ell+1)(2\ell+3)}}\tau
_{\ell+1}^P-\tau _{\ell-1}^P}\right] =C_0(\eta )\tau _\ell^{(0)}(0)+C_1(\eta
){\tau
_\ell^{(0)}}^{\prime }(0)+C_2(\eta ){\tau _\ell^{(0)}}^{\prime \prime }(0)\;.
\end{eqnarray}

\appendix{E}
\appendixtitle{Truncation Conditions} 
\label{truncation}

The Ellis-Treciokas-Matravers treatment \cite{ETMb} makes various general
statements about exact relativistic kinetic theory in the free streaming
case. Of these probably the most important are:

Firstly, if {\it any} four consecutive harmonics vanish, say those
with $l=L+1,L+2,L+3,L+4$, but those for $l=L$ are non-zero, then 
\begin{equation}
F_{\la A_l}\sigma _{a_{L+1}a_{L+2}\ra}=0\;.
\end{equation}
This means, as $F_{A_l}$ is non-zero, that the shear must vanish exactly: 
$\sigma _{ab}=0$. This results arises from the requirement that 
$\lim_{E\rightarrow\infty }F_{A_l}=0$.

Secondly, it can also be shown that if the first 3 multipole harmonics
are zero, {\it {\ i.e. }}, $l=1,2,3$, once again the shear is necessarily
zero: $\sigma _{ab}=0$: 
\begin{equation}
\sigma _{ab}\int_m^\infty E^5{\frac{\partial F}{\partial E}}dE=0\;.
\end{equation}
Thus in both cases the resulting spacetimes are highly restricted, 
and do not include generic perturbations.

A simplistic approach to linearisation will suggest that these relations and
their implications can be ignored in (linearised) almost-Friedmann-Lema\^{\i}tre universes,
because the equations leading to these results are second-order relations
and so can be dropped when linearising. However that argument is not
correct, if one carries out a careful linearisation procedure: indeed both
these statements will hold in almost-Friedmann-Lema\^{\i}tre universes also. This can be
seen as follows: although both $F_{A_l}$ and $\sigma _{ab}$ are at 
most O[1] (or $O(\epsilon )$) in the almost-Ehlers-Geren-Sachs sense 
\cite{SME},
there are no zero or first order terms in the relevant equations 
leading to the above results, to explicitly linearise with respect
to. Thus they cannot be dropped relative to larger (first order) 
terms in these equations, as there are no such terms; the first
non-zero terms are second-order, and hence these equations with 
these terms must be obeyed even if we carry out a (first-order) 
linearisation. Thus they are both at most O[2] equations, but are both
still valid in the almost-Friedmann-Lema\^{\i}tre universes\footnote{Although
$O(\epsilon^2) \ll O(\epsilon)$, this doesn't mean that 
$O(\epsilon^2)=0$ on its own, even though formally the notation 
$O(\epsilon^2) \simeq 0$ is often adopted. These are not equivalent.}.

What this means is that one must be very careful about any kind of 
truncation in the multipole hierarchy, even in the almost-Friedmann-Lema\^{\i}tre
universes - this includes not only the free-streaming case but the case with
a generalized Krook equation for the scattering term \cite{ETMb}. What
should be of interest is that in the matter dominated almost-Friedmann-Lema\^{\i}tre models with
scalar perturbations, any truncation leading to zero shear would suppress
the perturbations, reducing the dynamics to that of an {\it exact} Friedmann-Lema\^{\i}tre
model. This doesn't mean that one cannot consistently damp the higher 
moments out, it just means that they cannot be formally truncated 
-- when higher moments are ignored the consistency condition in the 
exact theory should still be taken into account. Where this 
makes a difference is, for example, 
on intermediate scales where one is tempted to drop everything 
with $l\ge 3$, {\it {\ i.e. }}, when one is close to tight-coupling. 
However, this is, strictly speaking a truncation and hence problematic. This 
is why a perturbative analysis in the Thompson scattering time is important; 
one can in this way consistently, without truncation, build up the entire 
multipole divergence equation hierarchy perturbatively - provided one has 
a meaningful sense of smallness, in this case the relaxation time. 
We do this in the case of small scales, in this way decoupling a
subset of the hierarchy from the full set of multipole divergence 
equations, as in the gauge-invariant  formulation of 
Hu-Sugiyama \cite{HS95a,HS95b}. 

\appendix{F}
\appendixtitle{Linking Different Expansions} 

The gauge invariant and covariant mode expansion (\ref{SQdef-modef}) and
(\ref{expn}) \cite{GE,cl2} in the almost-flat-Friedmann-Lema\^{\i}tre case ($K=0$) 
give the mode coefficient recursion relations for $\ell \ge 2$:
\begin{eqnarray}
-\dot{\tau}_\ell\simeq \frac ka\left[ {{\frac{(\ell+1)^2}
{{(2\ell+3)(2\ell+1)}}}\tau_{\ell+1}-\tau _{\ell-1}}\right]\;.
\label{lnk1}
\end{eqnarray}
Multiplying through by $\beta _\ell$, one finds
\begin{eqnarray}
-(\beta _\ell\tau _\ell)^{\dot{}}\simeq \frac ka\left[ {\
{\frac{(\ell+1)}{(2\ell+3)}}
(\beta _{\ell+1}\tau _{\ell+1})-{\frac \ell{(2\ell-1)}}(\beta
_{\ell-1} \tau _{\ell-1})}\right]\;.
\label{lnk2}
\end{eqnarray}
Changing from the proper time derivative in (\ref{lnk2}) to the
conformal time derivative $^{\prime }$ using $dt =ad \eta$ we find
\begin{eqnarray}
-(\beta _\ell\tau _\ell)^{\prime }\simeq k\left[
{{\frac{(\ell+1)}{(2\ell+3)}} (\beta_{\ell+1}\tau _{\ell+1})-
{\frac \ell{(2\ell-1)}}(\beta _{\ell-1}\tau_{\ell-1})} \right]\;.
\label{lnk3}
\end{eqnarray}
This can be immediately seen to be the same mode equation for $\ell>2$
as in Hu \& Sugiyama \cite{HS95a} and Wilson \& Silk \cite{W83} (eqn 7).
 
If now $\beta _\ell$ is replaced with $\alpha^{-1} _\ell(2 \ell+1)
=\beta _\ell$, we find that (\ref{lnk3}) can be rewritten
(on first multiplying through by $(2 \ell+1)^{-1}$) as
\begin{eqnarray}
-(2 \ell+1)(\alpha^{-1} _\ell\tau _\ell)^{\prime }
\approx k\left[ {(\ell+1)(\alpha _{\ell+1}^{-1} \tau
_{\ell+1})- \ell(\alpha _{\ell-1}^{-1} \tau _{\ell-1})}\right]\;.
\label{lnk4}
\end{eqnarray}
This can be recognized as the form of the $\ell>2$ free-streaming integrated
Boltzmann equation of Ma \& Bertschinger \cite{MB} ({\it cf}. $\ell$-th mode
equation, 49 or 50) or Seljak \& Zaldarriaga \cite{SeZalb} equation
3d. Once again, it may be useful to remind the reader of our
nomenclature, $\tau _\ell$ are the mode coefficients (from the mode 
expansion) while $\tau _{A_\ell}$ are the multipole coefficients 
(from the multipole expansion). This distinction is not made in the 
Bardeen-variable gauge-invariant treatments based on the Friedmann-Lema\^{\i}tre mode expansion.       
 
The solution to the mode coefficients with respect to the time-like
integration in the flat ($K=0$) almost-Friedmann-Lema\^{\i}tre case, are spherical Bessel
functions. This should not be surprising given that the recursion
relation in the linear Friedmann-Lema\^{\i}tre case is merely a projection from an initial
section onto a sphere around here and now, modified as a result of 
Robertson-Walker expansion.
   
\appendix{G}
\appendixtitle{Almost-Friedmann-Lema\^{\i}tre Equations}

Here some of the results arising from the 
Einstein field equations are listed, these are presented in \cite{MGE}.

\section{Main Almost-Friedmann-Lema\^{\i}tre equations} \label{almost-Friedmann-Lemaitre-efe}

We give the almost-Friedmann-Lema\^{\i}tre constraint, propagation and perturbation equations
\cite{MES,EHB89}.
Using the $1+3$ covariant dynamical equations \cite{HvEl}
\cite{E_varenna}, in geometrized units : $c=8\pi G=1$. With 
the projection tensors $U_{~b}^a=-u^au_b$, 
$h_{~b}^a=\delta _{~b}^a+u^au_b$, $\dot{h}^{\la ab \ra}=0$ 
and $0=\D_aU_{bc}=\D_ah_{bc}$, where 
the totally projected spatial derivative is $\D_a$. 
The covariant derivative of $u^a$ is
\begin{eqnarray}
\nabla _bu_a=-u_aA_b+\D_au_b:=-u_aA_b+\sigma
_{ab}+Hh_{ab}+\epsilon _{abc}\omega ^c,
\end{eqnarray}
where we have written the vorticity in terms of the vorticity vector,
$\varepsilon _{abc}=\eta _{abcd}u^d,$ and the shear 
$\sigma _{ab}=\sigma _{<ab>}$.
We consider a cosmological model with matter (dust) and radiation. 
Then the stress-tensor given by 
\begin{eqnarray}
T^{ab} =  (\rho_M + \rho_R) u^a u^b + 2 q^{(a}u^{b)} 
+ \pi^{ab} + p h^{ab}\;,
\end{eqnarray}
where energy densities $\rho_M$ and $\rho_R$ are the energy densities
of the matter and radiation respectively, $q^a$ is the total 
energy flux, $\pi_{ab}$ is the anisotropic pressure of the radiation.

We linearise about a Robertson-Walker model as explained previously 
(see section 3.2 and \cite {BE}). The convention 
$A=B$ implies that $A$ equals $B$ in the exact theory, 
while $A\simeq B$ indicates that $A$ equals $B$ to at least 
O[1] in the almost-Friedmann-Lema\^{\i}tre sense, and lastly $A\approx B$ is 
retained to indicate that $A$ equals $B$ in relation to some 
other specific smallness parameter such as the relaxation
time or ratio of radiation density to matter density. 
The almost-Friedmann-Lema\^{\i}tre Gauss-Codacci relation (Hamiltonian constraint) is
\begin{equation}
^3 R \simeq 2 \rho + \frac23 \Theta^2\;. 
\label{h-constraint}
\end{equation}
Using the Ricci identities, the linearised {\it propagation equations} are
\begin{eqnarray}
\dot{\Theta}+\frac 13\Theta ^2- (\div A)+\frac 12(\rho +3p) &\simeq
&-\frac 32B\;,\label{ray-e2} \\
\dot{\omega}_{\left\langle a\right\rangle }+2H\omega _{\left\langle
a\right\rangle }+\frac 12\mbox{curl}A_a &\simeq &0\;, \\
\dot{\sigma}_{\left\langle ab\right\rangle }+2H\sigma _{\left\langle
ab\right\rangle }-\D_{\left\langle a\right. }A_{\left. b\right\rangle
}+E_{\left\langle ab\right\rangle } &\simeq &\frac 12\pi _{\left\langle
ab\right\rangle }\;,
\label{EFE-dots}
\end{eqnarray}
and the {\it constraint equations} are given by
\begin{eqnarray}
(\div \sigma) ^a-\frac 23\D^a\Theta -\c
\omega ^a &\simeq &q^a\;,  
\label{mom-flux1} \\
\D_a\omega ^a &\simeq &0\;, \\
H_{\left\langle ab\right\rangle }-\c \sigma _{\left\langle
ab\right\rangle }-\D_{\left\langle a\right. }\omega _{\left. b\right\rangle }
&\simeq &0\;.
\end{eqnarray}
Using the second Bianchi identities, the linearised {\it propagation
equations} are
\begin{eqnarray}
\dot{\rho}+(\rho + p)\Theta &\simeq &-3HB-\D^aq_a\;, 
\label{EFE-ec}\\
(\rho +p)A_a+\D_ap &\simeq &-\D_aB-\dot{q}_a-4Hq_a-\D^b\pi _{ab}\;,
\label{EFE-mc}
\end{eqnarray}
\begin{eqnarray}
\dot{E}^{\left\langle ab\right\rangle }+3HE^{\left\langle ab\right\rangle }-%
\mbox{curl}H^{\left\langle ab\right\rangle }+\frac 12(\rho +p)\sigma
^{\left\langle ab\right\rangle } &\simeq &-\frac 12\dot{\pi}^{\left\langle
ab\right\rangle }-\frac 12H\pi ^{\left\langle ab\right\rangle }-\frac
12\D^{\left\langle a\right. }q^{\left. b\right\rangle }\;,
\label{E-dot} \\
\dot{H}^{\left\langle ab\right\rangle }+3HH^{\left\langle ab\right\rangle }+
\mbox{curl} E^{\left\langle ab \right\rangle }&\simeq& \frac 12\mbox{curl}\pi
^{\left\langle ab\right\rangle }\;, 
\label{H-dot}
\end{eqnarray}
and the {\it constraint equations} are:
\begin{eqnarray}
(\div E)_a -\frac 13\D_a\rho &\simeq &-Hq_a-\frac 12\D^b\pi _{ab}\;,
\label{div-E} \\
(\div H)_a-(\rho +p)\omega _a &\simeq &-\frac 12\mbox{curl}q_a\;.
\label{div-H}
\end{eqnarray}
The background Friedmann-Lema\^{\i}tre equations are:
\begin{eqnarray}
\rho &=& 3 H^2 + {3 K \over a^2}\;, 
\label{Friedmann-Lemaitre_E}\\
\dot{H} &=& - H^2 - \ts {1 \over 6} (\rho + 3 p), \label{Friedmann-Lemaitre_F}\\
\dot{\rho} &=& - 3 H (\rho + p)\;.  
\label{Friedmann-Lemaitre_R}
\end{eqnarray}
\section{Source Terms} \label{app-source-terms}
{}From \cite{MGE}, the source terms in 
the almost-Friedmann-Lema\^{\i}tre multipole divergence equations are:
\begin{eqnarray}
{\cal B}_0(x) \simeq - \frac13 \D^a \tau_a(x)\;,~~ 
{\cal B}_a(x) \simeq \D_a \ln T(x) + A_a(x)\;, ~~
{\cal B}_{ab}(x) \simeq \sigma_{ab}(x). 
\label{efe2.1b}
\end{eqnarray}
In the case of scalar perturbations in a matter dominated
almost-Friedmann-Lema\^{\i}tre universe \cite{MGE}, using the mode expansion \cite{GE},
the electric part of the Weyl tensor $E_{ab}$
(gravitational tidal effects), the acceleration $A_a$ and 
anisotropic pressure $\pi_{ab}$ can be related to their corresponding
potentials:  
\begin{eqnarray}
 E_{ab} = \bar \Phi(k,t) Q_{ab} ,~~~ A_a = \Phi_u(k,t) Q_a,~~~
 \pi_{ab} = \Phi_{\pi}(k,t) Q_{ab}\;. 
\label{scalar-potential1}
\end{eqnarray}
Using the Mode functions $Q$, $Q_a$ and $Q_{\langle ab \rangle}$, 
taking care of the additional wavenumber factors which get introduced
{\footnote{Once again one should not confuse the Fourier coefficients of the 
potentials here with those defined from $E_{ab} = \D_{\la ab \ra}
\Phi(x^i)$, $A_a = \D_a \Phi_u(x^i)$ and $\pi_{ab} = \D_{\la ab \ra} 
\Phi_{\pi}(x^i)$. }} \cite{GE}, the following decomposition will be used: 
\begin{eqnarray}
&&(\D^{b} E_{ab})^{k} = \Phi(k,t) \D^b Q_{ab}\;, ~~(\D^b \pi_{ab})^{k} =
\Phi_{\pi} \D^b Q_{ab}\;,\\ 
&&{(\D_a \ln \rho_{\!_M})}^{k} = \delta^{\!_M}(k,t) Q_a 
= + \frac{k}{a} \Delta(k,t) Q_a\;, \\
&&{(\D_a \ln \rho_{\!_R})}^{k} = 4 \ts\frac{k}{a} \delta T(k,t) Q_a,  
\label{scalar-potential2}
\end{eqnarray}
It then follows that \footnote{To see where these factors come from 
notice that $a^2 \D^c \D_{ac} Q = (-k^2 + 2K) \D_a Q$ $\Rightarrow$
$a^2 \D^c \D_{\la ac \ra }Q = \frac23 (- k^2 + 3 K) \D_a Q$ (after removing
the trace) for the general $l-th$ order relations see \cite{GE}.
$(-\lambda)^{-l} \D_{\la A_l \ra} Q = Q_{A_l}$ to find that
$\D^b Q_{ab} = - \frac23 (ak)^{-1} ( - k^2 + 3 K) Q_a$, where as
before $\D^a \D_a Q = - \lambda^2 Q$ and $\lambda = {k \over a}$ 
\cite{BDE92,GE}.}
\begin{eqnarray}
{\cal B}_0(k,t) \approx - \frac 13 {k \over a} \tau _1(k,t)\;, 
\label{efe2.1a} \\
{\cal B}_1(k,t) \approx \frac 23 (\rho^{-1}(t) \bar \Phi(k,t))
\frac{k}{a} \left( {k^2 - 3K \over k^2}\right)\;, 
\label{efe2.2a} \\
{\cal B}_2(k,t) \approx  - 2 [\rho^{-1}(t) \bar \Phi(k,t)]^{.}\;.
\label{efe2.3a}
\end{eqnarray}
In the case of matter domination we then find the relationship 
between the Newtonian potential and the matter density gradients: 
\begin{eqnarray}
2 \frac{a}{k} (k^2 - 3 K) \bar \Phi(k,t) \simeq
(a^2 \rho_{\!_M}) \delta^{\!_M} \approx \frac{k}{a} 
(a^2 \rho_{\!_M}) \Delta(k,t)\;. 
\label{aFriedmann-Lemaitre_md01} 
\end{eqnarray}
Replacing $a^2 \rho_{\!_M}$ in terms of the density parameter and the
curvature constant from the Freidmann equation (See Appendix A) for both 
the $K=0$  ($\Omega_0 = +1$) and $K<0$  ($\Omega_0 < 1$) matter 
dominated cases respectively, we obtain
\begin{eqnarray}
\bar \Phi(k,t) &&\approx \frac32 (H_0^2 \Omega_0) D(\eta) \Delta(k,\eta_0)
\\ &&\approx \frac32 \frac{a}{k} H_0^2 \delta^{\!_M}\;,
\label{aFriedmann-Lemaitre_md02} \nonumber \\
 \bar \Phi(k,t) 
&&\approx \frac32 \left[ {{1 \over (k^2- 3K)} 
{K \Omega_0 \over (\Omega_0-1)}} \right] D(\eta) \Delta(k,\eta_0) \\
&&\approx 
\frac32 \frac{k}{a} {1 \over (k^2- 3K)} 
{K \Omega_0 \over (\Omega_0-1)} \delta^{\!_M} \nonumber \;. 
\label{aFriedmann-Lemaitre_md03}
\end{eqnarray}
These are then the equations that will be used for 
the slow decoupling and free-streaming eras.

\section{The Temperature Monopole}

Using the energy conservation equations ($\ell=0$ multipole divergence 
equation) \footnote{Notice that $\D_a (\ln T)^{\dot{}} 
\simeq (\D_a \ln T)^{\dot{}} + H (\D_a (\ln T) + A_a)$
where $\D_a = h_{ab} \nabla^b$ and 
$-\dot{u}_a (\ln T)^{\cdot} \simeq H \dot{u}_a$. It is also 
well known that $\D^a (\D_a \ln T)^{\dot{}} 
\simeq (\D^2 \ln T)^{\dot{}} + H (\D^2 \ln T)$.}  
\begin{eqnarray}
(\D_a \ln T)^{\dot{}} + H (\D_a \ln T + A_a)  + \frac13
\D_a \Theta \simeq + \frac13 \D_a  (\D^c \tau_c)\;, 
\label{MDE_mono}
\end{eqnarray}
and taking another spatial covariant derivative, we obtain
one of key equations in the primary source calculation: 
\begin{eqnarray}
(\D^2 \ln T)^{\dot{}} + 2 H (\D^2 \ln T) - \frac13 \D^2 (\D^c
\tau_c) \simeq - \frac12 \left( {\D_a \D_b
\sigma^{ab} - \D_ a q^a} \right) - H (\D^a A_a)\;.  
\end{eqnarray}

\section{The Adiabatic Condition} \label{sec-adb}

The entropy perturbation $S_a$,
for a radiation-dust almost-Friedmann-Lema\^{\i}tre universe, is given by 
\begin{eqnarray}
 S_a \simeq \ts\frac14 {\D_a \ln \rho_{\!_R}} 
- \ts\frac13 {\D_a \ln \rho_{\!_M}}\;,
\end{eqnarray}
and this gives the relation used in \cite{dunsby}:
\begin{eqnarray}
\D^b E_{ba} \simeq \left( { \ts\frac14 {\D_a \ln \rho_{\!_R}} - S_a } 
\right) \rho_{\!_M}\;.
\end{eqnarray}
The adiabatic condition is then characterized by $S_a =0$
{\footnote{By adiabatic we mean that the comoving 
entropy density is constant.}}. This can either be written 
as $\D_a \rho_{\!_R} = R\D_a \rho_{\!_M}$, 
$\ts\frac{k}{a}\delta T \D^a Q_a = + \frac13 \delta^{\!_M} \D^a Q_a$
and $\delta T(k,t) = \ts\frac{1}{3}\Delta(k,t)$.

Provided the initial conditions produced, for example after inflation, 
lead to adiabatic perturbations, these perturbations
will remain adiabatic until decoupling, however after decoupling generic
density perturbations do not satisfy the adiabatic condition
$S_{(rm)}=0$. This is due to the fact that the average velocity
of the radiation does not proceed along geodesics, while the matter does. 
Thus any perturbation that starts off adiabatic at last scattering will
not remain so \cite{dunsby}.

\section{Almost--Friedmann-Lema\^{\i}tre Scalar Perturbations}

In the case of scalar perturbations, the almost-Friedmann-Lema\^{\i}tre Einstein field 
equations formatter domination reduce to the following CGI
perturbation equations: 
\begin{eqnarray} 
\dot{\sigma}_{ab} + 2 H \sigma_{ab} + E_{ab} &\approx& 0
~~\iff (a^2 \sigma_{ab})^{\cdot} \approx - a^2 E_{ab}\;, \\
\dot{E}_{ab} + 3 H E_{ab} + \frac12 \rho_{\!_M} \sigma_{ab}  &\approx& 0\;, 
~~\iff (\rho_{\!_M}^{-1} E_{ab})^{\cdot} \approx - \frac12 \sigma_{ab}\;.
\end{eqnarray}
Taking the time derivative of the above equations, we obtain
\begin{eqnarray}
(a^2 \sigma_{ab})^{\ddot{}}+ H(a^2 \sigma_{ab})^{\dot{}} &\approx& 
+ \frac12 a^2 \rho_{\!_M} \sigma_{ab}\;, \\ 
(\rho_{\!_M}^{-1} E_{ab}){\ddot{}} 
+ 2 H (\rho_{\!_M}^{-1} E_{ab})^{\dot{}} &\approx&
\frac12 E_{ab}\;. 
\label{prop-ele}
\end{eqnarray}
To see how these relate to evolution equations for the density
gradient, we take the spatial divergence of (\ref{prop-ele}) \cite{BDE92}.

A useful consequence of the above relations is that for matter
domination\footnote{A further consequence of
$(a^3 E_{ab}){}^{\cdot} \approx - \frac12 a^3 \rho \sigma_{ab}$  
and $\rho = 3 H_0^2 \Omega_0 a^{-3}$ 
for $a_0=+1$; is that $(a^3 E_{ab})^{\cdot} \approx - \frac32 H_0^2
\Omega_0^2 \sigma_{ab}$.}, the monopole equations take on a simple form:
\begin{eqnarray}
(D_a \ln T)^{\cdot} + H (D_a \ln T) 
+ \frac13\D_a (D^c \tau_c) \approx - \frac12 \D^b \sigma_{ab}\;. 
\label{md-monoT}
\end{eqnarray}
To summarize, the matter dominated limit lead to the following 
simple relationships between the dynamical variables and the 
electric part of the Weyl tensor:
\begin{eqnarray}
A_a &\approx& 0\;, \\
\sigma_{ab} &\approx& - \frac23 (H_0^2 \Omega_0)^{-1}
 u^c \nabla_c \left( {a^3 E_{ab}} \right)\;, \\
\D_a \ln \rho &\approx& (H_0^2 \Omega_0) \D^b \left({a^3 E_{ab}}\right)\;, \\
\frac23 \D^a \Theta &\approx& 
- \frac23 (H_0^2 \Omega_0)^{-1} \left[ { \left({a^3 \D_b E^{ab}}\right)^{\cdot}
 +  H (a^3 \D_b E^{ab} )} \right]\;.
\end{eqnarray}
Using
\begin{eqnarray}
E_{ab} = \sum_k {\bar \Phi} Q_{ab} \equiv D_{\la a} \D_{b \ra} \Phi_{\!_E}(x)
&&\mbox{and}~~~~ \D_a \Phi_{\!_E}(x) = 
\sum_k \ts\frac{k}{a}\Phi_{\!_E}(k,t) Q_a
\end{eqnarray}
we obtain
\begin{eqnarray}
\bar \Phi(k,t) \approx - \ts{k^2 \over a^2} \Phi_{\!_E}(k,t).
\label{afrw-1}
\end{eqnarray}
Note that in \cite{dunsby}, $E_{ab} = \sum_k (k^2/a^2) \Phi_k Q_{ab}$ 
which is based on the notation of Kodama and Sasaki \cite{KS}.  The 
relationship between $\Phi_k$ and the potential used here 
is $\bar \Phi = (k^2/a^2) \Phi_k$.

The evolution equation for the Newtonian like potential follows from 
(\ref{prop-ele}) \cite{EBH90,EHB89}:
\begin{equation}
(\rho_{\!_M}^{-1} \bar \Phi)^{\ddot{}} 
+ 2 H (\rho_{\!_M}^{-1} \bar \Phi)^{\cdot} \approx 
\ts\frac12 \bar \Phi\;.
\label{potential-ev}
\end{equation}
In terms of the conformal time derivative ($dt = a d \eta$) which 
we denote by a prime `$\prime$' this becomes:
\begin{equation}
(\rho_{\!_M}^{-1} \bar \Phi)'' + H (\rho_{\!_M}^{-1} \bar \Phi)' 
\approx \ts\frac12 a \bar \Phi.
\end{equation}
On rearranging (\ref{potential-ev}) using (\ref{afrw-1}) and 
 (\ref{Friedmann-Lemaitre_E}-\ref{Friedmann-Lemaitre_R}) we find the evolution equation for $\Phi_E$:
\begin{equation}
\left[ {\ddot{\Phi}_{\!_E} + 4 H \dot{\Phi}_{\!_E}} \right] 
\approx \left[ {\ts\frac12 \rho_{\!_M} - \dot{H} - 3 H^2 } \right] \Phi_{\!_E} 
\approx \left( {\ts{2 K \over a^2}} \right) \Phi_{\!_E}.
\end{equation}
which can be simplified to give
\begin{equation}
(a \Phi_{\!_E})^{\cdot \cdot}(t,k) 
+ 2 H (a \Phi_{\!_E})^{\cdot}(t,k) \approx \ts\frac32 H_0^2 \Omega_0
a^{-2} \Phi_{\!_E}(t,k).
\end{equation}
Notice that for $K=0$ we obtain the usual equation
$\ddot{\Phi}_E \approx - 4 H \dot{\Phi}_E$ for dust as used in 
\cite{dunsby}. This gives the well known result 
\begin{eqnarray}
\Phi_{\!_E}(k,t) 
\approx \Phi^+_{\!_E}(k,0) 
+ \Phi^-_{\!_E}(k,t)t^{-5/3}\;.
\end{eqnarray}
Considering the constant mode only, we then have : 
\begin{eqnarray}
 (\bar \Phi \rho_{\!_M}^{-1}) \approx (3 H_0^2 \Omega_0)^{-1} a 
[k^2 \Phi_{\!_E}(k,0)] \approx (3 H_0^2 \Omega_0)^{-1} a [k^2 \Phi_A(k,0)], 
\label{pI-pII}
\end{eqnarray}
It then follows that $(\bar \Phi \rho_{\!_M}^{-1})^{\cdot} 
\sim k^2 \dot{a}$ and $\Phi^{\prime}_A \approx 
- \Phi^{\prime}_H \approx 0$.  Hence we recover the standard 
result that the `potential fluctuations'
$\Phi_{\!_E}$ are time independent for the $K=0$ matter dominated dust 
scenario.

We can then  write $a \Phi_A(k,0) = D/a (a \Phi_A(k,0)) = (D/a) \Phi_A(k,t)$
where $D$ is the the linear growth factor. The flat dust case is 
recovered using $D=a$. 
\begin{eqnarray}
(\bar \Phi \rho_{\!_M}^{-1}) 
\approx (3 H_0^2 \Omega_0)^{-1} {D \over a} [k^2 \Phi_A(k,t)].
\label{phi-flat}
\end{eqnarray}

\appendix{H}
\appendixtitle{The Einstein Field Equations in the Energy Frame}
In the energy frame $\tilde{q}_a=0$, one finds the following 
useful constraints and a simple form for the evolution equation 
for the electric part of the Weyl tensor: 
\begin{eqnarray}
(\div \tilde{\sigma})_a &\simeq& + \frac23 \D^a{ \tilde{\Theta}}\;,
 \label{efe_tc_e1} \\
(\div E)_a &\simeq& + \frac13 \D_a (\rho + \rho_R), \label{efe_tc_e2} \\
(\rho_M + \rho_R + p) \tilde{A}_a &\simeq& - \D_a p, \label{efe_tc_e3} \\
\dot{E}^{ab} + 3 H E^{ab} &\simeq& - \frac12 (\rho_M + \rho_R + p) 
{\tilde \sigma}^{ab}\;.
\label{efe_tc_e4}
\end{eqnarray}
These are very similar to those found in the case of matter domination
\cite{MGE}. The useful features that arises when using the energy frame 
are : Firstly, we have a simply relationship between the expansion 
perturbation and the shear (\ref{efe_tc_e1}). Secondly, the Newtonian 
like potential can be related to the density of the matter 
and radiation content (\ref{efe_tc_e2}). Thirdly, by taking the 
divergence of (\ref{efe_tc_e3}) we find that\footnote{Note that this 
can also be written as $\D^2 (\ln \rho_R) \simeq \ts {4 \over 3}(1+R) 
(\div \tilde{A})$ using $\D^2 (\ln \rho_R) = 4 \D^2 (\ln T)$.}
\begin{equation}
(\rho_M + \frac43 \rho_R) (\div \tilde{A}) \simeq - \D^2 \rho_R ~~\Rightarrow  
\D^2 \rho_R \simeq - (1 + R^{-1}) \rho^{-1} (\div \tilde{A})\;,
\end{equation}
from which we can find an evolution equation for $(\div A)$ from the
perturbation equations for the radiation energy density.
Finally, equation (\ref{efe_tc_e4}) gives the relationship 
between the shear and the Newtonian potential.

\section{On Relating  $a^2 \rho$ to the Curvature and $\Omega$}

The Friedmann equation in a Friedmann-Lema\^{\i}tre universe is 
\begin{eqnarray}
H^2 + {\frac{K }{a^2}} \simeq \frac13 \rho ~\Rightarrow~
a^2 H^2 + K = \frac13 a^2 \rho\;.  
\label{1a}
\end{eqnarray}
Using definition of the density parameters: 
\begin{eqnarray}
\Omega = {\frac{\rho}{3 H^2}}\;, ~\mbox{and} 
~~\Omega_{\!_I} = {\frac{\rho_{\!_I}}{3 H^2}}\;. 
\end{eqnarray}
we can show that 
\begin{eqnarray}
 a^2 \rho \simeq 3 a^2 H^2 \Omega\;.  
\label{2a}~\mbox{and}~~a^2 \rho_{\!_I} \simeq 3 a^2 H^2 
\Omega_{\!_I} \label{1b}.
\end{eqnarray}
Eq. (\ref{1a}) and eqn.(\ref{1b}) can be used to deduce that 
\begin{eqnarray}
a^2 H^2 + K = a^2 H^2 \Omega, 
\Rightarrow a^2 H^2 = {\frac{K}{(\Omega -1)}}.
\end{eqnarray}
Using this and (\ref{1b}) 
we obtain 
\begin{eqnarray}
a^2 \rho \simeq {\frac{3 K \Omega}{(\Omega-1)}}\;. 
\label{2c}
\end{eqnarray}
\appendix{I}
\appendixtitle{The Correlation Function} \label{sec-corr-fn}
If $\la {(\delta N /N)^2} \ra$ is the square of the variance in
the number of objects in a volume V, then the correlation function, the
excess probability over a random variable of finding an object within a
distance $\chi$ of a given object \cite{Peb80}, is given by
\begin{eqnarray}
\xi (\chi) = {d \over dV} \left[ {V(\chi) \la {\left( {\delta N \over N}
\right)} \ra_{V(\chi)} }\right]\;,
\end{eqnarray}
where $V(\chi)$ is the volume enclosed with a radius $\chi$ for flat
universe ($K=0$). The power spectrum ${\cal P}(k)$ is related
{\footnote{Care should be taken with the normalization convention.}}
to the correlation function for a given distribution \cite{KSper}:
\begin{eqnarray}
\xi(\chi) = {1 \over 2 \pi^2} \int k^2 dk {\cal P}(k) {\mbox{sin} k \chi 
\over k \chi}~~ \iff ~~ {\cal P}(k) = 4 \pi \int \chi^2 d \chi \xi(\chi) 
{\mbox{sin}k \chi \over k \chi}. 
\end{eqnarray}
In an open universe the definition for the correlation function can still
be retained. To see how, one first notices that in a flat universe
$\la {(\delta N / N)^2} \ra_{V(\chi)} \sim \la {N} \ra_V^{-1}
\sim V^{-\alpha}$, where $\la {N} \ra_V$ is the average number of
objects in a volume $V$. In a space of constant negative curvature the
volume enclosed by a sphere of radius $\chi$ is $V(\chi)= \pi ( 
\mbox{sinh}(2 \chi)
- 2 \chi)$ so that 
\begin{equation}
\xi(\chi) \sim V^{-\alpha} \sim (\mbox{sinh}(2 \chi) - 2 \chi)^{-\alpha}.
\end{equation}
The power spectrum for a power law correlation function in the volume
in an open universe is then 
\begin{eqnarray}
{\cal P}(k) = { 1 \over 2 \pi^2} B \int \mbox{sinh}^2 \chi d \chi { \mbox{sin} k
\chi 
\over k \mbox{sinh} \chi}
{1 \over (\mbox{sinh}(2 \chi) - 2 \chi)^{\alpha}}\;,
\end{eqnarray}
where $B$ is the normalization constant. This diverges for small $\chi$
so a small scale cut-off is necessary \cite{KSper}. To relate this to 
the power spectrum today on scales measured by galaxy surveys, the power 
spectrum is multiplied by the square of a transfer function $T^2(k)$. 
The power spectrum can then be normalized to $\sigma_8$.

\appendix{J}
\appendixtitle{Power Spectrum Normalization} \label{sec-ps-norm}
There are two possible normalization schemes which one can follow. 
\begin{enumerate}
\item On considering a power spectrum of the form
\begin{equation}
{\cal P}(k) = {\cal A}(k \eta_0)^{n-1}\;, \label{power_sA}
\end{equation}
where $\eta_0 \simeq 3 t_0 \simeq 2 H_0^{-1}$ for $\Omega_0 =1$
gives the conformal time today, the scale factor can be normalized and
${\cal A}$ is one way of expressing the amplitude of scalar
perturbations since it is related to the dimensionless scale of
matter fluctuations at horizon crossing, $\lambda_H$, by
\cite{Peb82,AbottW84,BondE89,White94}:
\begin{equation}
\lambda_H^2 = \frac{4}{\pi} {\cal A}\;.
\end{equation}
\item The alternative scheme is to consider a power spectrum 
of the form 
\begin{equation}
{\cal P}(k) = {\cal B} k^n T^2(k)\;, 
\label{power_sB}
\end{equation}
where the transfer function $T(k) \sim +1$ on large scales. This
means that if the fluctuations arise purely from the Sachs-Wolfe
effect (potential fluctuations) near decoupling, ${\cal B}$ 
and ${\cal A}$ can be related at $n=+1$ \cite{White94} 
\begin{eqnarray}
{\cal P}(k) = 2 \pi^2 \eta_0^4 {\cal A} k T^2(k).
\end{eqnarray}
What is often used is ${\cal P}(k) = 2.5 \times 10^{16} {\cal A} T^2(k)$.
The units of ${\cal A}$ are Mpc$^2$ and of $T^2$ (Mpc$^3$). For 
{\it standard\hs CMD}
it is the convention to use the parametrized transfer 
function  \cite{BondE84}:
\begin{eqnarray}
T(k) = \left[ {1 + \left({ a k + (b k)^{\frac32}
+ (c k )^2} \right)^{\nu}} \right]^{\frac{1}{\nu}}\;,
\label{transfer-fn-efs}
\end{eqnarray}
where 
\begin{eqnarray}
a = 6.4 \Gamma^{-1} h^{-1} \mbox{Mpc}\;, ~~
b = 3.0 \Gamma^{-1} h^{-1} \mbox{Mpc}\;, ~~
c = 1.7 \Gamma^{-1} h^{-1} \mbox{Mpc}\;, ~~\nu = 1.13\;.
\end{eqnarray}
Now the shape function $\Gamma$ can be given as approximately
$\Gamma \simeq \Omega_0 h$. (by choosing $h=0.5$ and $\Omega_B=0.05$
the shape function is given as $\Gamma = 0.48$). 
 
Large scale flows provide a measure of the power spectrum in as much
as the variance of the velocity field sphere of radius $x_f$,
$v^2_{rms}(x_f)$ can be expressed as an integral over the power
spectrum. On small scales (clusters of galaxies) the power
spectrum is normalize to $\sigma_8$, the variance of the galaxy
distribution  on scales of $x_f= 8 h^{-1} \mbox{Mpc}$ \cite{DaviesPe87}:
\begin{eqnarray}
\sigma_8^2 = \frac{1}{b_{\rho}^2} = \frac{1}{2 \pi^2} \int k^2 dk {\cal
P}(k) T^2 (k) W^2( k x_f )\;,
\end{eqnarray}
where $b_{\rho}$ is the $x_f$ scale `bias' such that $\sigma_8^m =
\sigma_8 / b_{\rho}$ and the appropriate variance and $W(x) = 3 (\sin x - x
\cos x) {1/x^3}$ is the top-hat function. It should also be pointed out 
that it is more convenient to use the form 
\begin{eqnarray}
\sigma_8^2 = \int_0^{\infty} {dk \over k} A(k \eta_0)^{n+3} T^2(k) \left( {
 3 j_1(k x_f) \over k x_f}\right)\;.
\end{eqnarray}
The variance of galaxies possibly biased to the matter ($\delta_{gal} = b
\delta_{\rho}$) is roughly unity on the scale of $8 h^{-1}$ Mpc. Interestingly
enough the {\it standard-CDM} normalization from COBE seems to give 
$\sigma_8 \simeq 1.3$ which is seems to imply that it is not 
correct to assume a pure Sachs-Wolfe-HZ power spectrum or even a 
$n=1.15$ Sachs-Wolfe one. An appropriate table of {\it standard-CDM} 
normalizations is provided in \cite{BSW}.
\end{enumerate}
\appendix{K}
\appendixtitle{The Open Almost-Friedmann-Lema\^{\i}tre Case} \label{open-FRW}
\section{Extending the Integral Solution to the Open Case} 
In the main body of this paper we keep to the almost-flat-Friedmann-Lema\^{\i}tre case for
clarity; the extension to the open case is straight forward. The point is
that we shown that the standard results can be recovered from the CGI
approach. Given that the generic linear
Friedmann-Lema\^{\i}tre models are  well treated in the standard literature, we 
provide only an outline for the open case. The essence of the 
open solution is given via:
\begin{eqnarray}
\tau_\ell^{\prime} &+& k \left[ {{(\ell+1)^2 \over (2 \ell+3) (2 \ell +1)}
\left({1 - {K \over k^2} ((\ell+1)^2-1)}\right) \tau_{\ell+1} 
- \tau_{\ell-1}} \right] \nonumber \\ &\approx& - \kappa^{\prime} \tau_{\ell} 
- \left[ {a {\cal B}_0 \delta_{\ell 0} + (a{\cal B}_1 +
\kappa^{\prime} v_{\!_B}) \delta_{\ell 1} + a{\cal B}_2 \delta_{\ell2}} \right]\;.
\label{open-ext}
\end{eqnarray}
The left hand side, the homogeneous case, is solved
in very much that same way as for the flat case.

The flat radial eigenfunctions are found from (\ref{open-eigen}):
\begin{eqnarray}
{\frac d{d\chi }} j_\ell (k\chi )={\frac \ell {(2\ell+1)}}kj^{\ell - 1}(k\chi )
-{\frac{\ell + 1}{(2 \ell + 1)}}k j^{\ell+1}(k\chi )\;,  
\label{BI-002}
\end{eqnarray}
where $Q_{lm}=j_\ell(k\chi ){\cal Y}_{\ell m}^{A_\ell} O_{A_\ell}$. For 
the general almost-Friedmann-Lema\^{\i}tre (linearised Friedmann-Lema\^{\i}tre) models \cite{HS95a,GSS,SZ97a} 
{\footnote{Notice that $(\nu^2 +(\ell+1)^2) /(\nu^2+1)
=(1 - \ell(\ell+2)(K/k^2))$. }: 
\begin{eqnarray}
{\frac d{d\chi }}X_\nu ^\ell={\frac \ell{(2\ell+1)}}kX_\nu
^{\ell-1}-{\frac{(\ell+1)}{(2\ell+1)%
}}k\left[ {1-\ell(\ell+2){\frac K{k^2}}}\right] X_\nu ^{\ell+1}\;.
\label{open-eigen}
\end{eqnarray}
Thus, in order to construct the open solution, we just replace 
the homogeneous solution for $\nu ^2+1=k^2/(-K)$ 
\cite{GSS,W83} after reading off the
solution using the recursion relation:
\begin{equation}
\tau _\ell^{(0)}(k,\eta )=(2\ell+1)\beta _\ell^{-1}X_\nu ^\ell(\eta )\;.
\label{hopen-soln}
\end{equation}
Of course one needs to be careful to redefine the wavenumber. The integral 
solution (\ref{damp-soln-1}) is then used in (\ref{open-ext}), \ie, the 
solutions (\ref{hopen-soln}) are substituted into (\ref{int_soln_damping}) 
to recover the open integral solutions. 

One carries out the same treatment as for $K=0$ but using (\ref{open-eigen}) 
and (\ref{mibec-1}-\ref{mibec-3}). Alternatively one can re-define the mode 
expansion $M_{\ell}[Q]$ \cite{GE}. 

All that remains is to solve the evolution equations for the scalar 
perturbations, the coefficients $C_I(\eta,k)$ now include curvature 
terms when written out in terms of the ${\cal B}$ terms (one uses the open
recursion relation instead of the flat). In turn, these terms, {\cal B},
will also pick up curvature terms when written out in terms of the 
perturbation variables. When the curvature starts to dominate the evolution, 
one gets an additional ISW contribution (which will be similar to the late 
ISW effect for a $\Lambda$ dominated flat model).
\subsection{Extending the Power Spectrum to the Open Matter-dominated Case}
We consider the primary anisotropy term:
\begin{eqnarray}
{\tau^{SW}_\ell(\eta_0) \beta_\ell \over(2\ell+1)} &\simeq& 
\frac{a}{k} {\cal B}_1(\eta_*)
{ X_\ell^{\nu}(\Delta \eta_*)}, \label{temp_01} \\
{\cal B}_1(\eta_E) &\simeq& + \frac23 (ak)^{-1} \left( { 
\bar \Phi(k,\eta_*) \rho^{-1} } \right) (k^2 - 3 K)\;. 
\label{temp_03}
\end{eqnarray}
This gives the Sachs-Wolfe effect in the open case.

For spaces of constant negative curvature in an almost-Friedmann-Lema\^{\i}tre model:
\begin{eqnarray}
\la {\bar \Phi(k^a) \bar \Phi(k'^a)} \ra = \left[ {
(2 \pi)^3 \frac32 {K \over 3K - k^2} 
{\Omega_0 \over (\Omega_0-1)}} \right]{\cal P}(k)\;.
\label{open_power3}
\end{eqnarray}
Using (\ref{temp_01}-\ref{temp_03}) with $K<0$ we find
\begin{eqnarray}
\tau_l(\eta_0) {\beta_\ell \over (2\ell+1)} \simeq -\frac23 
\bar \Phi(k,\eta_*) \rho^{-1}_* \frac{1}{k^2} (3 K-k^2) 
X_\ell^{\nu}(\Delta \eta)\;.
\end{eqnarray}
It follows that (noticing that $\Delta \eta_* = \chi$)
\begin{eqnarray}
{|\tau_\ell|^2 \beta^2_\ell \over (2\ell+1)^2} 
\approx (2 \pi)^3 \left[ { \frac{\rho^{-1}_0 K}{k^2} 
{\Omega_0 \over (\Omega_0-1)}} \right]^2 {\cal P}(k)\;.
\end{eqnarray}
Then we can use 
\begin{eqnarray}
\la {\tau_{A_\ell} \tau^{A_\ell}} \ra =
{2 \over \pi} \beta_\ell \int k^2 dk \Xi_n^2 
|\tau_\ell|^2 j_\ell^2(k \chi)\;,
\end{eqnarray}
for
\begin{eqnarray}
\Xi_\ell^2 =  \prod_{n=1}^\ell {n^2 \over (2n+1)(2n-1)} 
\left[ {1 - {K \over k^2} (n^2-1)} \right]\;,
\end{eqnarray}
and (\ref{flat-power}) to find the angular correlation function. 
It is more straight-forward to redefine the mode function $G_l[Q]$ in
the mode function expansion, that is we use $M_\ell[Q] = (\Xi_\ell)^{-1}
G_\ell[Q]$ instead of $G_\ell[Q]$. This then allows us to retain the 
flat like form for the angular correlation function (\ref{flat-power}), 
however we must retain the flat mode eigenfunction normalization:
\begin{eqnarray}
C_l = {16 \pi^2} \left[ {\frac{K}{\rho^{-1}} 
{\Omega_m \over (\Omega_m-1)}} \right]^2  
\int \frac{d \nu}{\nu^2} {\cal P}(\nu) j^2_l(\nu \chi)\;, 
\label{Cl-open}
\end{eqnarray}
where $\nu^2 = k^2 -1$. 
 
The angular power spectrum can then be normalized to a given structure
formation theory such as {\it standard-CDM}. It would seem that the 
favoured model (by current observational limits) is the {\it $\Lambda$ CDM}
model \cite{CHLineweaver}.

Also, in the open case from (\ref{Cl-open} and \ref{Cl-CDM-smls}):
\begin{eqnarray}
D_{\ell} = \left( {K \Omega_m \over a^2 H^2 (\Omega_m-1)} \right)^2 \int 
{d \nu \over \nu^2} {\cal P}^* (\nu) j_{\ell}^2 (\nu \chi)\;.
\end{eqnarray}
Here ${\cal P}^*$ (\ref{pstar}) is defined as before, while the mode 
expansion has been carried out in terms of $M_{\ell}[Q]$ rather 
than $G_{\ell}[Q]$ \cite{GE}.

\end{document}